\documentclass[a4paper,11pt]{article}
\pdfoutput=1
\usepackage{jheppub} 

\usepackage[T1]{fontenc}
\usepackage{amssymb}
\usepackage{amsmath}
\usepackage{color}
\usepackage{xspace}
\usepackage{appendix}
\usepackage{float}
\usepackage{subfigure}
\usepackage{graphicx}
\usepackage{hyperref}
\usepackage{chngcntr}
\usepackage{diagbox}
\usepackage{wasysym}
\usepackage{multirow}
\usepackage[utf8x]{inputenc} 
\newcounter{JW}

\newcommand{\bqa}{\begin{eqnarray}}
\newcommand{\eqa}{\end{eqnarray}}
\newcommand{\nn}{\nonumber}

\newcommand{\plaat}[3]{\raisebox{#3pt}{\epsfig{figure=./figures/#1.pdf,
width=#2\textwidth}}}

\def\as{\alpha_s}

\def\mglong{{\sc\small MadGraph5\_aMC@NLO}}
\def\mgshort{{\sc\small MG5\_aMC}}

\chardef\MyArticleWithColor=\pdfcolorstackinit page direct{0 g}

\title{The gluon-fusion production of Higgs boson pair: N$^3$LO QCD corrections and top-quark mass effects}
\author[a]{Long-Bin Chen,}

\affiliation[a]{School of Physics and Electronic Engineering, Guangzhou University, Guangzhou 510006, China}
\author[b]{Hai Tao Li,}
\affiliation[b]{Theoretical Division, Los Alamos National Laboratory, Los Alamos, NM, 87545, USA}
\author[c]{Hua-Sheng Shao,}

\affiliation[c]{Laboratoire de Physique Th\'eorique et Hautes Energies (LPTHE), UMR 7589, Sorbonne Universit\'e et CNRS, 4 place Jussieu, 75252 Paris Cedex 05, France}
\author[d]{Jian Wang}
\affiliation[d]{School of Physics, Shandong University, Jinan, Shandong 250100, China}

\emailAdd{chenlb@gzhu.edu.cn }
\emailAdd{haitaoli@lanl.gov}
\emailAdd{huasheng.shao@lpthe.jussieu.fr}
\emailAdd{j.wang@sdu.edu.cn}

\abstract{
    The Higgs boson pair production via gluon fusion at high-energy hadron colliders, such as the LHC, is vital in deciphering the Higgs potential and in pinning down the electroweak symmetry breaking mechanism. We carry out the next-to-next-to-next-to-leading order (N$^3$LO) QCD calculations in the infinite top-quark mass limit and present predictions for both the inclusive and differential cross sections, albeit the differential distributions other than the invariant mass distribution of the Higgs boson pair are approximated at N$^3$LO. Such corrections are indispensable in stabilising the perturbative expansion of the cross section in the strong coupling $\alpha_s$. At the inclusive level, the scale uncertainties are reduced by a factor of four compared with  the next-to-next-to-leading order (NNLO) results. Given that the inclusion of the top-quark mass effects is essential for the phenomenological applications,
we use several schemes to incorporate the N$^3$LO results in the infinite top-quark mass limit and the next-to-leading order (NLO) results with full top-quark mass dependence, and present  theoretical predictions for the (differential) cross sections in the proton-proton collisions at the centre-of-mass energies $\sqrt{s}=13,14,27$ and $100$ TeV. 
    Our results provide one of the most precise theoretical inputs for the analyses of the Higgs boson pair events.
}

\begin{document}
\maketitle
\flushbottom

\section{Introduction}

In view of the null results in the beyond the Standard Model (BSM) searches so far at colliders, it seems that a realistic way of looking for new physics in the future is to precisely study the nature of the Higgs sector. Any small deviation with respect to the Standard Model (SM) predictions would indicate the signal of  new physics. In particular, the electroweak symmetry breaking mechanism remains to be understood. It can be deciphered by specifying the form of the Higgs potential. In the SM, such a potential is  determined by two SU(2)$_L\times$U(1)$_Y$ gauge invariant renormalisable operators constructed from a single Higgs SU(2)$_L$ doublet
$H=\left( H^+ , H^0\right)^{T}$, 
i.e.
\begin{eqnarray}
V(H)=-\mu^2H^\dagger H+\lambda^{\rm SM} \left(H^\dagger H\right)^2,\quad \mu^2>0,\quad \lambda^{\rm SM}>0.
\end{eqnarray}
This Higgs potential has a well-known shape of a ``Mexican hat''. The spontaneous symmetry breaking happens after the Higgs field captures a non-vanishing vacuum expectation value $v$, which is  related to the Fermi constant $G_F$ via $v=\left(\sqrt{2}G_F\right)^{-1/2}=\left(\mu^2/\lambda^{\rm SM}\right)^{1/2}$. The quantum fluctuation of the real scalar field  around the minimum value of the potential $V(H)$ at $H_0=\left( 0 ,v/\sqrt{2}\right)^T$ represents  a physical Higgs boson $h$. The  Higgs boson mass at tree-level is given by $m_h^2=2\mu^2$, and the Higgs self-interactions become
\begin{eqnarray}
V(h)=\frac{m_h^2}{2}h^2+\lambda^{\rm SM} v h^3+\frac{1}{4}\lambda^{\rm SM} h^4.
\end{eqnarray}
One can see that the Higgs potential in the SM is fully determined by $G_F$ and $m_h$, whose values have been measured precisely~\cite{Tanabashi:2018oca}. Therefore, independent measurements on the Higgs trilinear and quartic couplings are  very important to test the SM Higgs sector. In fact, several UV-complete new physics models predict modifications of the Higgs potential and the Higgs trilinear coupling $\lambda_{hhh}$~\cite{deFlorian:2016spz,DiMicco:2019ngk}. Some of them (see e.g. refs.~\cite{Kanemura:2002vm,OConnell:2006rsp,Jurciukonis:2018skr}) can possess very different $\lambda_{hhh}$ value from the SM expectation $\lambda^{\rm SM}_{hhh}=\lambda^{\rm SM}=m_h^2/2v^2$ but still have SM-compatible Higgs interactions with the massive gauge bosons and fermions. The measurement of the Higgs self couplings seems the only way to understand the dynamics of electroweak symmetry spontaneously breaking.

The Higgs trilinear coupling can be either directly probed via the Higgs boson pair production or indirectly constrained by using the loop effects in the precision observables (e.g. the single Higgs boson signal strengths at the LHC~\cite{Gorbahn:2016uoy,Degrassi:2016wml,Bizon:2016wgr,Maltoni:2017ims,DiVita:2017eyz,Degrassi:2019yix} or at an e$^+$e$^-$ collider~\cite{McCullough:2013rea}, the electroweak oblique parameters~\cite{Kribs:2017znd}, or the $W$ boson mass and the effective sine~\cite{Degrassi:2017ucl}). The existing direct measurements of the Higgs pair cross sections at the LHC only loosely bound $\lambda_{hhh}$~\cite{Aad:2019uzh,Sirunyan:2018two} due to the low statistics. The current best constraint $-5<\lambda_{hhh}/\lambda_{hhh}^{\rm SM}<12$ at 95\% confidence level (CL) is from the ATLAS collaboration with 36.1 fb$^{-1}$ Run-2 data~\cite{Aad:2019uzh}. The situation will be largely improved at the phase of the HL-LHC with 3000 fb$^{-1}$ integrated luminosity~\cite{Cepeda:2019klc}. Meanwhile,  novel analysis techniques (e.g. new kinematic variables~\cite{Kim:2018cxf} or machine learning~\cite{Capozi:2019xsi}) have been proposed to expedite the discovery. In addition, the envisaged future hadron colliders, like the FCC-hh, are expected to be the ultimate precision machines for determining $\lambda_{hhh}$~\cite{Contino:2016spe}, strongly gaining from both the 20 times bigger cross section and the higher integrated luminosity. 

Although this process is mainly limited by the low statistics at the moment, the continuous measurements at the LHC are still quite valuable, because even the loose bounds can already exclude some new physics models or corner the parameter space, which predicts the enhanced yields of $pp\rightarrow hh$ (see e.g. \cite{Cacciapaglia:2017gzh}). The indirect constraints on $\lambda_{hhh}$ from the single-Higgs data have been set in the range $\lambda_{hhh}/\lambda_{hhh}^{\rm SM} \in [-3.2,11.9]$ at 95\% CL with  79.8 fb$^{-1}$ Run-2 data by  ATLAS~\cite{ATL-PHYS-PUB-2019-009}. These constraints are already comparable with the direct ones and impact the final bounds with the combinations of the direct and indirect measurements~\cite{ATLAS:2019pbo}. As opposed to the direct bounds, the improvements of the indirect bounds are limited by the systematics and thus will be harder at the HL-LHC. Nevertheless, these indirect approaches feature different systematics than  direct measurements and can be thought as independent cross checks. On the other hand, the extraction of the quartic Higgs self-coupling from the triple Higgs production is much difficult (though not hopeless) at hadron colliders, because the corresponding cross sections are three orders of magnitude smaller than the double-Higgs production~\cite{Maltoni:2014eza}.

\begin{figure}[h]
\centering
 \includegraphics[width=0.48\linewidth]{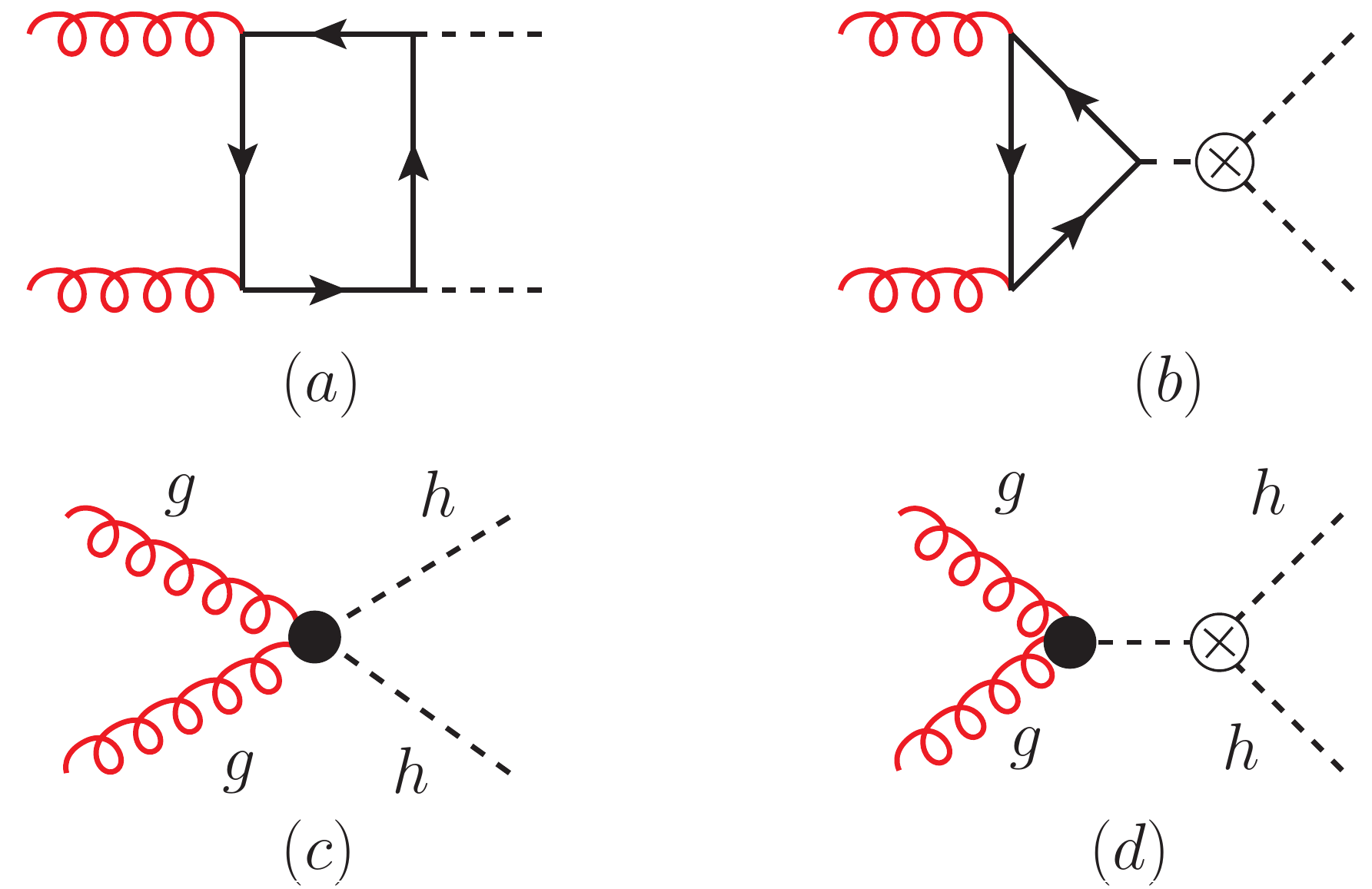}
  \caption{The LO Feynman diagrams of the process  $gg\to hh$ with full top-quark mass dependence (first row) and in the infinite top-quark mass limit (second row).} 
  \label{fig:lo}
\end{figure}

Similar to the single Higgs hadroproduction, the dominant di-Higgs production channel at a high-energy hadron collider is via the gluon-gluon fusion (ggF)~\cite{Baglio:2012np,Frederix:2014hta,DiMicco:2019ngk}. Other channels are at least one order of magnitude lower in their yields. Due to the absence of the tree-level interactions between the Higgs boson and gluons in the SM, the leading order (LO) cross section $\sigma(gg\rightarrow hh)$ was computed from one-loop amplitude squared~\cite{Eboli:1987dy,Glover:1987nx,Plehn:1996wb}, where two representative LO Feynman diagrams can be seen in the first row of figure \ref{fig:lo}. Further improvements of the fixed-order perturbative calculations without any approximation are quite challenging. The full next-to-leading order (NLO)  QCD calculations involving complicated two-loop Feynman integrals were carried out only recently~\cite{Borowka:2016ehy,Borowka:2016ypz,Baglio:2018lrj,Davies:2019dfy} thanks to the new advances of the numerical approaches~\cite{Borowka:2015mxa,Li:2015foa,Dick2004493}. The NLO results were complemented with soft-gluon resummation~\cite{Ferrera:2016prr} or parton-shower (PS) effects~\cite{Heinrich:2017kxx,Jones:2017giv,Heinrich:2019bkc}. The ggF NLO predictions are plagued with the large theoretical uncertainties from the scale variations~\cite{Borowka:2016ehy} and the top-quark mass scheme dependence~\cite{Baglio:2018lrj}. Moreover, at NLO+PS, some differential distributions (e.g. the distribution at large transverse momentum of the Higgs boson pair) differ significantly by adopting different matching schemes~\cite{Heinrich:2017kxx} or shower scales~\cite{Jones:2017giv}.

Instead of starting from the loop-induced process, one can also carry out the heavy top-quark mass $m_t$ expansions in the amplitudes. We refer to the leading expansion term in $1/m_t^2$ as the infinite top-quark mass limit $m_t\to +\infty$. In such an approximation, the two Higgs bosons can be generated by the two gluon scatterings at tree level (see the second row of figure \ref{fig:lo}), which makes the higher-order perturbative calculations more feasible. The first NLO computation in the $m_t\to +\infty$ limit was performed two decades ago~\cite{Dawson:1998py}. Next-to-next-to-leading order (NNLO) was also available~\cite{deFlorian:2013uza,deFlorian:2013jea,Grigo:2014jma,deFlorian:2016uhr}, and recently we have presented the first next-to-next-to-next-to-leading order (N$^3$LO) calculation~\cite{Chen:2019lzz}. Besides these fixed-order results, the soft-gluon resummation effects are also considered in refs.~\cite{Shao:2013bz,deFlorian:2015moa,deFlorian:2018tah}. In spite of the success of improving the perturbative accuracy in the cross section calculations, it is widely acknowledged that the $m_t\to +\infty$ approximation is insufficient for the phenomenological applications. Many theoretical efforts have been devoted to investigate the finite $m_t$ corrections to this approximation~\cite{Grigo:2013rya,Frederix:2014hta,Maltoni:2014eza,Grigo:2015dia,Degrassi:2016vss,Grazzini:2018bsd,Davies:2019xzc}.
Moreover, there are also many well-motivated attempts to evaluate the involved two-loop $gg\rightarrow hh$ amplitudes in the analytic forms by taking other approximations (e.g. in the small top-quark mass~\cite{Davies:2018ood,Davies:2018qvx}, the small Higgs transverse momentum~\cite{Bonciani:2018omm} and the small Higgs mass~\cite{Xu:2018eos} limits).

The primary goal of this paper is to extend our previous N$^3$LO results in ref.~\cite{Chen:2019lzz} and to include the top-quark mass effects for the phenomenological applications. The remaining context is organised as follows. In section~\ref{sec:n3lo}, after the description of our method, we  provide the validation of our calculations as well as the extensive numerical results in the infinite top-quark mass limit. We  take into account the finite $m_t$ effects at N$^3$LO based on the NLO QCD results with full $m_t$ dependence in section \ref{sec:n3lomt}. The conclusion is drawn in section \ref{sec:summary}. Additional results and some technical details can be found in the appendices. The hard functions, in particular the new one-loop analytic expressions, are shown in appendix \ref{app:hard}. An NLO model and the $R_2$ Feynman rules are described in appendix \ref{app:model}. The renormalisation scale dependence in the N$^3$LO results is discussed in appendix \ref{app:scale}. Finally,  appendix \ref{app:addplots}  collects the additional plots for the differential distributions.

\section{N$^3$LO corrections in the infinite top-quark mass limit\label{sec:n3lo}}

\subsection{Effective Lagrangian and Wilson coefficients}

The interactions between the Higgs bosons and gluons are mainly generated by top-quark loops, where two LO Feynman diagrams are shown in the first row of figure~\ref{fig:lo}. The effective Lagrangian in the infinite top-quark mass limit is obtained through integrating out the top-quark loop contribution (see the second row of figure~\ref{fig:lo}). For the Higgs boson pair production, the relevant effective Lagrangian can be written as  
\begin{align}\label{eq:effL}
 \mathcal{L}_{\rm eff}&=\frac{\as}{12\pi }\left[
  (1+\delta)\ln \left( 1+\frac{h}{v}\right) 
  -\frac{\eta}{2}\ln^2  \left( 1+\frac{h}{v}\right) 
 \right]
 G_{\mu\nu}^a G^{a~\mu\nu}\\\nonumber
 &=-\frac{1}{4} \left(C_h \frac{h}{v}-C_{hh}\frac{h^2}{2v^2}  \right)G_{\mu\nu}^a G^{a~\mu\nu}+\mathcal{O}(h^k,k\geq 3),
\end{align}
where $\alpha_s$ is the strong coupling and $G_{\mu\nu}^a$ is the gluon field strength tensor. On the right hand side of the second equation, $\mathcal{O}(h^k,k\geq 3)$ means that we have ignored terms involving more than two Higgs bosons in the effective Lagrangian. The Wilson coefficients $\delta$ and $\eta$, or equivalently $C_h$ and $C_{hh}$,  comprise the QCD radiative corrections of the top-quark loops. $C_h$ and $C_{hh}$ can be easily derived in terms of $\delta$ and $\eta$ as
\begin{eqnarray}
C_{h}&=&-\frac{\alpha_s}{3\pi}\left(1+\delta\right), \qquad C_{hh}=-\frac{\alpha_s}{3\pi}\left(1+\delta+\eta\right).
\end{eqnarray}
These Wilson coefficients can be perturbatively expanded in a series of $\alpha_s$,
\begin{eqnarray}
    \delta= \sum_{i=0}{\left(\frac{\alpha_s}{4\pi}\right)^i\delta^{(i)}}\,, & \quad & \eta=\sum_{i=0}{\left(\frac{\alpha_s}{4\pi}\right)^i\eta^{(i)}}\,, 
    \nonumber \\ 
    C_h= -\frac{\as}{3\pi } \sum_{i=0}{\left(\frac{\alpha_s}{4\pi}\right)^iC_{h}^{(i)}}\,, &\quad& 
    C_{hh}=- \frac{\as}{3\pi } \sum_{i=0}{\left(\frac{\alpha_s}{4\pi}\right)^iC_{hh}^{(i)}}\,.
\end{eqnarray}
Their four-loop analytic expressions  are already known in the literature~\cite{Kataev:1981gr, Inami:1982xt, Chetyrkin:1997iv, Chetyrkin:1997un, Chetyrkin:2005ia, Schroder:2005hy,Kniehl:2006bg,deFlorian:2013uza, Grigo:2014jma, Baikov:2016tgj,Spira:2016zna,Gerlach:2018hen}.
In our N$^3$LO calculations, the results up to three loops are needed. They are given in the on-shell top-quark mass scheme  by \cite{Spira:2016zna}
\begin{align}
    \delta^{(0)}=&0\,,  \qquad  \delta^{(1)} =  11 \, ,
    \nonumber \\ 
    \delta^{(2)} =&  L_t \left( 19 + \frac{16}{3} n_f \right) + \frac{2777}{18} - \frac{67}{6} n_f \, ,
    \nonumber \\ 
    \delta^{(3)} =&  L_t^2 \left( 209 + 46 n_f - \frac{32}{9} n_f^2 \right) + L_t \left( \frac{4834}{9} + \frac{2912}{27} n_f + \frac{77}{27} n_f^2 \right) 
    \nonumber \\ & -\frac{2761331}{648} + \frac{897943\zeta_3}{144}  + \left( \frac{58723}{324} - \frac{110779\zeta_3}{216} \right) n_f - \frac{6865}{486} n_f^2 \,, 
\end{align}
and 
\begin{align}
    \eta^{(0)} =& 0\, , \qquad \eta^{(1)} = 0 \, ,
    \nonumber \\
    \eta^{(2)} =& \frac{32 n_f }{3}+\frac{70}{3}\, ,
    \nonumber \\
    \eta^{(3)} =&      L_t \left( -\frac{128 n_f^2}{9}+\frac{1528 n_f}{9}+\frac{2356}{3} \right)+
\frac{154 n_f^2}{27}+\frac{4324 n_f}{27}+\frac{5332}{27}\, ,
\end{align}
where $L_t=\ln(\mu_R^2/m_t^2)$, $m_t$ is the top-quark pole mass, $\mu_R$ is the renormalisation scale and $n_f$ is the number of the light-quark flavours. 
The expressions of $C_h$ and $C_{hh}$ are
\begin{eqnarray}
    C_h^{(0)}&=1,\quad C_{hh}^{(0)}&=C_h^{(0)}, \\\nonumber
    C_h^{(i)}&=\delta^{(i)},\quad C_{hh}^{(i)}&=C_h^{(i)}+\eta^{(i)}, i\geq 1.
\end{eqnarray}

\subsection{Breakdown in three channels}

\begin{figure}[hbt!]
    \centering
    \includegraphics[scale=0.3]{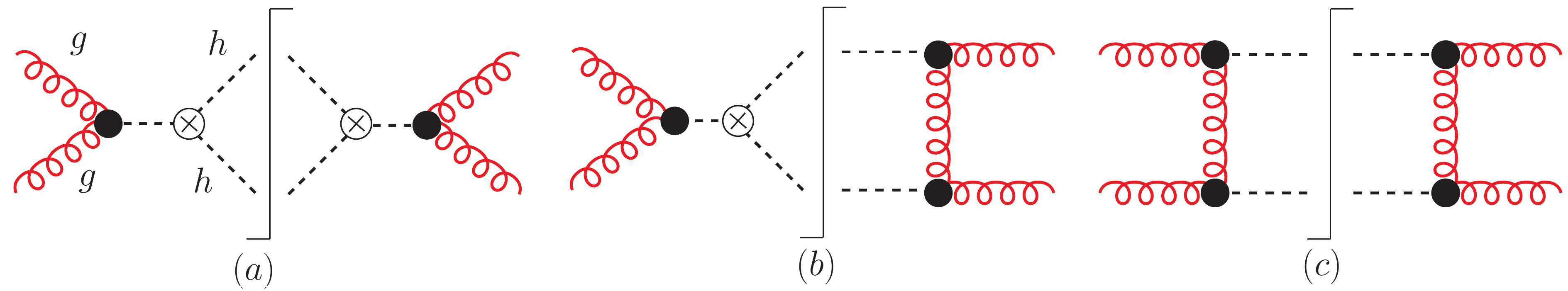}
    \vspace{-0.2cm}
    \caption{Representative Born cut-diagrams for the Higgs boson pair production in the effective theory. The cross section can be classified by the number of effective vertices between the two Higgs bosons and gluons.  }
    \label{fig:FeynDia}
\end{figure}

The ggF  Higgs boson pair production in the infinite top-quark mass limit with the effective Lagrangian defined in eq.(\ref{eq:effL}) can be divided into three  channels according to the number of effective vertices at the squared amplitude level. Three representative Born cut-diagrams are shown in figure~\ref{fig:FeynDia}. There are two (class-$a$), three (class-$b$) and four (class-$c$) effective vertices insertions respectively. In other words, the double-Higgs (differential) cross section can be decomposed into
\begin{align}
    d\sigma_{hh} = d\sigma^{a}_{hh} + d\sigma^{b}_{hh} + d\sigma^{c}_{hh} .
\end{align}
Because there are at least one $\alpha_s$ power in the Wilson coefficients $C_h$ and $C_{hh}$, their Born cross sections contribute to different $\alpha_s$ orders, which are summarised in table~\ref{tab:my_label}. The lowest orders of class-$a$, -$b$ and -$c$ are $\mathcal{O}(\alpha_s^2)$, $\mathcal{O}(\alpha_s^3)$ and $\mathcal{O}(\alpha_s^4)$ respectively, which means that they contribute to LO, NLO and NNLO parts of the Higgs boson pair cross section. For the purpose of N$^3$LO calculations in the present paper, we need to calculate N$^3$LO, NNLO and NLO corrections to the class-$a$, -$b$ and -$c$ part, respectively.

\begin{table}[hbt!]
    \centering
    \begin{tabular}{|c|c|c|c|c|}
    \hline
        &  LO  & NLO  & NNLO  & N$^3$LO  \\
    \hline
    total  &  $\mathcal{O}(\alpha_s^2)$  &   $\mathcal{O}(\alpha_s^3)$ &   $\mathcal{O}(\alpha_s^4)$  & $\mathcal{O}(\alpha_s^5)$
     \\
    \hline
     class-$a$  &  $\mathcal{O}(\alpha_s^2)$  &   $\mathcal{O}(\alpha_s^3)$ &   $\mathcal{O}(\alpha_s^4)$  & $\mathcal{O}(\alpha_s^5)$
     \\
    \hline
     class-$b$   &  0 &   $\mathcal{O}(\alpha_s^3)$ &   $\mathcal{O}(\alpha_s^4)$  & $\mathcal{O}(\alpha_s^5)$
     \\
    \hline
     class-$c$   & 0 &  0 &   $\mathcal{O}(\alpha_s^4)$  & $\mathcal{O}(\alpha_s^5)$
     \\
     \hline
    \end{tabular}
    \caption{The perturbative orders in $\alpha_s$ for different classes at the amplitude squared level.
    We call the $\mathcal{O}(\alpha_s^3)$ contribution in class-$b$ as the LO in this class
    though it is an NLO correction to the cross section of Higgs pair production.
    The same rule applies to the class-$c$ part. }
    \label{tab:my_label}
\end{table}

\subsection{Methodology and validation}

\subsubsection{The class-$a$ part\label{sec:classa}}

We have two approaches to compute NNLO (i.e. up to $\mathcal{O}(\alpha_s^4)$) cross section in the class-$a$ part. The first one is that we can perform a fully-differential NNLO calculation based on the $q_T$-subtraction method, which was originally proposed in ref.~\cite{Catani:2007vq}.~\footnote{With $q_T$-subtraction method, tremendous works have been done at the NNLO accuracy~\cite{Catani:2007vq,Catani:2009sm,Catani:2011qz,Gao:2012ja,Cascioli:2014yka,Ferrera:2014lca,Gehrmann:2014fva,
Boughezal:2015dva,Boughezal:2015aha,Gaunt:2015pea,Boughezal:2015ded,Campbell:2016jau,Campbell:2016yrh,Grazzini:2016swo,Grazzini:2016ctr,Boughezal:2016wmq,Berger:2016oht,deFlorian:2016uhr,Li:2016nrr,Li:2017lbf}. 
Through solving the renormalisation equations up to N$^3$LO, the small $q_T$ cross section has also been studied at N$^3$LO for certain processes~\cite{Chen:2018pzu,Cieri:2018oms, Billis:2019vxg} with constant terms missing at three loops in the collinear sector.} In this paper, we will use the 
$q_T$-subtraction method in the framework of the soft-collinear effective theory (SCET)~\cite{Bauer:2000ew,Bauer:2000yr,Bauer:2001ct,Bauer:2001yt,Beneke:2002ph}. In this approach, the class-$a$ (differential) cross section can be further divided into
\begin{align}
d\sigma_{hh}^{a} = d\sigma_{hh}^{a}\Big|_{p_T^{hh}<p_T^{\rm veto}} + d\sigma_{hh}^{a}\Big|_{p_T^{hh}>p_T^{\rm veto}},
\label{eq:cutoffa}
\end{align}
where $p_T^{hh}$ is the transverse momentum of the Higgs pair system, i.e. $q_T=p_T^{hh}$. The first (second) term on the right-hand side of eq.(\ref{eq:cutoffa}) is imposed the kinematic cut $p_T^{hh}<p_T^{\rm veto}$ ($p_T^{hh}>p_T^{\rm veto}$).

The first piece $d\sigma_{hh}^{a}\Big|_{p_T^{hh}<p_T^{\rm veto}}$ is computed with the aid of the transverse-momentum resummation formalism in SCET. The cross section of this part is factorised as a convolution of the hard, beam and soft functions 
\begin{align}\label{eq:qt_smalla}
   \frac{d\sigma^{a}_{hh}}{dp_T^{hh}} =  H^{a}\otimes B_g \otimes B_g \otimes S \times \left( 1+  \mathcal{O}\left(\frac{\left(p_T^{hh}\right)^2}{Q^2}\right)\right),
\end{align}
where we have ignored the power-suppressed terms $\mathcal{O}\left(\left(\frac{p_T^{hh}}{Q}\right)^2\right)$. Such a factorisation formalism holds when $p_T^{hh}$ is sufficiently smaller than the hard scale $Q$, which is derived by studying  the IR  behaviour of QCD. The transverse-momentum dependent (TMD) gluon beam function $B_g$ is universal in the sense that it is independent of the process but only relies on the species of the initial state (i.e. gluon). The soft function $S$ is also the same for all processes only involving colourless final states with the gluon-gluon initial state. The calculations of the TMD beam and soft functions can be carried out with a rapidity regulator, while the physical results are independent of the choice of such a regulator. The two-loop analytic results for these TMD beam and soft functions can be found in~\cite{Gehrmann:2012ze, Gehrmann:2014yya,Luebbert:2016itl,Echevarria:2016scs,Luo:2019hmp, Luo:2019bmw},
and the N$^3$LO results have been obtained very recently~\cite{Li:2016ctv,Luo:2019szz}. On the other hand, the hard function $H^{a}$ is process dependent. The detailed discussions about the hard functions can be found in  appendix~\ref{app:hard}.

Due to the non-vanishing transverse momentum of the Higgs pair system $p_T^{hh}>p_T^{\rm veto}$, only the events with additional jets will be maintained in the second piece of eq.(\ref{eq:cutoffa}) $d\sigma_{hh}^{a}\Big|_{p_T^{hh}>p_T^{\rm veto}}$. In our case, the NNLO computation of class-$a$ requires us to calculate the NLO corrections to a Higgs pair plus a jet with two effective vertices insertions. Such a task can be carried out by using the automated simulation framework \mglong\ (\mgshort\ henceforth)~\footnote{Let us briefly describe the framework here. The computations of one-loop amplitudes are carried out in the module {\sc\small MadLoop}~\cite{Hirschi:2011pa,Alwall:2014hca} by exploiting {\sc\small Collier}~\cite{Denner:2016kdg} package, while the real-emission parts are evaluated with the module {\sc\small MadFKS}~\cite{Frederix:2009yq,Frederix:2016rdc} by using the FKS infrared (IR) subtraction method~\cite{Frixione:1995ms,Frixione:1997np}.}~\cite{Alwall:2014hca} with an NLO Universal FeynRules Output (UFO) model~\cite{Degrande:2011ua} based on the SM Lagrangian and the effective Lagrangian eq.(\ref{eq:effL}). The details about the model, in particular the analytic expressions of the rational $R_2$ terms, can be found in appendix \ref{app:model}. Due to the different $\alpha_s$ orders in the three Born  classes, we need the recent development~\cite{Frederix:2018nkq} in \mgshort\ that is capable of handling mixed-order scenarios.

\begin{figure}[hbt!]
\centering
\includegraphics[width=.55\columnwidth]{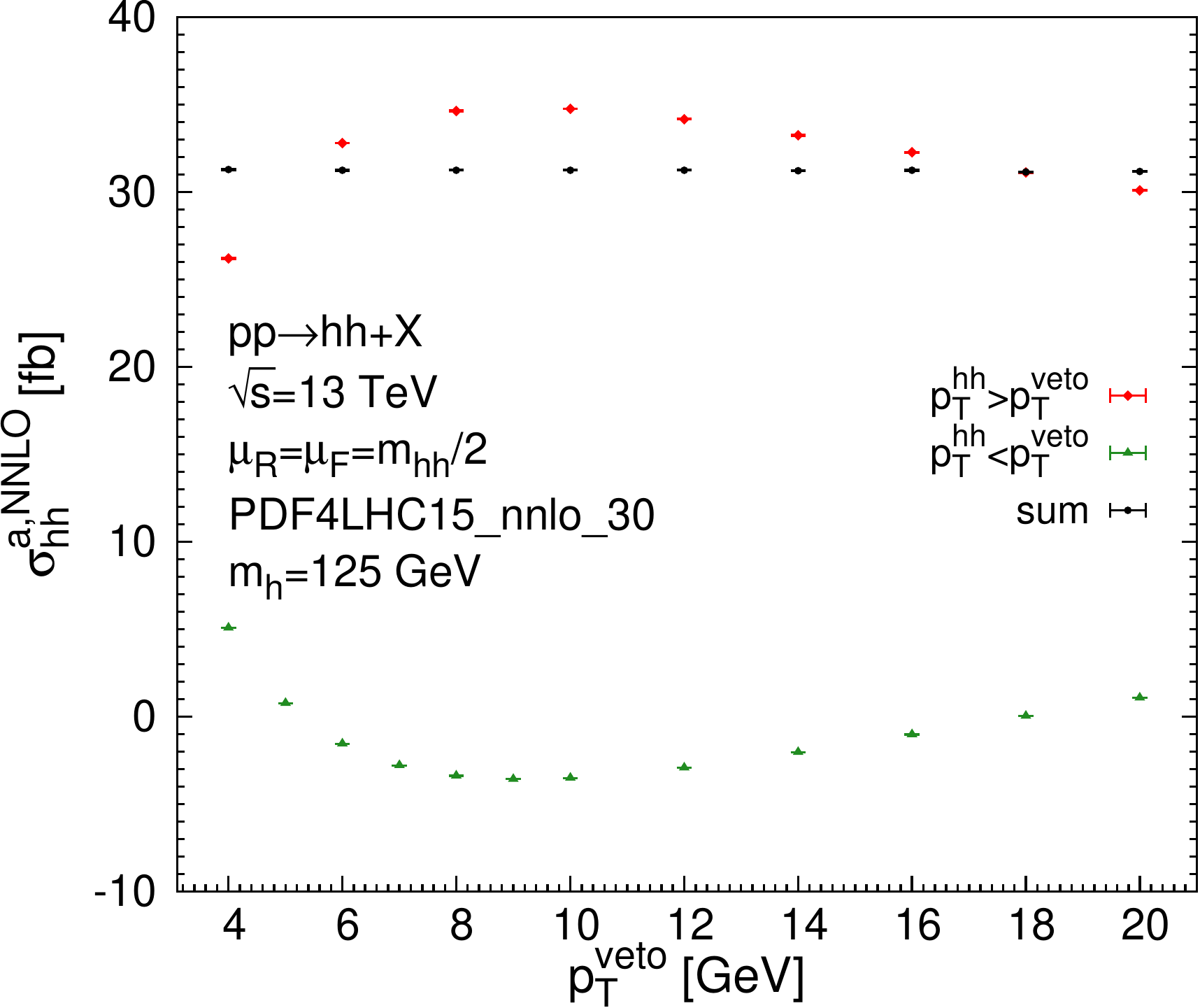}
\caption{The $p_T^{\rm veto}$ dependence of the NNLO cross section for the class-$a$ at $\sqrt{s}=13$ TeV LHC. The error bars denote the Monte Carlo integration uncertainties.}
\label{fig:ptvetoa}
\end{figure}

Within the $q_T$-subtraction approach, the independence on $p_T^{\rm veto}$ in the finite cross section should always be guaranteed when $p_T^{\rm veto}$ is approaching zero. We have explicitly checked this in the NNLO class-$a$ cross section $\sigma_{hh}^{a,{\rm NNLO}}$ shown in figure~\ref{fig:ptvetoa}.

Alternatively, the class-$a$ cross section can be related to the single Higgs production cross section, because they share exactly the same topology in the infinite top-quark mass limit. In the di-Higgs case, the class-$a$ part can be viewed as the production of an off-shell Higgs boson from ggF and its decay  into two on-shell Higgs bosons. The  off-shell Higgs boson has an invariant mass of the final-state Higgs boson pair $m_{hh}$. The explicit relation is
\begin{align}
    \frac{d\sigma^{a}_{hh}}{dm_{hh}} = f_{h\to hh} 
    \bigg(\frac{C_{hh}}{C_h} - \frac{6 \lambda_{hhh} v^2}{m_{hh}^2-m_h^2} \bigg)^2 
    \times \left(\sigma_{h}\big|_{m_h\to m_{hh}}\right)\, ,
    \label{eq:htohh}
\end{align}
where the function $f_{h\to hh}$ accounts for the phase space factor mapping from the single Higgs production to the Higgs pair production,  
\begin{align}
    f_{h\to hh} = \frac{\sqrt{m_{hh}^2- 4 m_h^2}}{16 \pi^2 v^2} ,
\end{align}
and $\sigma_{h}$ denotes the cross section for the single Higgs boson production. The replacement $m_h\rightarrow m_{hh}$ in eq.(\ref{eq:htohh}) means that the cross section is calculated with the Higgs mass  $m_{hh}$. In the first parentheses of the right-hand side of eq.(\ref{eq:htohh}), $\frac{C_{hh}}{C_h}$ accounts for the Wilson coefficient difference in figure~\ref{fig:lo}c, while the second term takes into account the propagator of the off-shell Higgs and the Higgs self-coupling in figure~\ref{fig:lo}d.
Such a method has already been used in the previous NNLO calculation of the ggF di-Higgs production in ref.~\cite{deFlorian:2013jea}.

\begin{figure}[hbt!]
\centering
\includegraphics[width=.48\columnwidth]{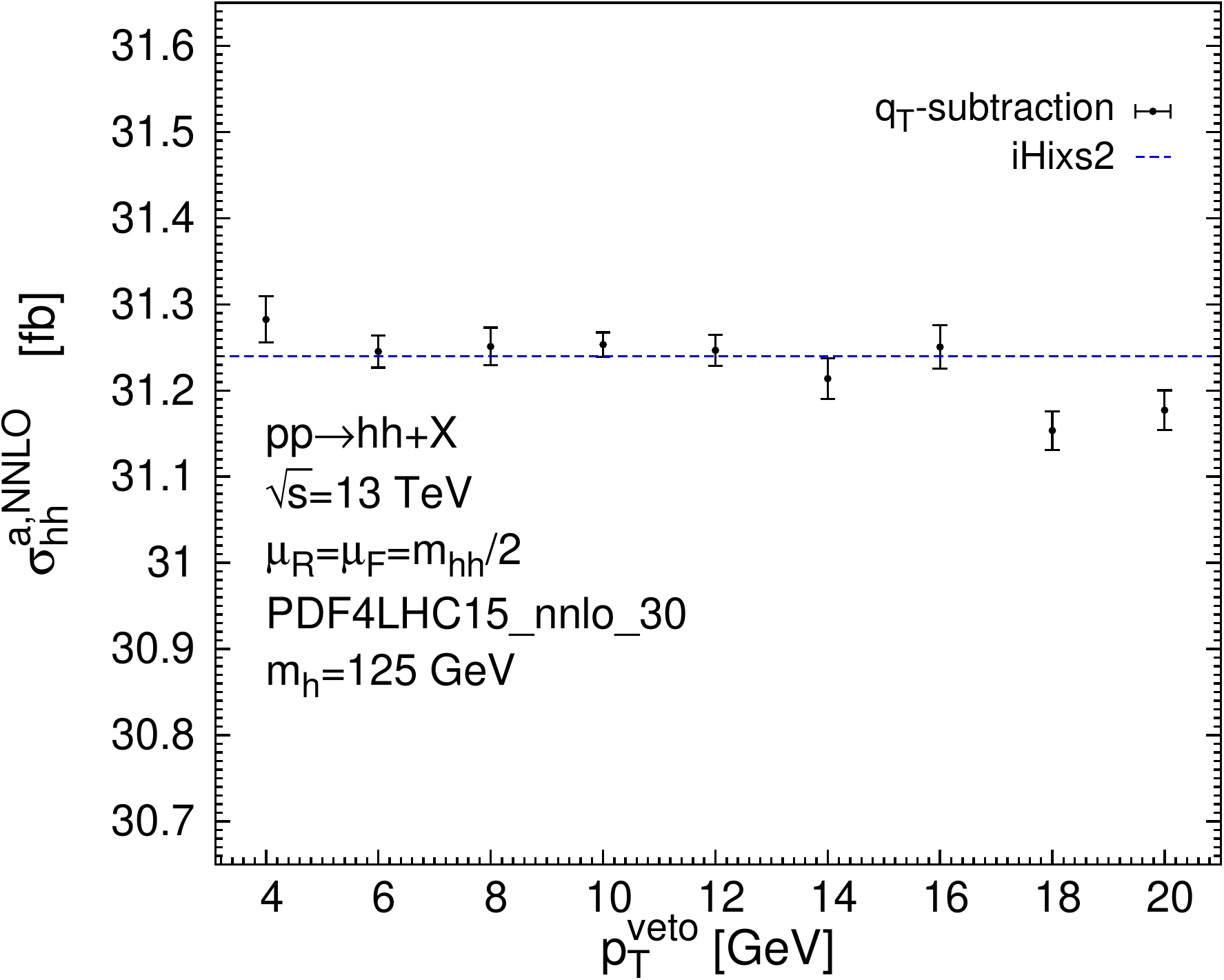}
\includegraphics[width=.48\columnwidth]{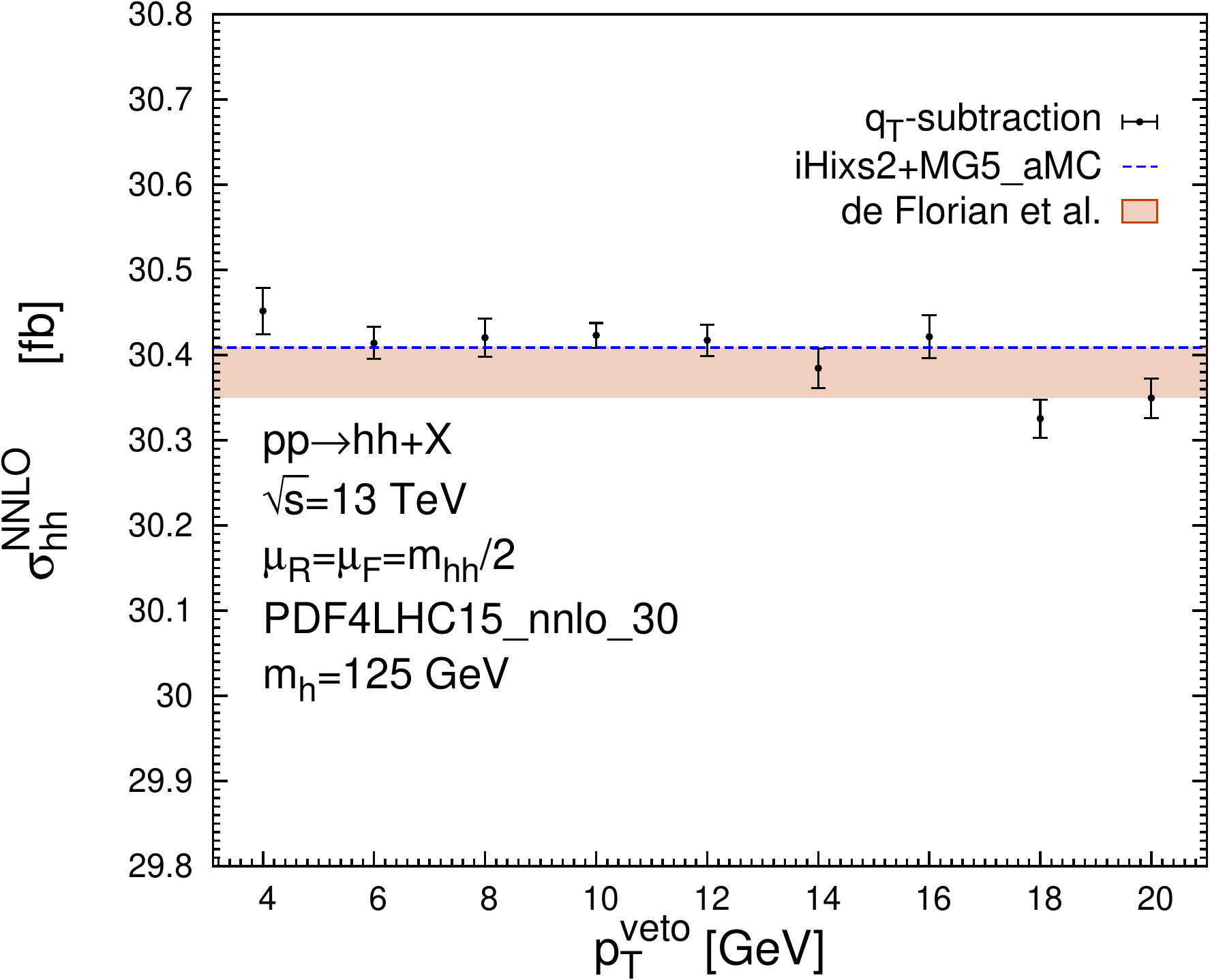}
\caption{The comparisons of the NNLO class-$a$ cross sections from two different approaches (left panel) and of the NNLO $hh$ cross sections from three different calculations (right panel) at $\sqrt{s}=13$ TeV LHC. The error bars denote the Monte Carlo integration uncertainties. In the right panel, the band represents the Monte Carlo integration error quoted in ref.~\cite{deFlorian:2016uhr}.}
\label{fig:ptvetoaNNLO}
\end{figure}

We have compared the results with the above two independent approaches for NNLO class-$a$ cross sections shown in the left panel of figure~\ref{fig:ptvetoaNNLO}. The calculation with $q_T$-subtraction matches the result by using eq.(\ref{eq:htohh}) and {\sc\small iHixs2} within the Monte Carlo integration errors when $p_T^{\rm veto}\leq 16$ GeV. Thus, we have validated eq.(\ref{eq:htohh}). After inclusion of class-$b$ and class-$c$ contributions ($\sigma^{\rm NNLO}_{hh}=\sigma^{a,{\rm NNLO}}_{hh}+\sigma^{b,{\rm NLO}}_{hh}+\sigma^{c,{\rm LO}}_{hh}$), we can compare our two calculations with the previous NNLO di-Higgs calculation in ref.~\cite{deFlorian:2016uhr}. 
As shown in the right panel of figure~\ref{fig:ptvetoaNNLO},  we have obtained perfect agreement with ref.~\cite{deFlorian:2016uhr}.

In order to compute the N$^3$LO class-$a$ cross section, we need to know the N$^3$LO cross section of $\sigma_{h}$. 
Since $\sigma_{h}$ is only known inclusively (i.e. total cross section) at N$^3$LO,
we  only perform the exact N$^3$LO calculations for the total inclusive cross sections and the invariant mass distributions of the class-$a$ part. In the present paper, we will use the public code {\sc\small iHixs2}~\cite{Dulat:2018rbf} to compute the N$^3$LO cross section $\sigma_{h}$. 

\subsubsection{The class-$b$ part\label{sec:classb}}

In order to achieve the N$^3$LO accuracy for the di-Higgs cross sections in the infinite top-quark mass limit, we have to calculate the NNLO QCD corrections to the class-$b$ part. The NNLO cross sections for the class-$b$ part were computed with the $q_T$-subtraction method similarly as described in the previous section, i.e. the differential cross section is decomposed into
\begin{align}
d\sigma_{hh}^{b} = d\sigma_{hh}^{b}\Big|_{p_T^{hh}<p_T^{\rm veto}} + d\sigma_{hh}^{b}\Big|_{p_T^{hh}>p_T^{\rm veto}},
\label{eq:cutoffb}
\end{align}
The two pieces $d\sigma_{hh}^{b}\Big|_{p_T^{hh}<p_T^{\rm veto}}$ and $d\sigma_{hh}^{b}\Big|_{p_T^{hh}>p_T^{\rm veto}}$ in eq.(\ref{eq:cutoffb}) can be computed  using the method described above  in the class-$a$ part. Therefore, we will refrain ourselves from describing them again except the hard function $H^b$ in the following equation
\begin{align}\label{eq:qt_smalla}
   \frac{d\sigma^{b}_{hh}}{dp_T^{hh}} =  H^{b}\otimes B_g \otimes B_g \otimes S \times \left( 1+  \mathcal{O}\left(\frac{\left(p_T^{hh}\right)^2}{Q^2}\right)\right).
\end{align}
The explicit expression of $H^b$ is shown in appendix \ref{app:hard}.

\begin{figure}[hbt!]
\centering
\includegraphics[width=.55\columnwidth]{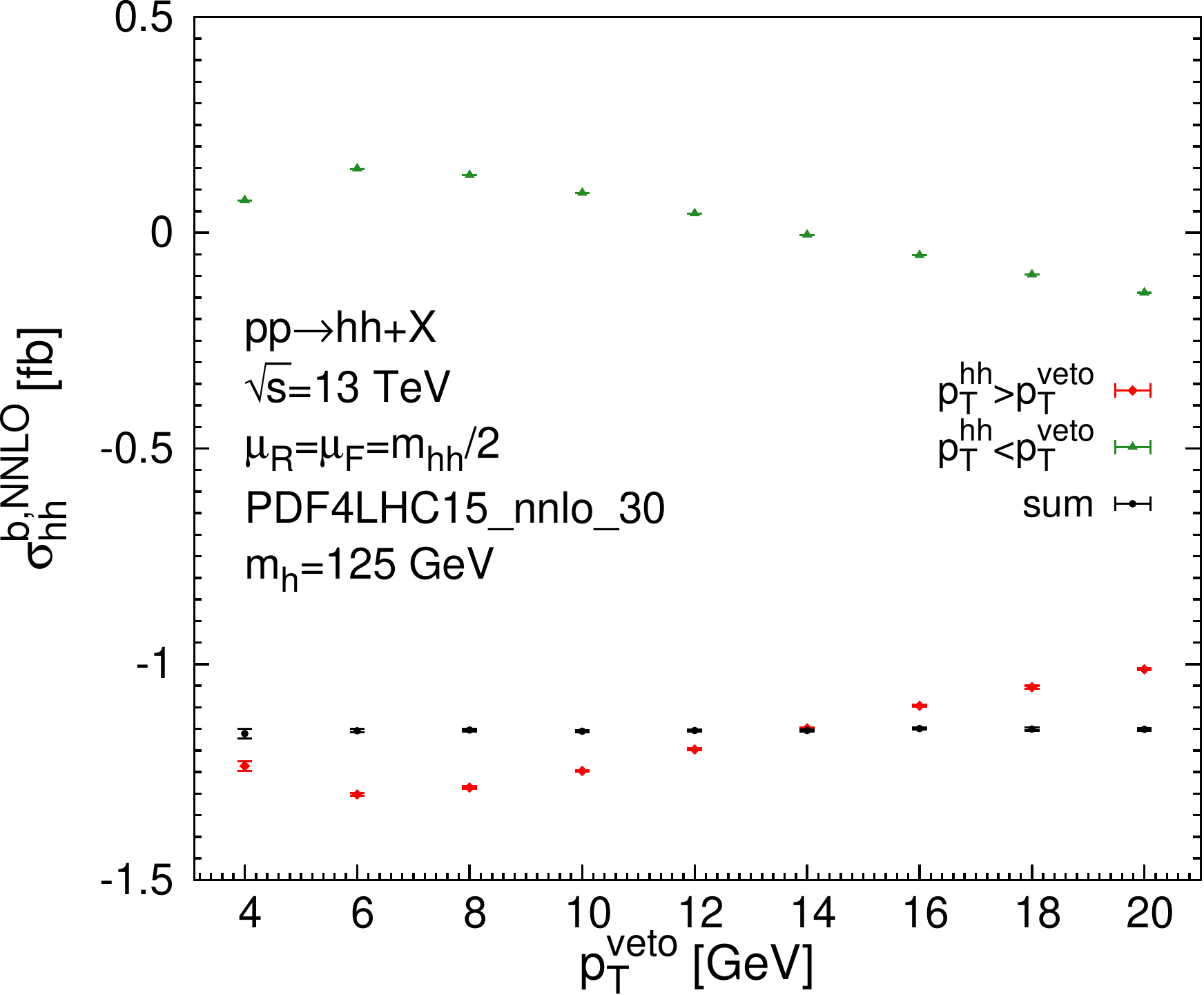}
\caption{The $p_T^{\rm veto}$ dependence of the total NNLO cross section for the class-$b$ at $\sqrt{s}=13$ TeV LHC. The error bars denote the Monte Carlo integration uncertainties.}
\label{fig:ptvetob}
\end{figure}

\begin{figure}[hbt!]
\centering
\includegraphics[width=.55\columnwidth]{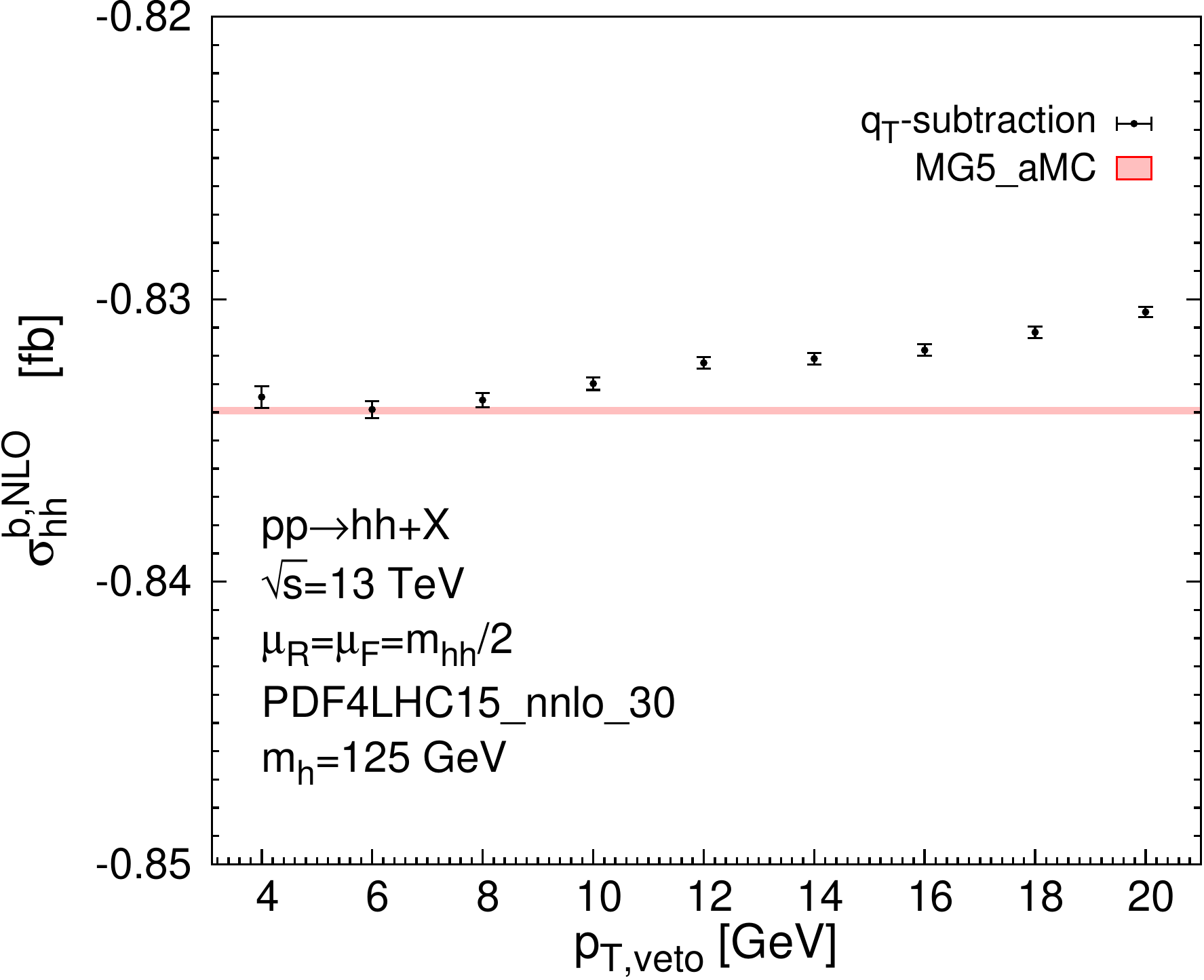}
\caption{The comparisons of the NLO class-$b$ cross sections from the $q_T$-subtraction method (error bars) and \mgshort~(red band) at $\sqrt{s}=13$ TeV LHC. The error bars and the band denote the Monte Carlo integration errors.}
\label{fig:ptvetobNLO}
\end{figure}

The $p_T^{\rm veto}$ independence of  $d\sigma_{hh}^{b}$ after summing the two pieces is explicitly verified in figure~\ref{fig:ptvetob}. As opposed to NNLO cross section of the class-$a$ part, we do not have a second independent cross section calculation for this part. The NLO cross section $\sigma_{hh}^{b,{\rm NLO}}$ however can be easily checked with \mgshort\ as shown in figure~\ref{fig:ptvetobNLO}. The perfect agreement below permille level is achieved when $p_T^{\rm veto}\leq 8$ GeV.

\begin{figure}[hbt!]
\centering
\includegraphics[width=.55\columnwidth]{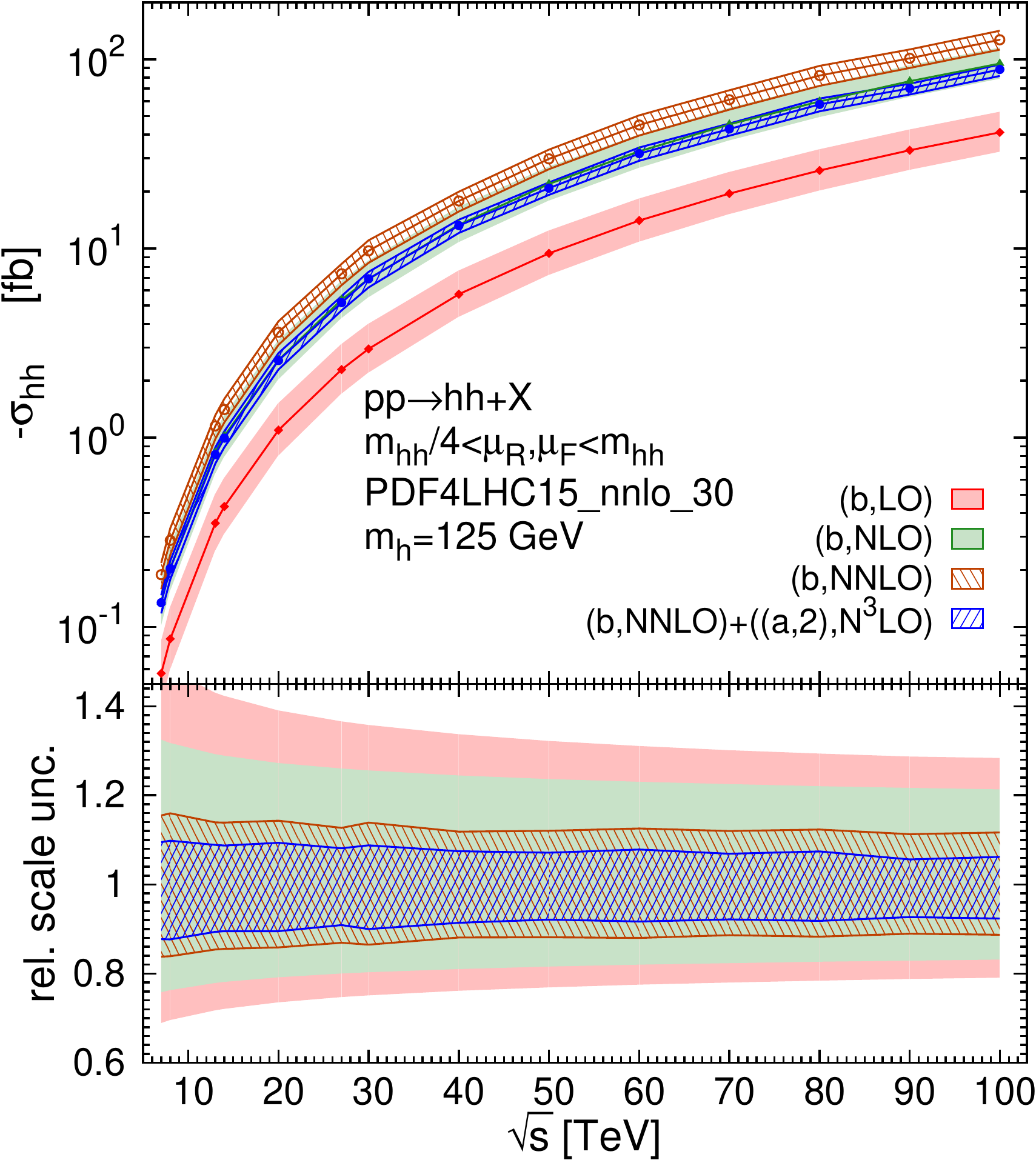}
\caption{The energy $\sqrt{s}$ dependence of the class-$b$ cross sections. They are LO (red), NLO (green), NNLO (brown hatched) and NNLO plus $\sigma_{hh}^{(a,2),{\rm N^3LO}}$ (blue hatched). The bands represent the scale uncertainties. In the lower panel, we have also shown their relative scale uncertainties.}
\label{fig:energybXS}
\end{figure}

At N$^3$LO, the renormalisation scale cancellation is guaranteed only when combining the class-$a$ and class-$b$ parts, which will be detailed in appendix \ref{app:scale}. It can serve as another powerful check to the NNLO class-$b$ cross section. The class-$a$ (differential) cross section can be decomposed into
\begin{eqnarray}
d\sigma_{hh}^{a} &=& d\sigma_{hh}^{(a,1)}+d\sigma_{hh}^{(a,2)},\nonumber\\
d\sigma_{hh}^{(a,1)} &\equiv & d\sigma_{hh}^{a}\bigg|_{C_{hh}\to C_h},\nonumber\\
d\sigma_{hh}^{(a,2)} &\equiv & d\sigma_{hh}^{a} - d\sigma_{hh}^{(a,1)},
\label{eq:classadecomp}
\end{eqnarray}
where $C_{hh}\to C_h$ means that we have replaced the Wilson coefficient $C_{hh}$ with $C_h$. The remaining renormalisation scale dependence in $d\sigma^{b,{\rm NNLO}}_{hh}$ can only be cancelled after combining with $d\sigma_{hh}^{(a,2),{\rm N^3LO}}$. In figure~\ref{fig:energybXS}, we have shown the class-$b$ cross sections multiplied by a factor of -1 from $\sqrt{s}=7$ TeV to $\sqrt{s}=100$ TeV in the upper panel. The relative scale uncertainties are displayed in the lower panel. We have indeed seen that the inclusion of $d\sigma_{hh}^{(a,2),{\rm N^3LO}}$ in the NNLO class-$b$ cross sections (the blue hatched) can further reduce the scale uncertainties.

\subsubsection{The class-$c$ part\label{sec:classc}}

\begin{figure}[hbt!]
\centering
\includegraphics[width=.55\columnwidth]{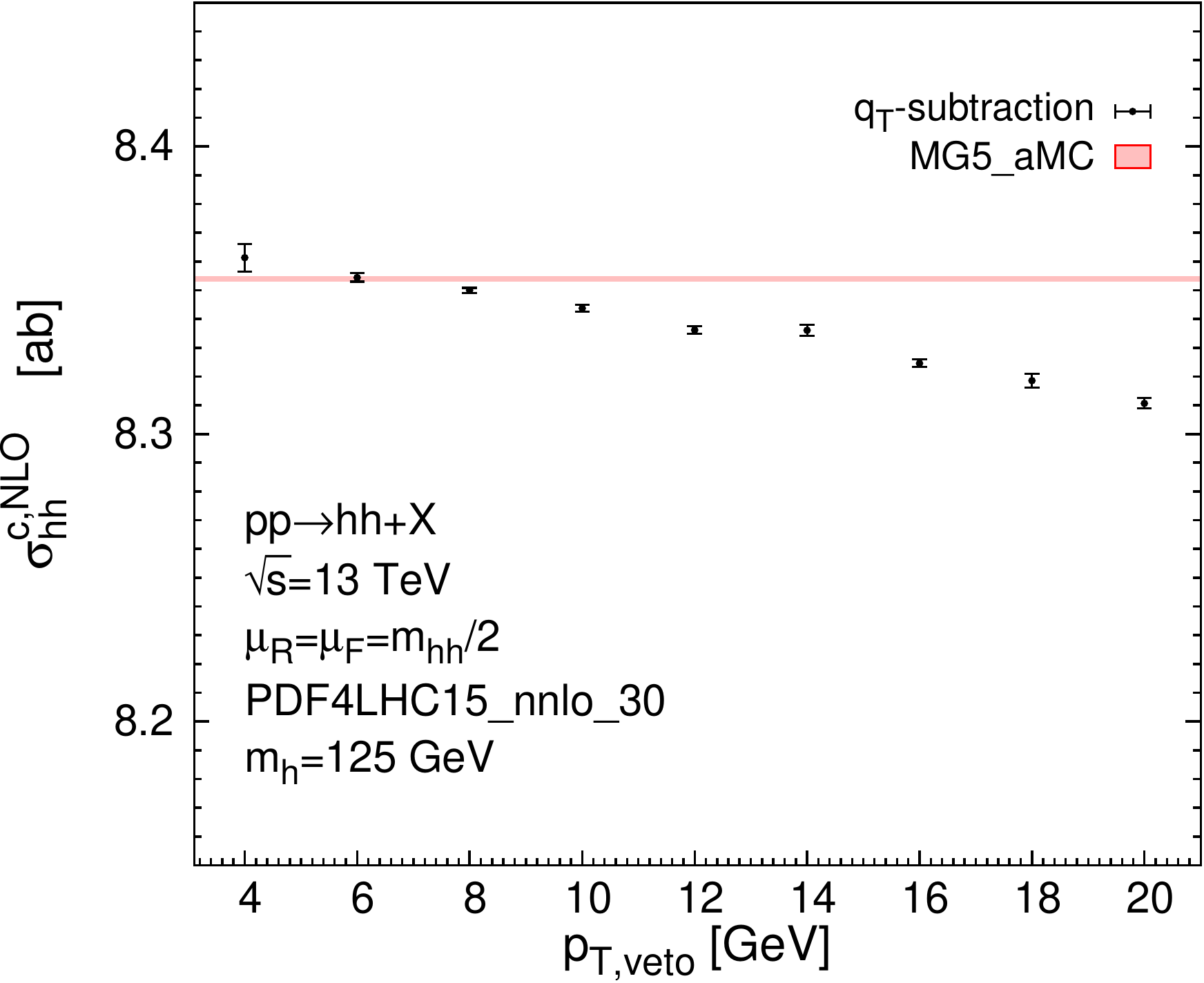}
\caption{The comparisons of the NLO class-$c$ cross sections from the $q_T$-subtraction method (error bars) and \mgshort~(red band) at $\sqrt{s}=13$ TeV LHC. The error bars and the band denotes the Monte Carlo integration errors.}
\label{fig:ptvetocNLO}
\end{figure}

We only need the NLO QCD corrections to the class-$c$ part in order to give N$^3$LO di-Higgs cross sections. The computations can be achieved with the full-fledged NLO techniques. We have compared the NLO cross sections for the class-$c$ part between the $q_T$-subtraction approach and the automated calculation by \mgshort\ in figure~\ref{fig:ptvetocNLO}. The perfect agreement is found when $p_T^{\rm veto}\leq 6$ GeV.

We have summarised the independent calculations we have performed with different approaches for the three classes contributing to various orders in table~\ref{tab:calcmethod}.

\begin{table}[hbt!]
    \centering
    \begin{tabular}{|c|c|c|c|}
    \hline
        & NLO  & NNLO  & N$^3$LO  \\
    \hline
 order  &   $\mathcal{O}(\alpha_s^3)$ &   $\mathcal{O}(\alpha_s^4)$  & $\mathcal{O}(\alpha_s^5)$\\\hline
     \multirow{3}{*}{a}  &   {\sc\small iHixs2} &   {\sc\small iHixs2}  & {\sc\small iHixs2}
     \\
     & $q_T$-subtraction & $q_T$-subtraction & \\
     & \mgshort\ & & \\
    \hline
     \multirow{2}{*}{b}   &   \multirow{2}{*}{-} &   $q_T$-subtraction  & $q_T$-subtraction
     \\
     & & \mgshort\ & \\
    \hline
     \multirow{2}{*}{c}   &  \multirow{2}{*}{-} &   \multirow{2}{*}{-}  & $q_T$-subtraction \\
     & & & \mgshort\ \\
     \hline
    \end{tabular}
    \caption{A summary of independent calculations we have performed at different orders and for different classes.}
    \label{tab:calcmethod}
\end{table}

\subsection{Results}

\subsubsection{Calculational setup\label{sec:setup}}

In our numerical calculations, we take $v=246.2$ GeV and the Higgs boson mass $m_h=125$ GeV. The top-quark pole mass, which enters only into the Wilson coefficients,  is $m_t=173$ GeV.  Unless it is explicitly specified, the trilinear Higgs coupling $\lambda_{hhh}$ is taken to be the SM value. We use the {\tt PDF4LHC15\_nnlo\_30} PDF~\cite{Butterworth:2015oua,Dulat:2015mca,Harland-Lang:2014zoa,Ball:2014uwa} available in the programme {\sc\small LHAPDF6}~\cite{Buckley:2014ana}, and the associated $\as$. The default central scale is chosen to be the invariant mass of the Higgs boson pair divided by 2, i.e. $\mu_0=m_{hh}/2$, and the scale uncertainty is evaluated through the 9-point variation of the factorisation scale $\mu_F$ and the renormalisation scale $\mu_R$ in the form of $\mu_{R,F}=\xi_{R,F} \, \mu_0$ with $\xi_{R},\xi_{F}\in \{0.5,1,2\}$. In the parts of ultilising the $q_T$-subtraction method, we will use $p_T^{\rm veto}=6$ GeV if $\sqrt{s}<27$ TeV and $p_T^{\rm veto}=10$ GeV if $\sqrt{s}\geq 27$ TeV. We have verified that the uncertainties due to the missing power-suppressed terms of $\left(\frac{p_T^{\rm veto}}{\mu_0}\right)^2$ are well below the Monte-Carlo integration errors.

\subsubsection{Inclusive total  cross sections}

\begin{figure}[ht]
    \centering
    \includegraphics[width=0.45\textwidth]{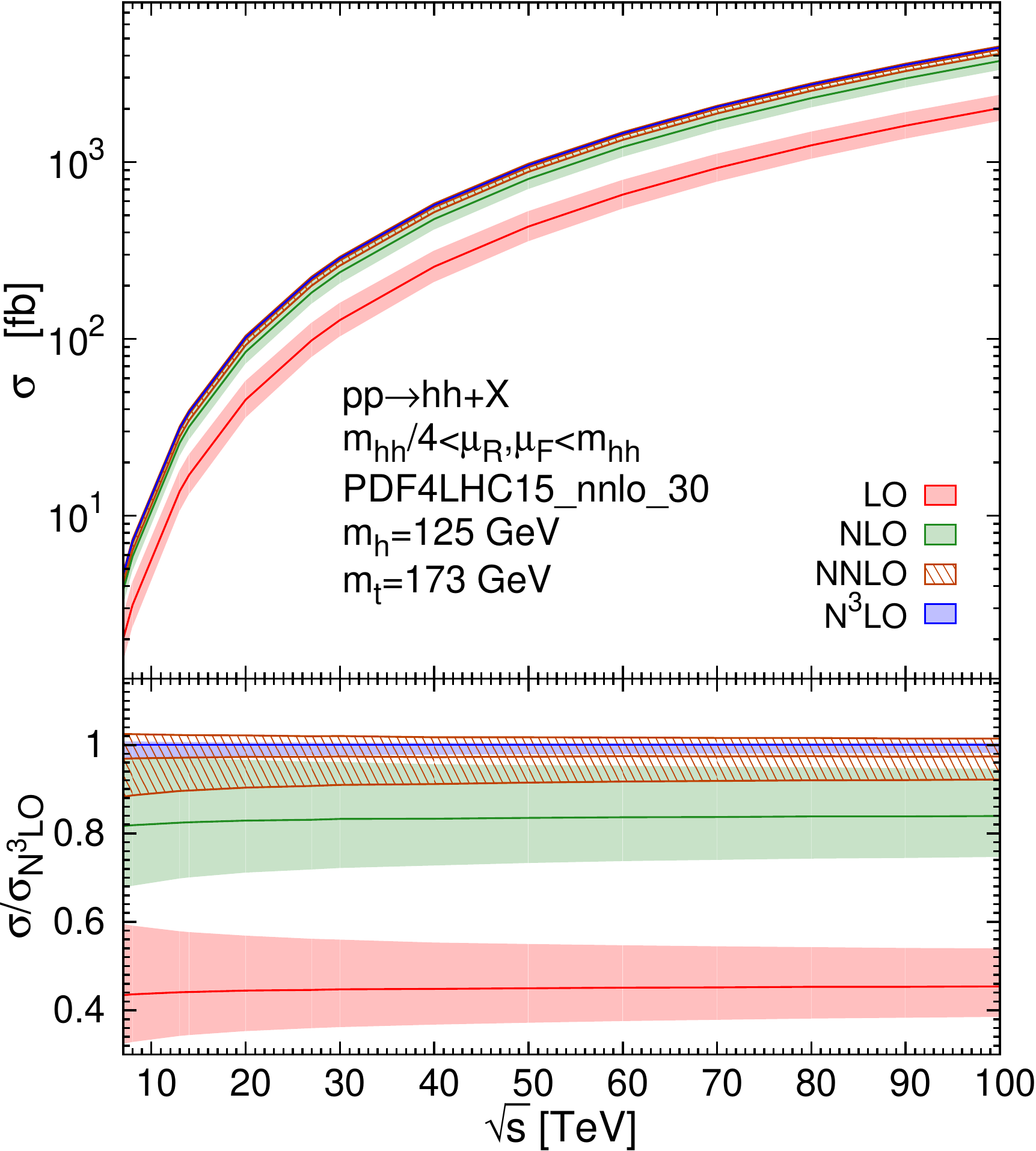}
    \includegraphics[width=0.45\textwidth]{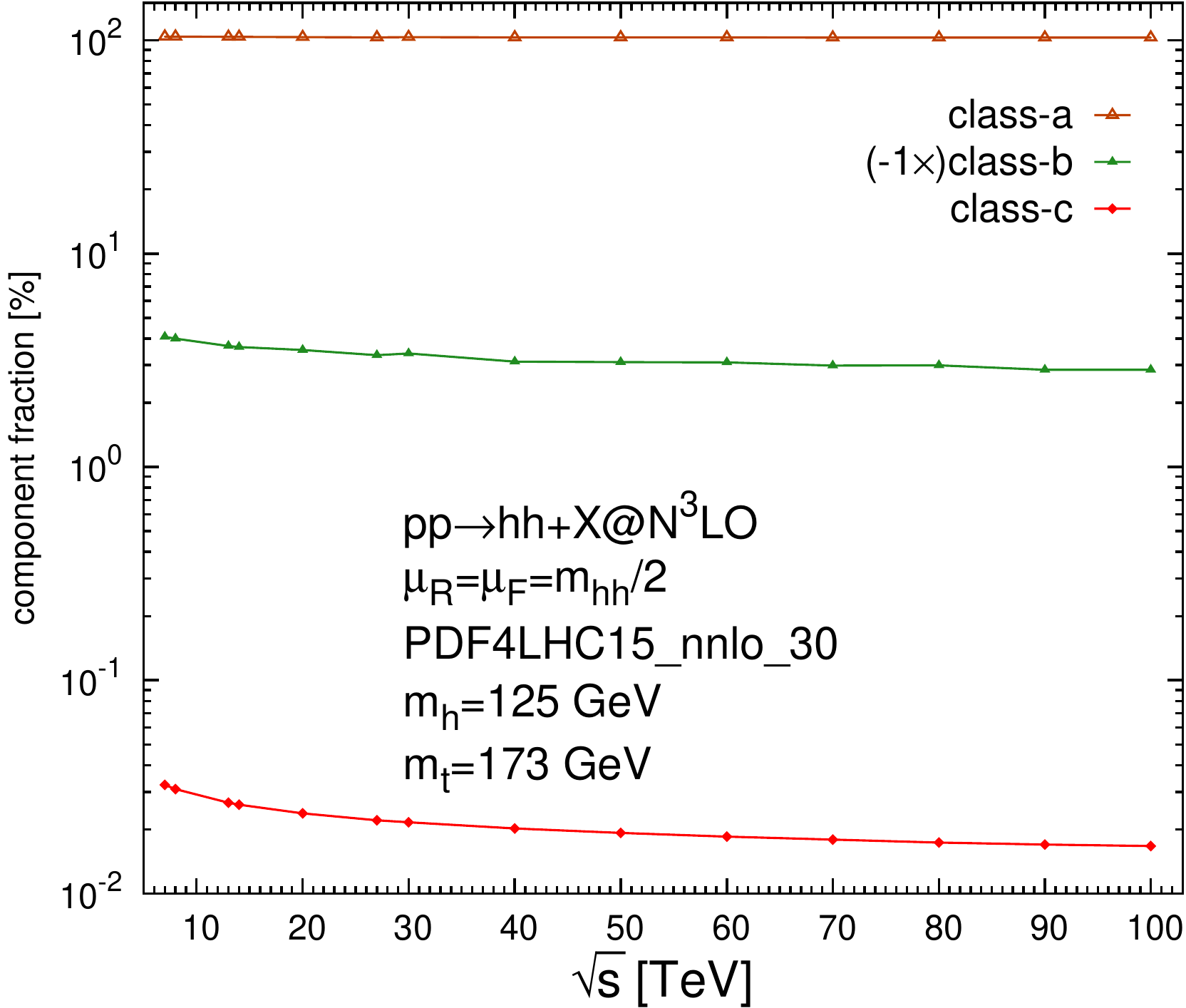}
    \vspace{0cm}
    \caption{The inclusive total cross sections (left) and the contribution breakdown of three classes (right) for the Higgs boson pair production in proton-proton collisions as a function of $\sqrt{s}$. 
    The bands represent the scale uncertainties.}
    \label{fig:sigma}
\end{figure}

We present the inclusive total cross sections from LO to N$^3$LO at  different centre-of-mass energies $\sqrt{s}=13,14,27,100$ TeV in table~\ref{tab:totxs}, where  the scale uncertainties are also shown. These particular energies are either the LHC energies ($13$ and $14$ TeV) or the nominated energies for the future hadron colliders~\cite{Cepeda:2019klc,Contino:2016spe}. The cross sections from $\sqrt{s}=7$ TeV to $\sqrt{s}=100$ TeV are also displayed in the left panel of figure~\ref{fig:sigma}, where the bands represent  the scale uncertainties. Similarly to the case of single Higgs production, the QCD corrections in the di-Higgs process are very prominent. The NLO QCD corrections increase the LO cross sections by  $87\%$  ($85\%$) at $\sqrt{s}=13~(100)$ TeV. The NNLO QCD corrections improve the NLO cross sections further by   $18\%$ ($16\%$), reducing the scale uncertainties by a factor of two to three to be below $8\%$. The N$^3$LO QCD corrections enhance the NNLO cross section by $3.0\%$ ($2.7\%$). The cross sections lie well within the scale uncertainty bands of the NNLO results, and the N$^3$LO scale uncertainties are less than $3\%$ and $2\%$ at 13 and 100 TeV respectively. In addition, the PDF parameterisation uncertainties are almost independent of the QCD corrections. Their relative sizes amount to $\pm 3.3\%, \pm 3.1\%, \pm 2.2\%$  and $\pm 1.4\%$ with respect to the central values at 13, 14, 27 and 100 TeV, overwhelming the remaining N$^3$LO scale uncertainties. We have also shown the contribution from three different classes separately in the right panel of figure~\ref{fig:sigma}, where  the class-$b$ contribution has been multiplied by a factor of -1  in order to make it visible in the frame. There is a strong hierarchy among the three classes. Typically, the class-$b$ part is only a few percent of the class-$a$, while the class-$c$ is a few percent of the class-$b$. Such a behaviour can be understood from the effective Lagrangian eq.(\ref{eq:effL}). One more effective vertex in the squared amplitude results in  one more factor of $\frac{\alpha_s}{3\pi}\sim 1\%$ suppression instead of the usual $\alpha_s$ suppression.


\begin{table}[h]
\centering
\begin{tabular}{|c|c|c|c|c|}
\hline
$\sqrt{s}$ & $13$ TeV & $14$ TeV & $27$ TeV & $100$ TeV\\\hline
LO & $13.80_{-22\%}^{+31\%}$ & $17.06_{-22\%}^{+31\%}$ & $98.22_{-19\%}^{+26\%}$ & $2015_{-15\%}^{+19\%}$ \\
NLO & $25.81_{-15\%}^{+18\%}$ & $31.89_{-15\%}^{+18\%}$ & $183.0^{+16\%}_{-14\%}$ & $3724_{-11\%}^{+13\%}$\\
NNLO & $30.41^{+5.3\%}_{-7.8\%}$ & $37.55^{+5.2\%}_{-7.6\%}$ & $214.2^{+4.8\%}_{-6.7\%}$ & $4322_{-5.3\%}^{+4.2\%}$ \\
N$^3$LO & $31.31^{+0.66\%}_{-2.8\%}$ & $38.65^{+0.65\%}_{-2.7\%}$ & $220.2^{+0.53\%}_{-2.4\%}$ & $4439^{+0.51\%}_{-1.8\%}$\\\hline
\end{tabular}
\caption{The inclusive total cross sections (in unit of fb) of Higgs boson pair production in the infinite top-quark mass limit at different centre-of-mass energies $\sqrt{s}$ from LO to N$^3$LO.
  The quoted relative uncertainties are from the 9-point scale variations. The errors due to the numerical Monte Carlo integration are well below 1\permil.}
\label{tab:totxs}
\end{table}


\begin{figure}[ht]
    \centering
    \includegraphics[width=0.55\textwidth]{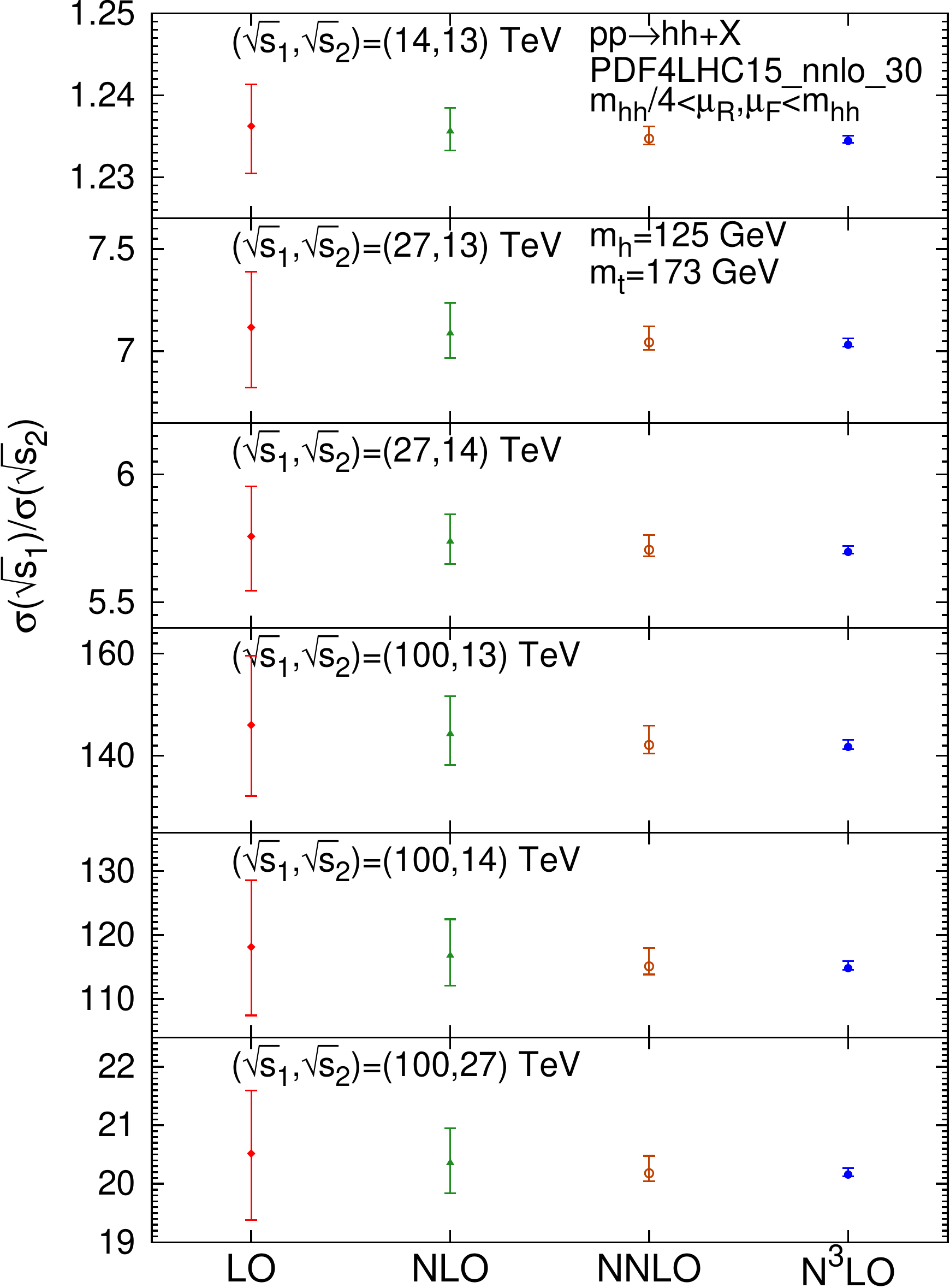}
    \vspace{0cm}
    \caption{The total cross section ratios between different $\sqrt{s}$ for the Higgs boson pair production in proton-proton collisions from LO to N$^3$LO. 
    The error bars represent the scale uncertainties.}
    \label{fig:sigmaratio}
\end{figure}

It was proposed in ref.~\cite{Mangano:2012mh} to use the ratios of cross sections with the same final state  between different centre-of-mass energies to perform precision studies (e.g. determining PDFs) and to improve the BSM sensitivities~\footnote{A similar idea but using different final states instead of different $\sqrt{s}$ was also introduced in ref.~\cite{Plehn:2015cta}.}. The success of such a programme relies on the large cancellations of theoretical systematic uncertainties in the ratios. In particular, the usually dominant scale uncertainties in the cross sections can be significantly reduced by fully correlating the renormalisation and factorisation scales between numerators and denominators. Such a reasonable working assumption, however, should be carefully checked when higher-order calculations become available. With the N$^3$LO calculations we have done, we can readily check such a hypothesis in the double-Higgs process. In figure~\ref{fig:sigmaratio}, we have plotted the cross section ratios in six different $\sqrt{s}$ pairs from LO to N$^3$LO. The scale correlation assumption in the cross section ratios is indeed verified in this process.

\begin{figure}[h]
    \centering
    \subfigure[$\sqrt{s}=13$ TeV]{\includegraphics[width=0.45\textwidth]{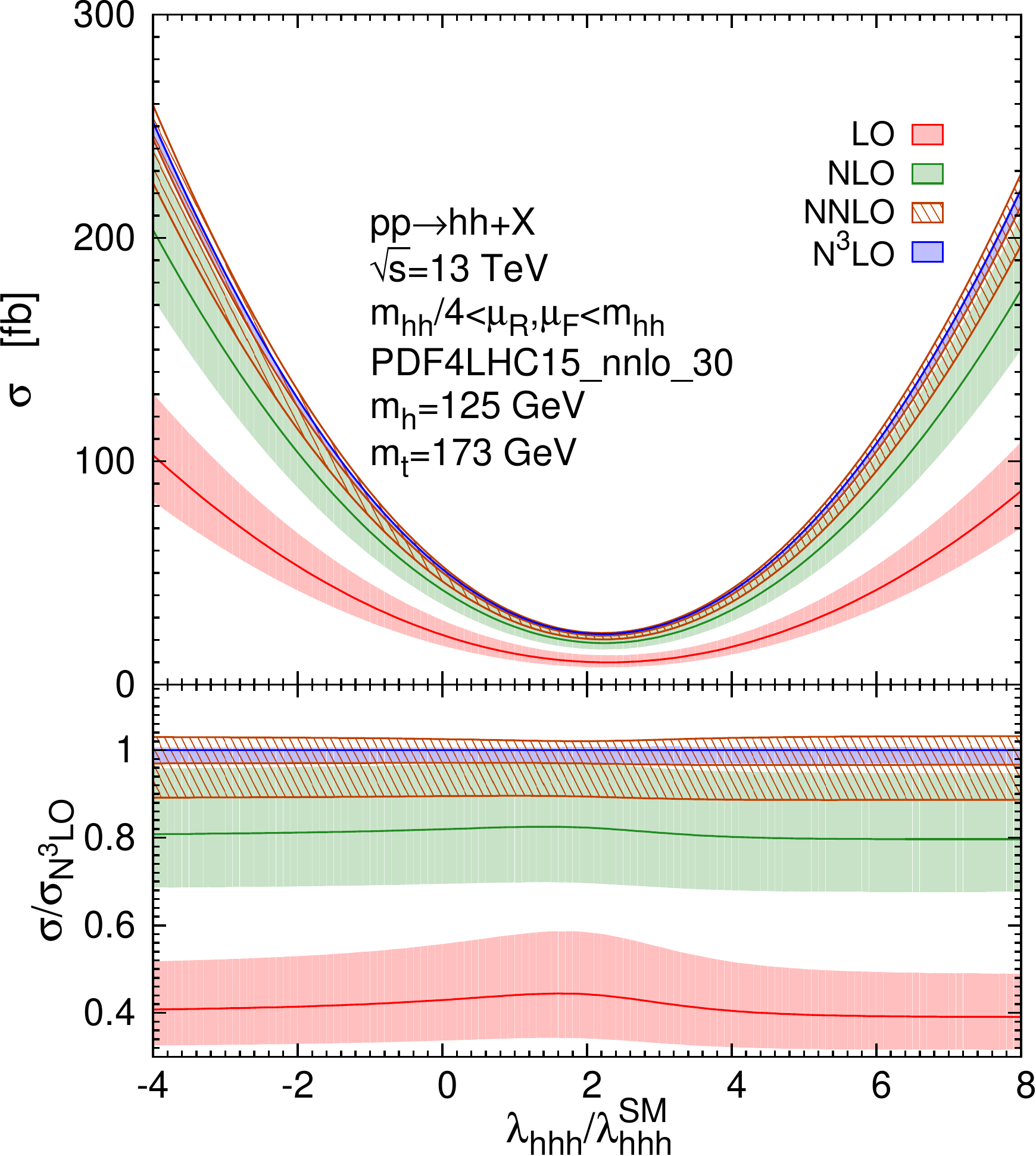}}
    \subfigure[$\sqrt{s}=14$ TeV]{\includegraphics[width=0.45\textwidth]{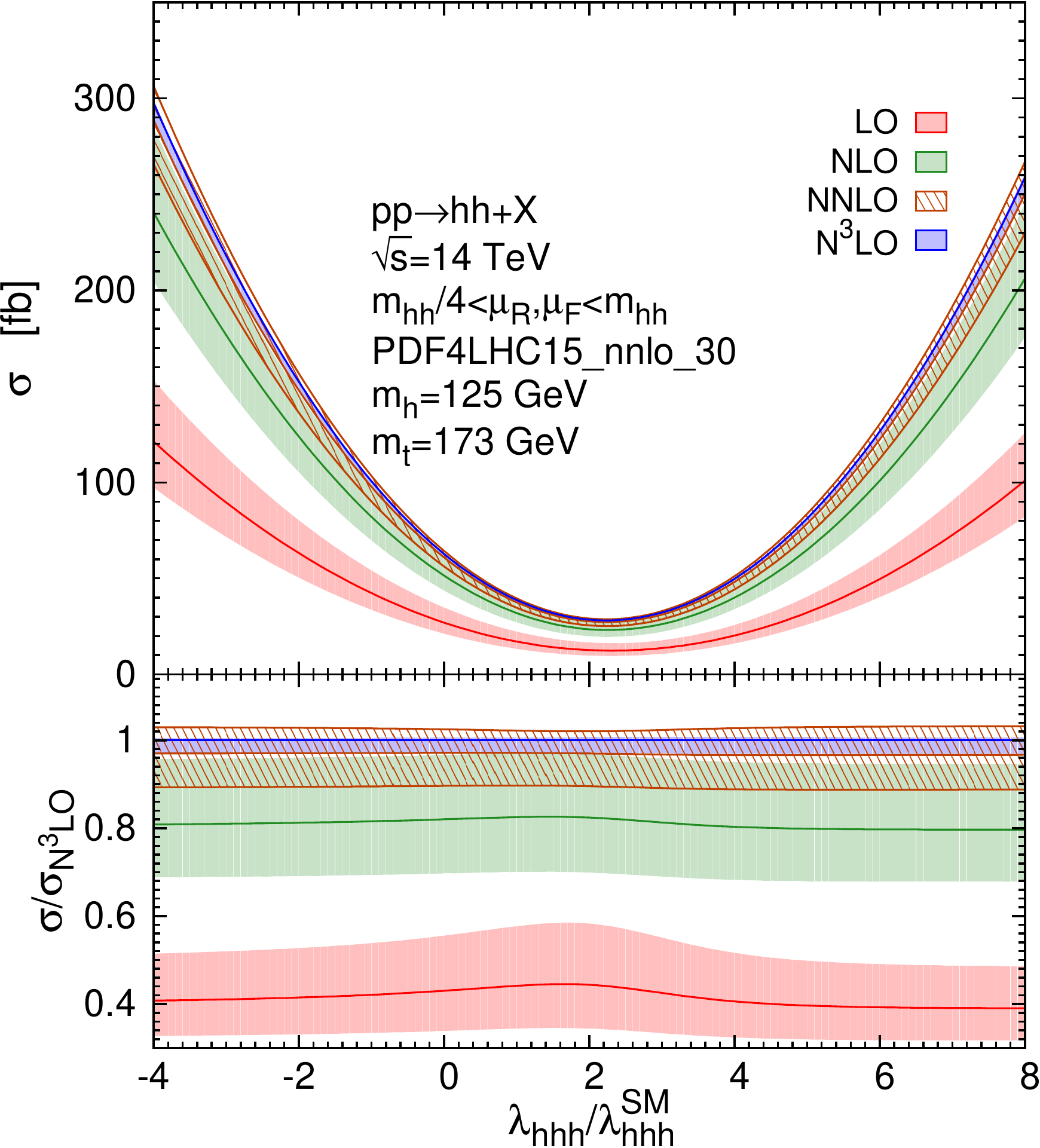}}\\
    \subfigure[$\sqrt{s}=27$ TeV]{\includegraphics[width=0.45\textwidth]{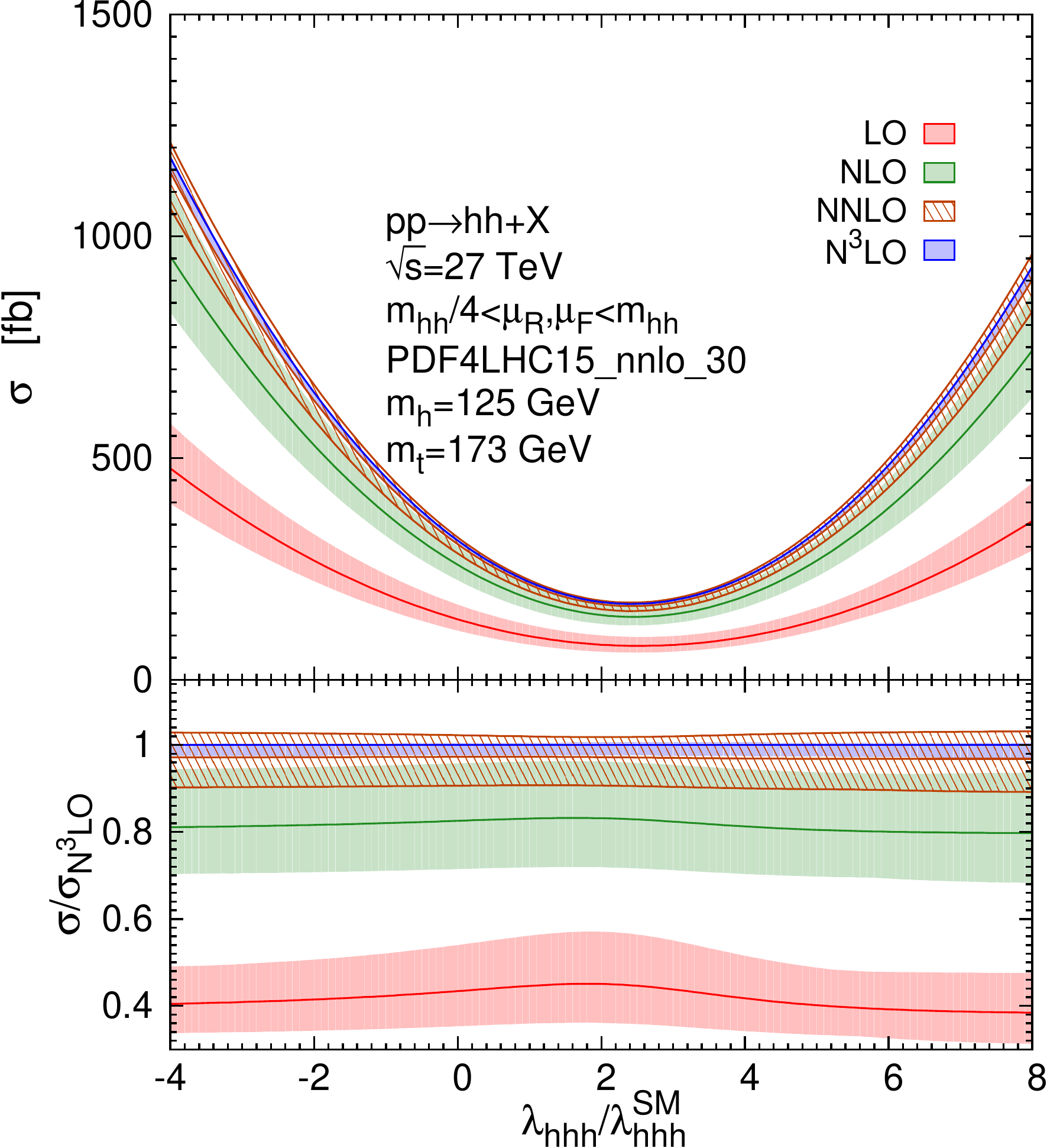}}
    \subfigure[$\sqrt{s}=100$ TeV]{\includegraphics[width=0.45\textwidth]{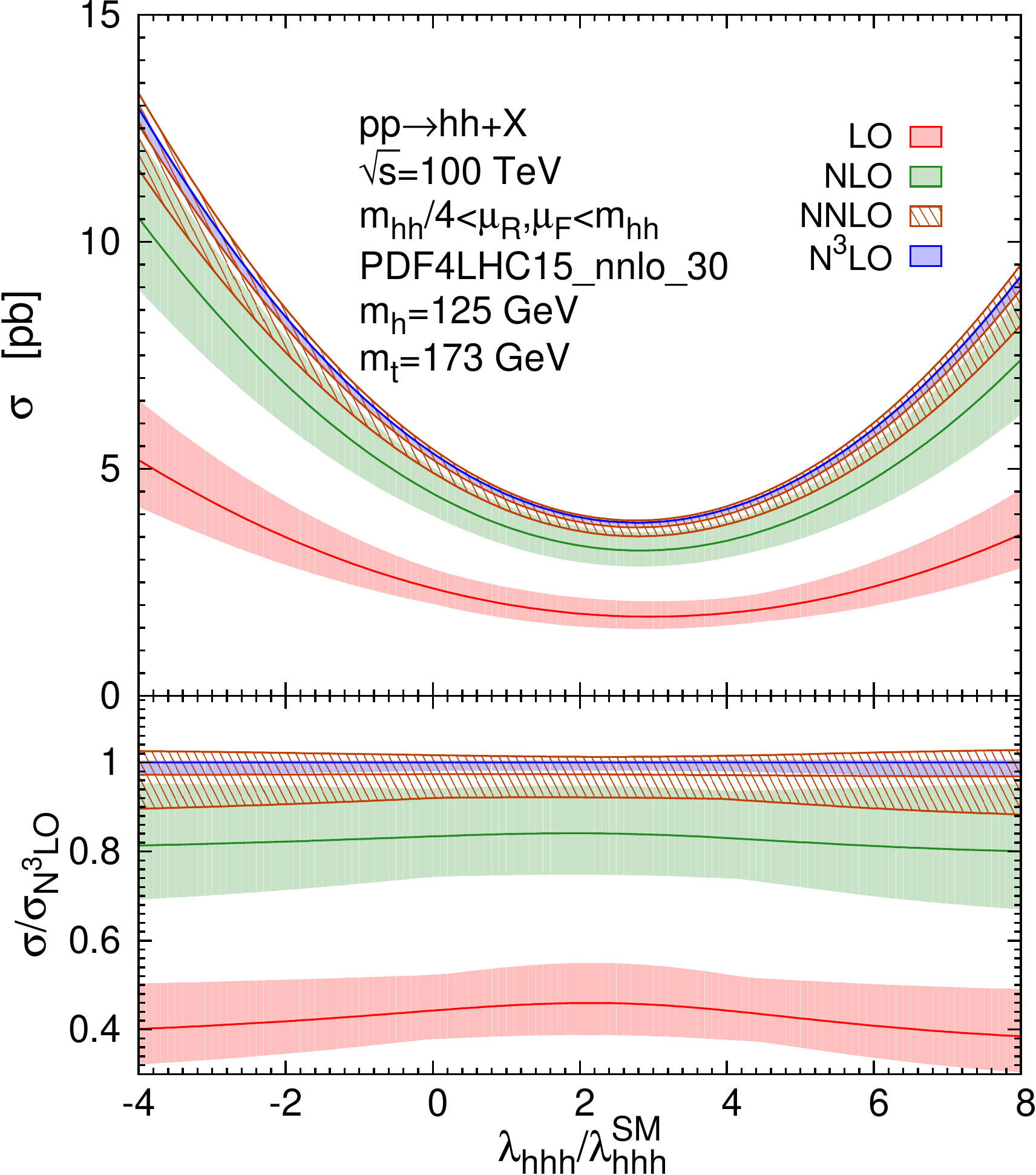}}
    \vspace{0cm}
    \caption{The $\lambda_{hhh}$ dependencies of the total inclusive cross sections for the Higgs boson pair production in proton-proton collisions with $\sqrt{s}=13,14,27,100$ TeV. The bands represent the scale uncertainties.  The red, green, brown and blue bands correspond to the LO, NLO, NNLO and N$^3$LO predictions, respectively. The bottom panel shows the ratios to the N$^3$LO distribution.}
    \label{fig:lambdaxs}
\end{figure}

Apart from the dependence on the collision energy, it is also very interesting to know how total cross sections vary when  $\lambda_{hhh}$ deviates from the SM value.
At four different centre-of-mass energies $\sqrt{s}=13,14,27,100$ TeV, we have varied $\kappa_{\lambda}=\lambda_{hhh}/\lambda_{hhh}^{\rm SM}$ from $-4$ to $8$ in figure~\ref{fig:lambdaxs}. The largest deconstruction between the $\lambda_{hhh}$-independent amplitude (e.g. from figure~\ref{fig:lo}c) and the $\lambda_{hhh}$-dependent amplitude (e.g. from figure~\ref{fig:lo}d) occurs when $\kappa_{\lambda}$ is close to $2$. The N$^3$LO corrections only marginally distort the NNLO predictions around $\kappa_{\lambda}=2$. This can be understood because the QCD radiative corrections to the above two kinds of different  amplitudes are not very different due to the same Lorentz structure shared between figure~\ref{fig:lo}c and figure~\ref{fig:lo}d.

\subsubsection{Invariant mass distributions\label{sec:mhhnomt}}

\begin{figure}[h]
    \centering
    \subfigure[$\sqrt{s}=13$ TeV]{\includegraphics[width=0.45\textwidth]{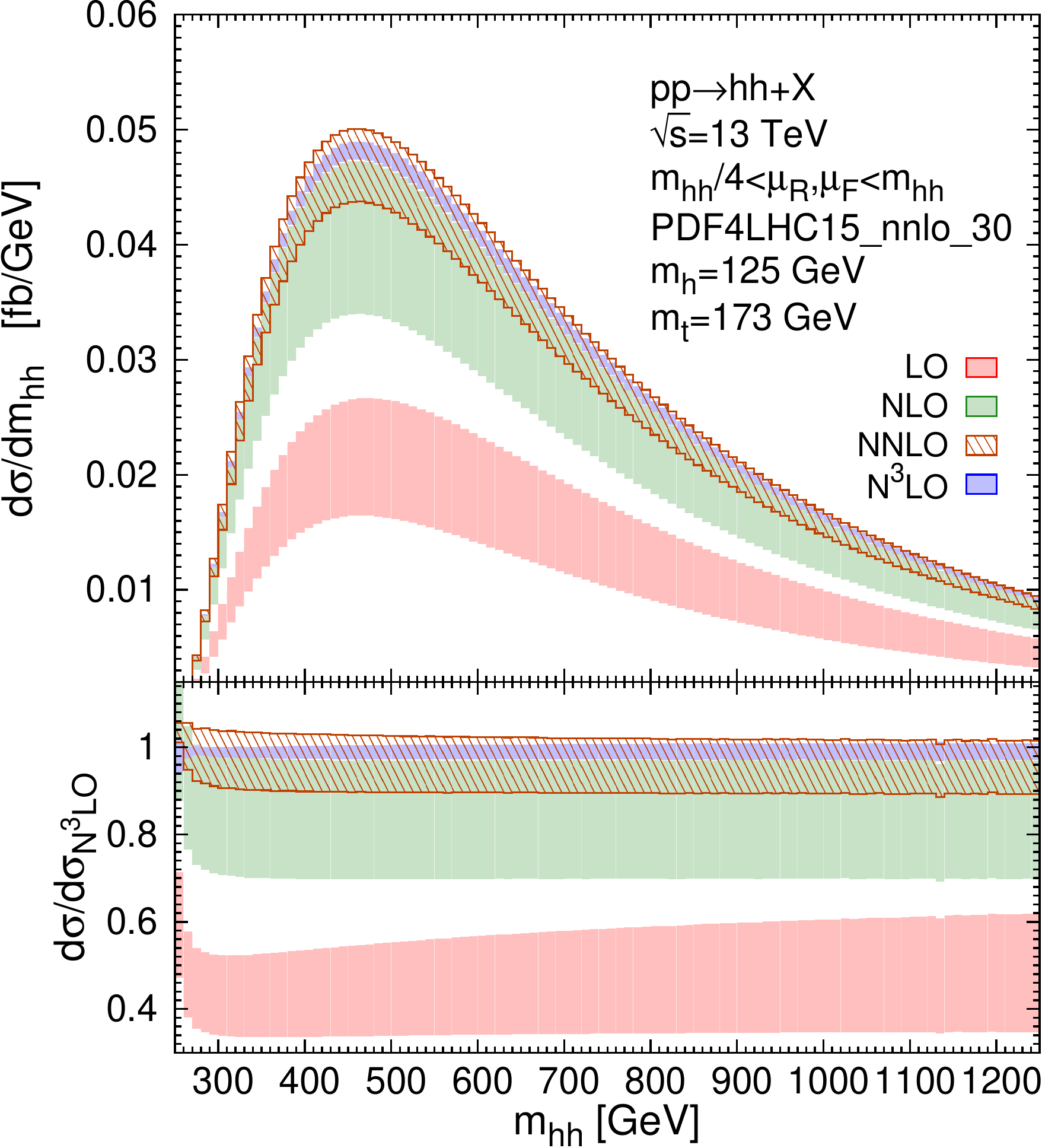}}
    \subfigure[$\sqrt{s}=14$ TeV]{\includegraphics[width=0.45\textwidth]{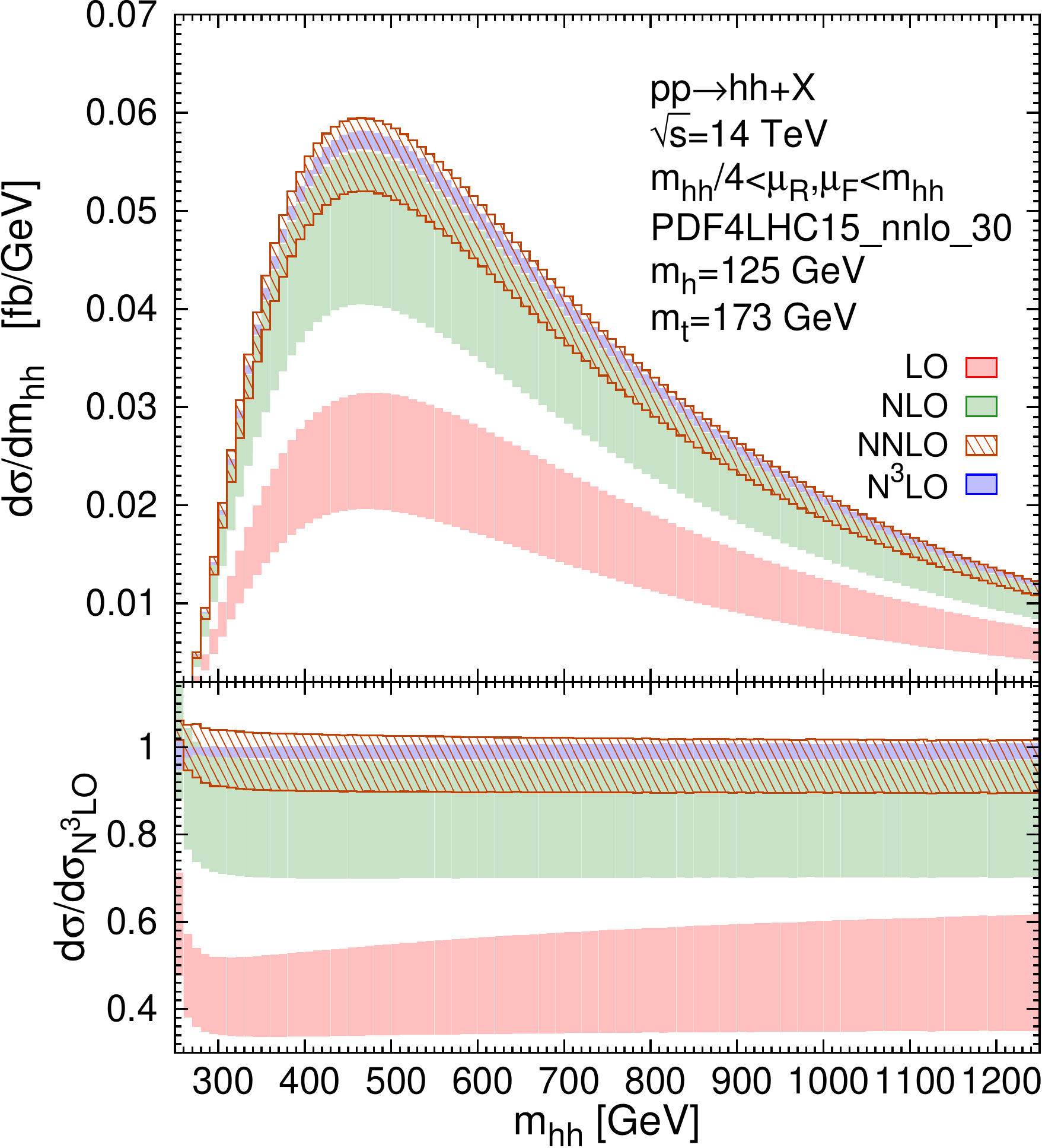}}\\
    \subfigure[$\sqrt{s}=27$ TeV]{\includegraphics[width=0.45\textwidth]{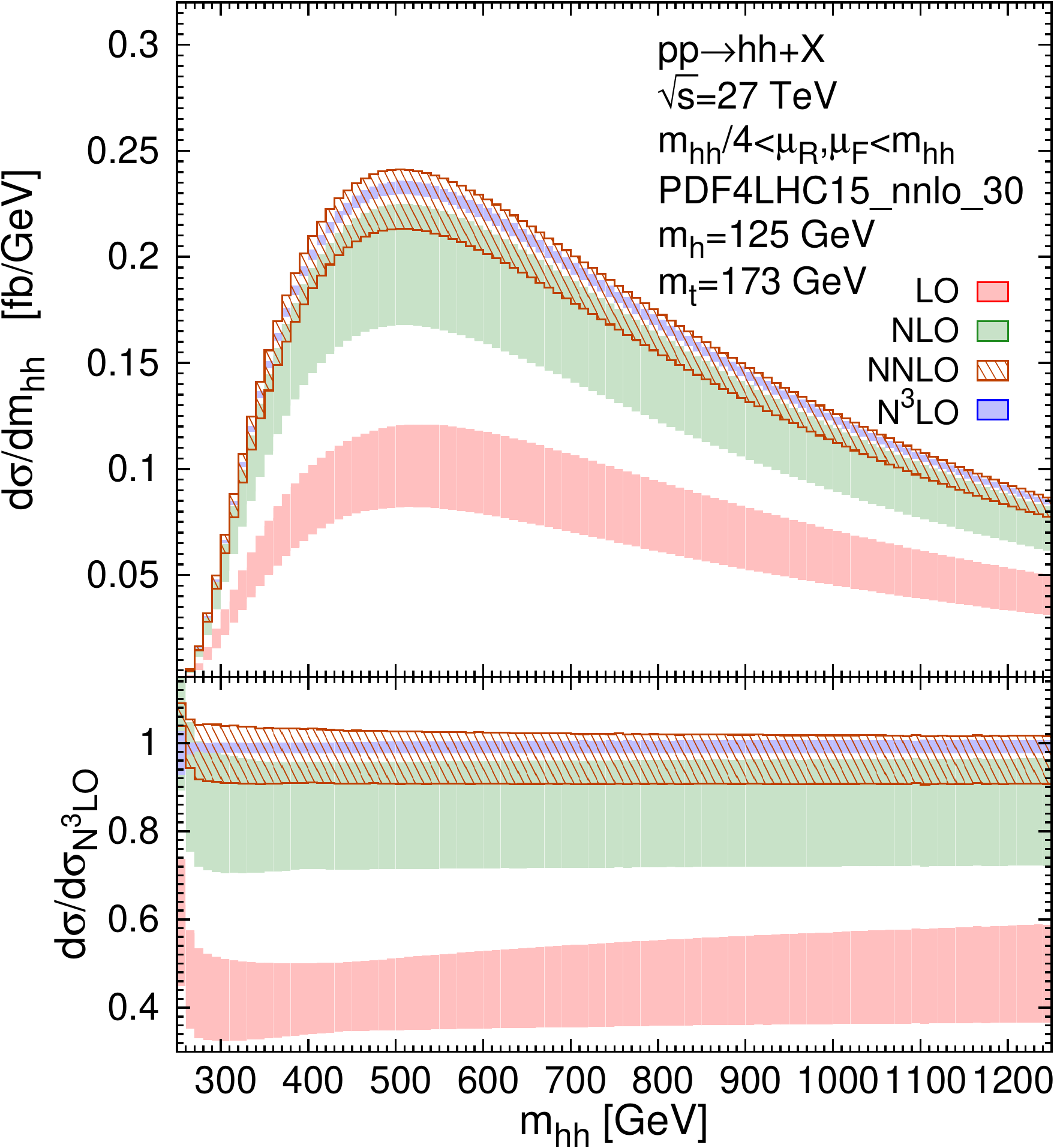}}
    \subfigure[$\sqrt{s}=100$ TeV]{\includegraphics[width=0.45\textwidth]{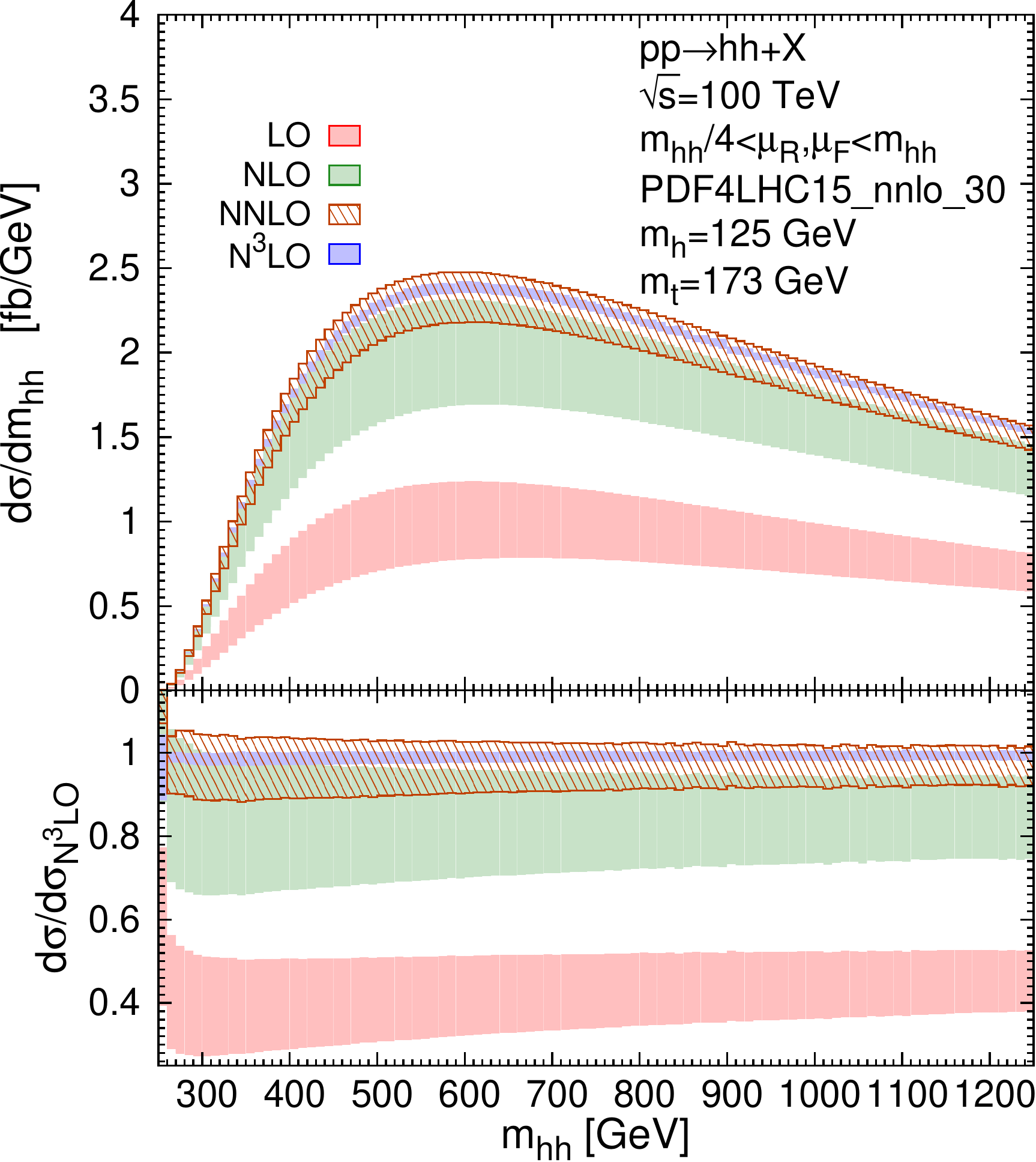}}
    \vspace{0cm}
    \caption{Invariant mass distributions for the Higgs boson pair production in proton-proton collisions with $\sqrt{s}=13,14,27,100$ TeV. The bands represent the scale uncertainties.  The red, green, brown and blue bands correspond to the LO, NLO, NNLO and N$^3$LO predictions, respectively. The bottom panel shows the ratios to the N$^3$LO distribution.  }
    \label{fig:mhh}
\end{figure}

Besides the total cross sections, we are also able to calculate the exact N$^3$LO results for the invariant mass $m_{hh}$ distributions, which are shown in figure~\ref{fig:mhh} with the 4 different energies $\sqrt{s}=13,14,27,100$ TeV. Due to the larger phase space, the $m_{hh}$ spectrum becomes harder when $\sqrt{s}$ increases. The inclusion of the N$^3$LO QCD corrections dramatically stabilises the perturbative calculations of the invariant mass differential distributions. The N$^3$LO corrections only marginally change the shapes, and the N$^3$LO results, which have very small scale uncertainties, are completely enclosed within the NNLO uncertainty bands. Such a feature consolidates that the perturbative expansions of the invariant mass differential cross sections are converging in $\alpha_s$ up to the fourth order.

\begin{figure}[h]
    \centering
    \subfigure[$\kappa_{\lambda}=-1$]{\includegraphics[width=0.45\textwidth]{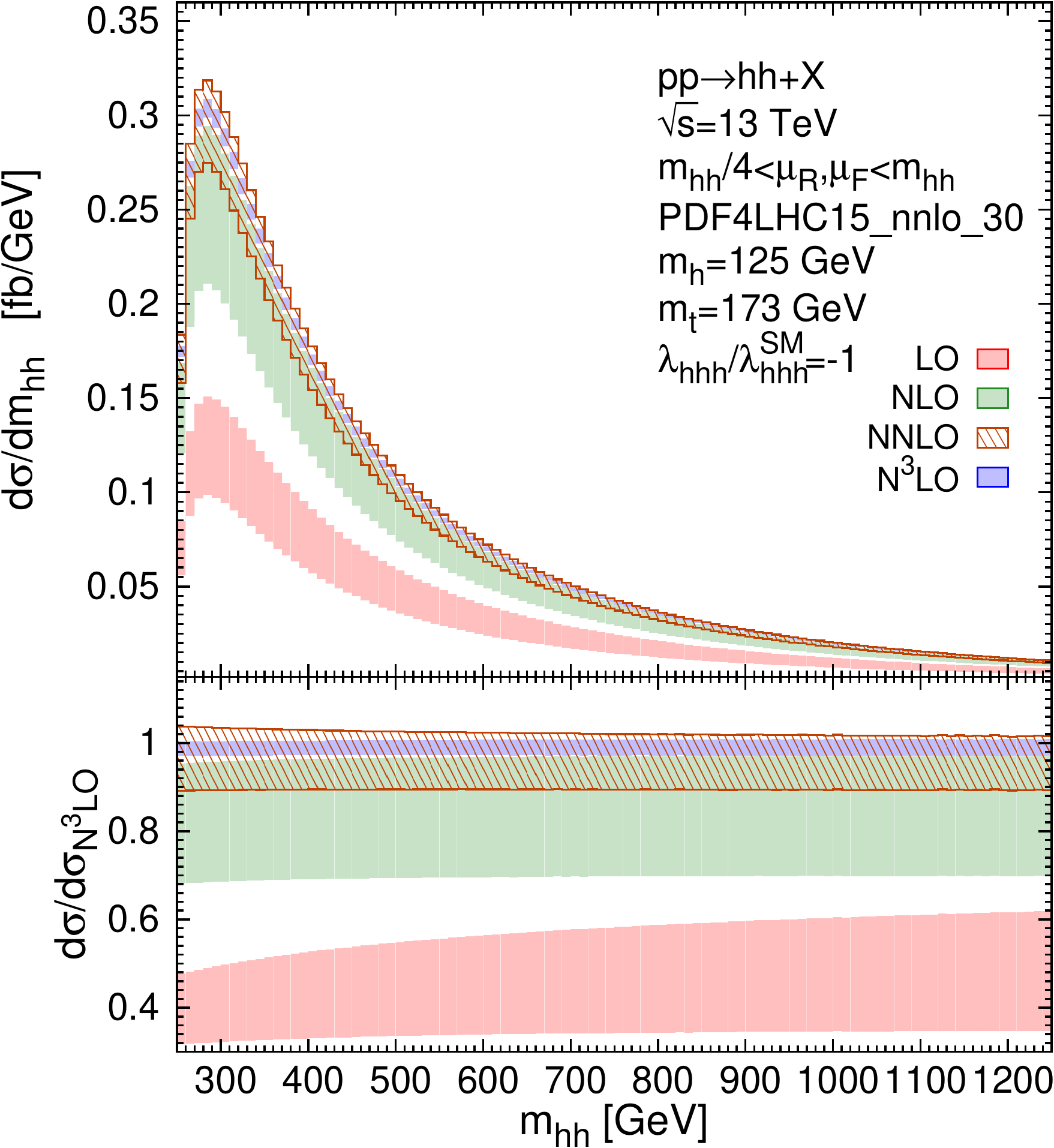}}
    \subfigure[$\kappa_{\lambda}=3$]{\includegraphics[width=0.45\textwidth]{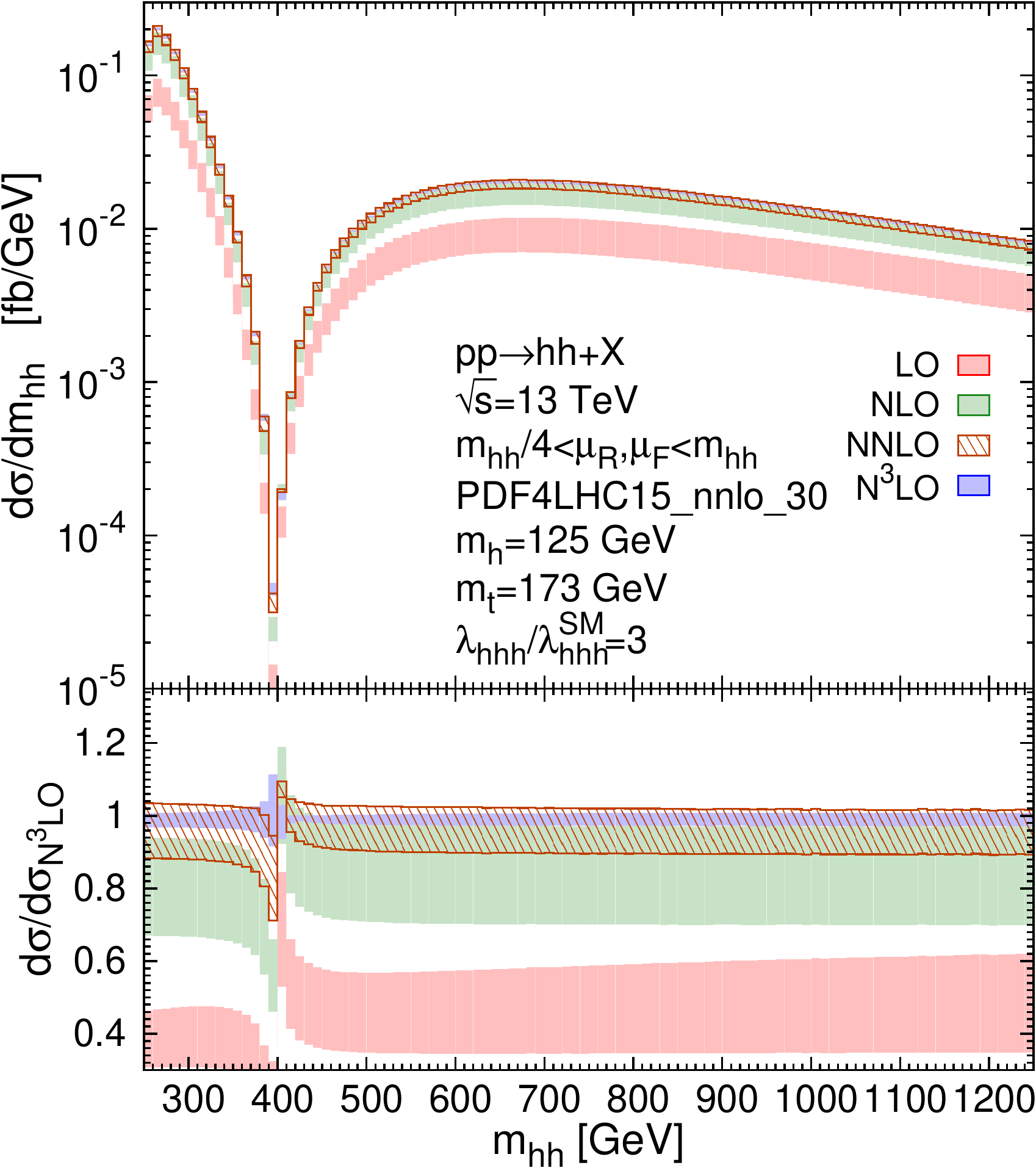}}\\
    \subfigure[$\kappa_{\lambda}=5$]{\includegraphics[width=0.45\textwidth]{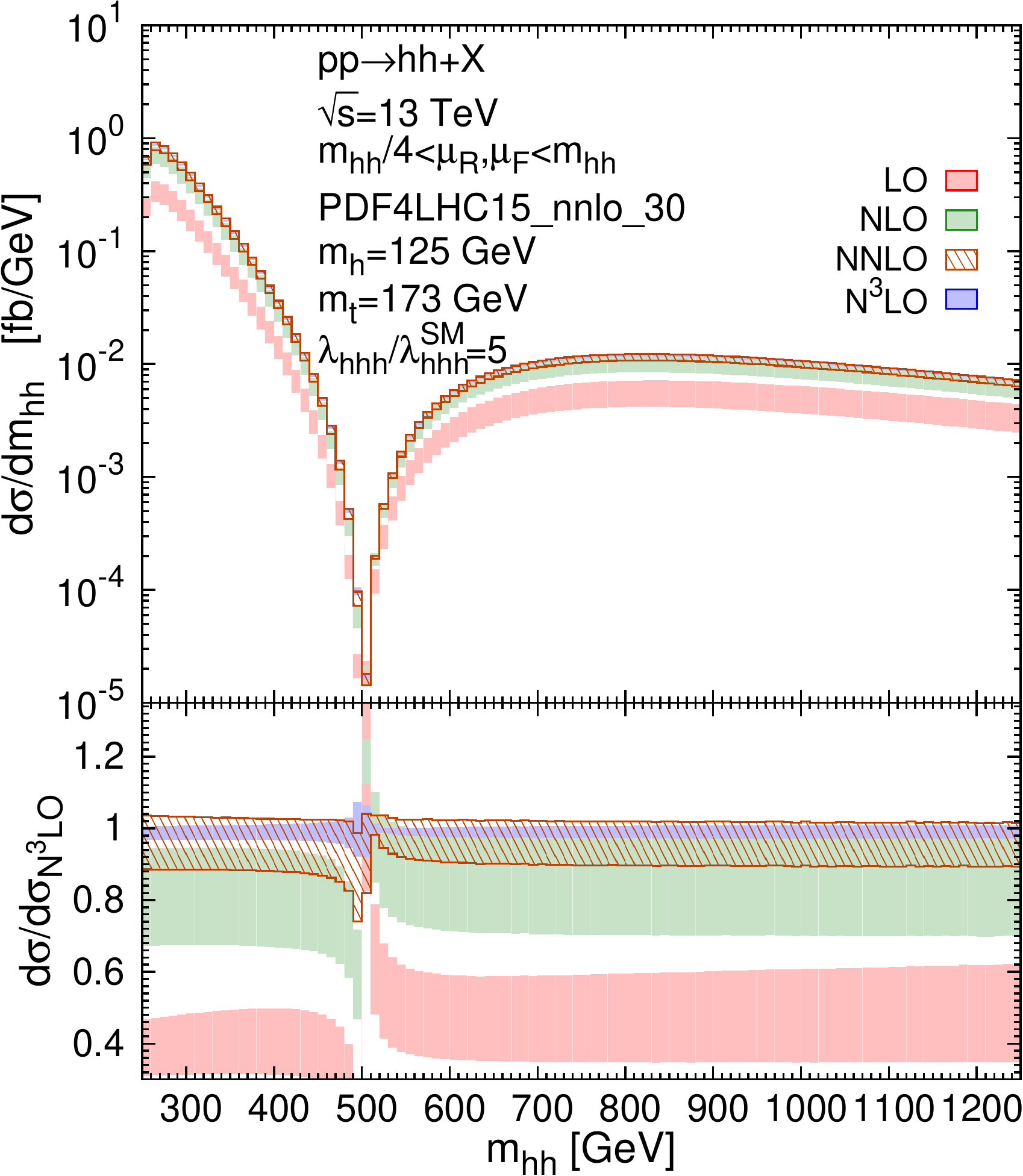}}
    \subfigure[N$^3$LO comparison]{\includegraphics[width=0.45\textwidth]{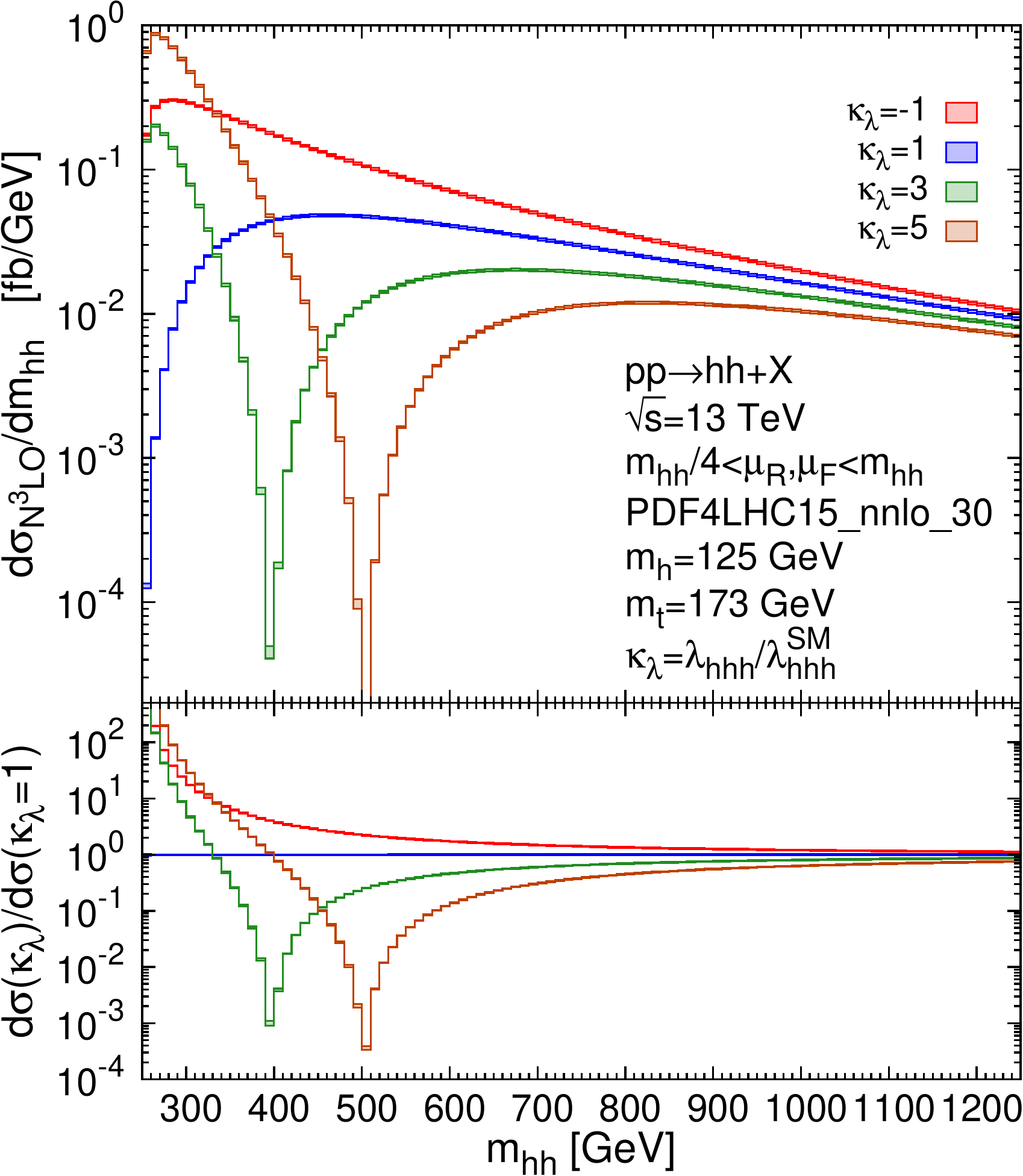}}
    \vspace{0cm}
    \caption{Invariant mass distributions for the Higgs boson pair production in proton-proton collisions at $\sqrt{s}=13$ TeV with different $\kappa_{\lambda}=\lambda_{hhh}/\lambda^{\rm SM}_{hhh}$.}
    \label{fig:mhhlambda}
\end{figure}

It is also very interesting to investigate how the invariant mass distribution changes with respect to the value of $\kappa_{\lambda}=\lambda_{hhh}/\lambda^{\rm SM}_{hhh}$. We have shown the LO to N$^3$LO distributions with $\kappa_{\lambda}=-1$ (upper left), $3$ (upper right) and $5$ (lower left) in figure~\ref{fig:mhhlambda}. In addition, the comparison of N$^3$LO $m_{hh}$ distributions with four values $\kappa_{\lambda}=-1,1,3,5$ is given in the lower right panel of figure~\ref{fig:mhhlambda}. The differential distribution dramatically changes when $\kappa_{\lambda}$ varies. 
This feature can be understood  qualitatively  by looking at eq.(\ref{eq:htohh}). $\left.\sigma_{h}\right|_{m_{h}\rightarrow m_{hh}}$ decreases monotonically when increasing $m_{hh}$, which explains the behaviour in the large invariant mass regime. At small $m_{hh}$ (i.e. $m_{hh}\rightarrow 2m_{h}$), the distribution is governed by the prefactors $f_{h\rightarrow hh}\propto \sqrt{m_{hh}^2-4m_h^2}$ and $\left(\frac{C_{hh}}{C_h}-\frac{6\lambda_{hhh}v^2}{m_{hh}^2-m_h^2}\right)^2\simeq \left(1-\kappa_{\lambda}\frac{3m_h^2}{m_{hh}^2-m_h^2}\right)^2$. Given the phase space boundary $m_{hh}\ge 2m_h$, the second prefactor is a monotonically decreasing (increasing) function of $m_{hh}$ when $\kappa_{\lambda}<0$ ($0<\kappa_{\lambda}\leq 1$).
If $\kappa_{\lambda}>1$, $\left(1-\kappa_{\lambda}\frac{3m_h^2}{m_{hh}^2-m_h^2}\right)^2$ monotonically decreases in the region $m_{hh}\in [2m_h,\sqrt{1+3\kappa_{\lambda}}m_h]$ and then monotonically increases when $m_{hh}>\sqrt{1+3\kappa_{\lambda}}m_h$. This explains the fact that
the suppression at threshold $m_{hh}\rightarrow 2m_{h}$ is more dramatic in the SM case $\kappa_{\lambda}=1$ than others. 
On the other hand, when $m_{hh}$ approaches  $\sqrt{1+3\kappa_{\lambda}}m_h$ ($395$ GeV for $\kappa_{\lambda}=3$ and $500$ GeV for $\kappa_{\lambda}=5$), a cancellation happens in $\left(1-\kappa_{\lambda}\frac{3m_h^2}{m_{hh}^2-m_h^2}\right)^2$, which results in the dip structures in figure~\ref{fig:mhhlambda} for the $\kappa_{\lambda}>1$ cases. These interesting features can be definitely used in the BSM searches via the di-Higgs final states~\cite{Capozi:2019xsi}.

\subsubsection{Other differential distributions}

In order to carry out N$^3$LO calculations for other differential distribution, we have to take some approximations, because the fully-differential N$^3$LO corrections to single Higgs production are still unknown. Therefore, at the moment, we have to approximate the N$^3$LO class-$a$ corrections for other differential cross sections. As we already mentioned in section \ref{sec:classb}, the class-$a$ differential cross sections can be divided into two pieces given in eq.(\ref{eq:classadecomp}). The second piece $d\sigma^{(a,2),{\rm N^3LO}}_{hh}$ is essential to cancel the remaining renormalisation scale dependence in $d\sigma^{b,{\rm NNLO}}_{hh}$. Both of them are in fact known fully differentially. For the first piece $d\sigma^{(a,1),{\rm N^3LO}}_{hh}$ (i.e. the class-$a$ cross sections by setting $C_{hh}=C_{h}$), we have the fully differential calculations for the NNLO class-$a$ cross sections with the $q_T$-subtraction method. Therefore, in our paper, we can define the approximated N$^3$LO (AN$^3$LO) differential distributions for other observable $O$ as
\begin{eqnarray}
\frac{d\sigma_{hh}^{\rm AN^3LO}}{dO}&=&\frac{d\sigma_{hh}^{\rm (a,1),NNLO}}{dO}\frac{\sigma_{hh}^{\rm (a,1),N^3LO}}{\sigma_{hh}^{\rm (a,1),NNLO}}+\frac{d\sigma^{\rm (a,2),N^3LO}_{hh}}{dO}+\frac{d\sigma_{hh}^{\rm b,NNLO}}{dO}+\frac{d\sigma_{hh}^{\rm c,NLO}}{dO}.\label{eq:N3LOapprox}
\end{eqnarray}
The $(a,1)$ piece is simply multiplying a global K factor $\frac{\sigma_{hh}^{\rm (a,1),N^3LO}}{\sigma_{hh}^{\rm (a,1),NNLO}}$ assuming no kinematic dependence. Such an assumption is more-or-less justified given the extremely flat K factor found in the rapidity distributions of the single Higgs process~\cite{Dulat:2018bfe}. In contrast, the exact fully-differential predictions are achievable for other three pieces at $\mathcal{O}(\alpha_s^5)$. Our calculations can certainly be improved as long as the fully-differential N$^3$LO calculation of the single-Higgs process is available.

\begin{figure}[h]
    \centering
    \includegraphics[scale=.33,draft=false,trim = 0mm 0mm 0mm 0mm,clip]{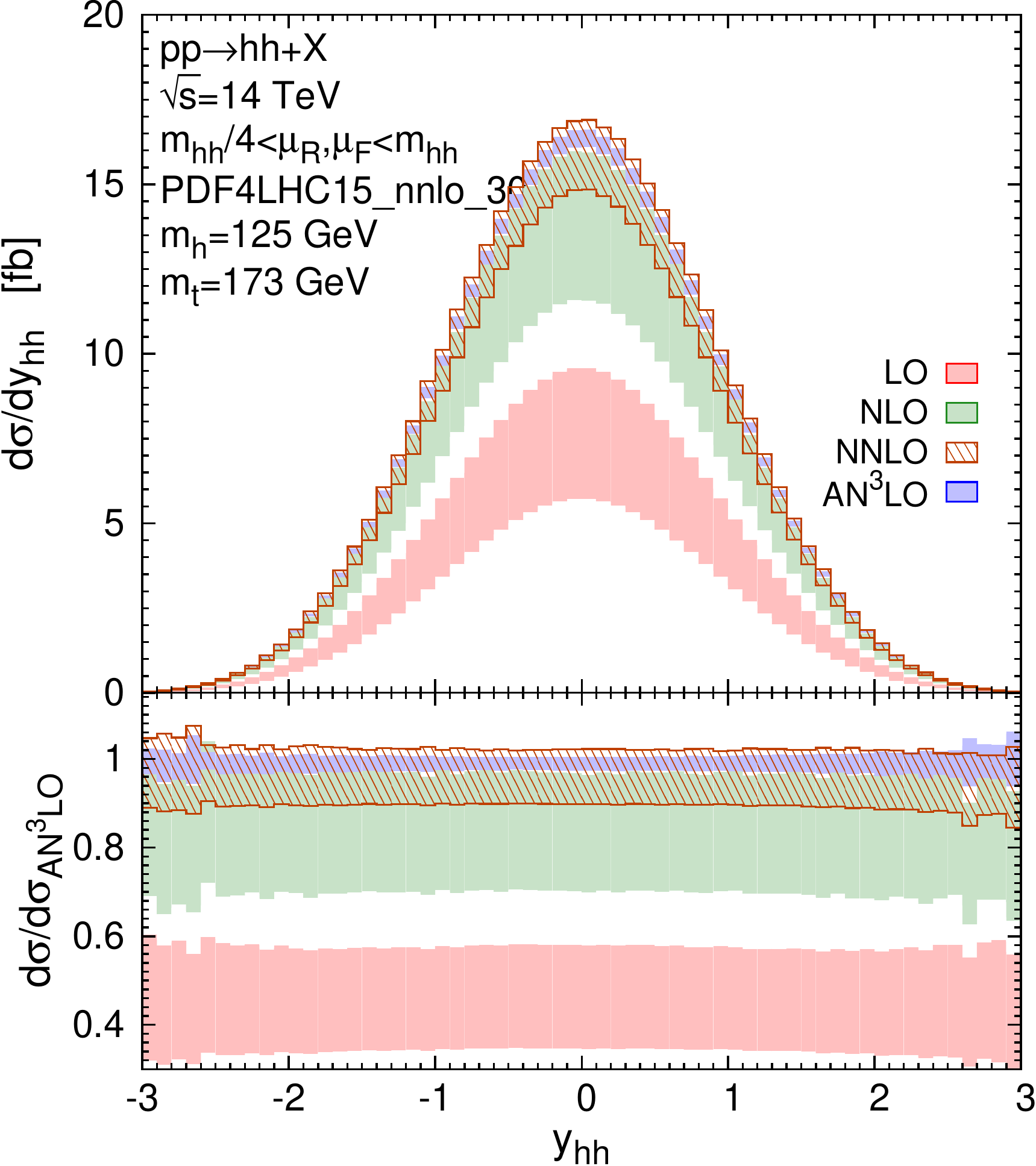}
    \includegraphics[scale=.33,draft=false,trim = 0mm 0mm 0mm 0mm,clip]{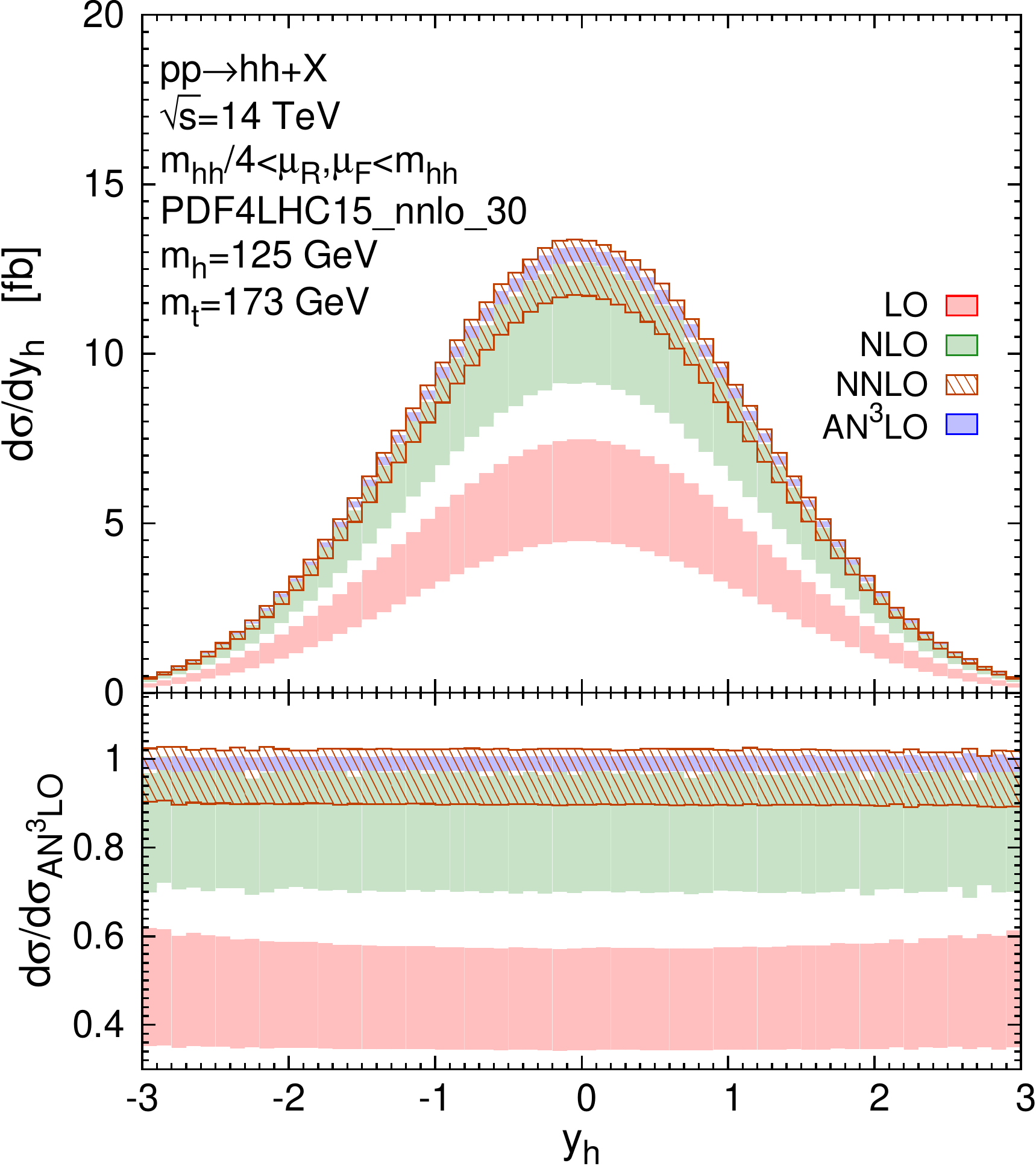}\\
    \includegraphics[scale=.33,draft=false,trim = 0mm 0mm 0mm 0mm,clip]{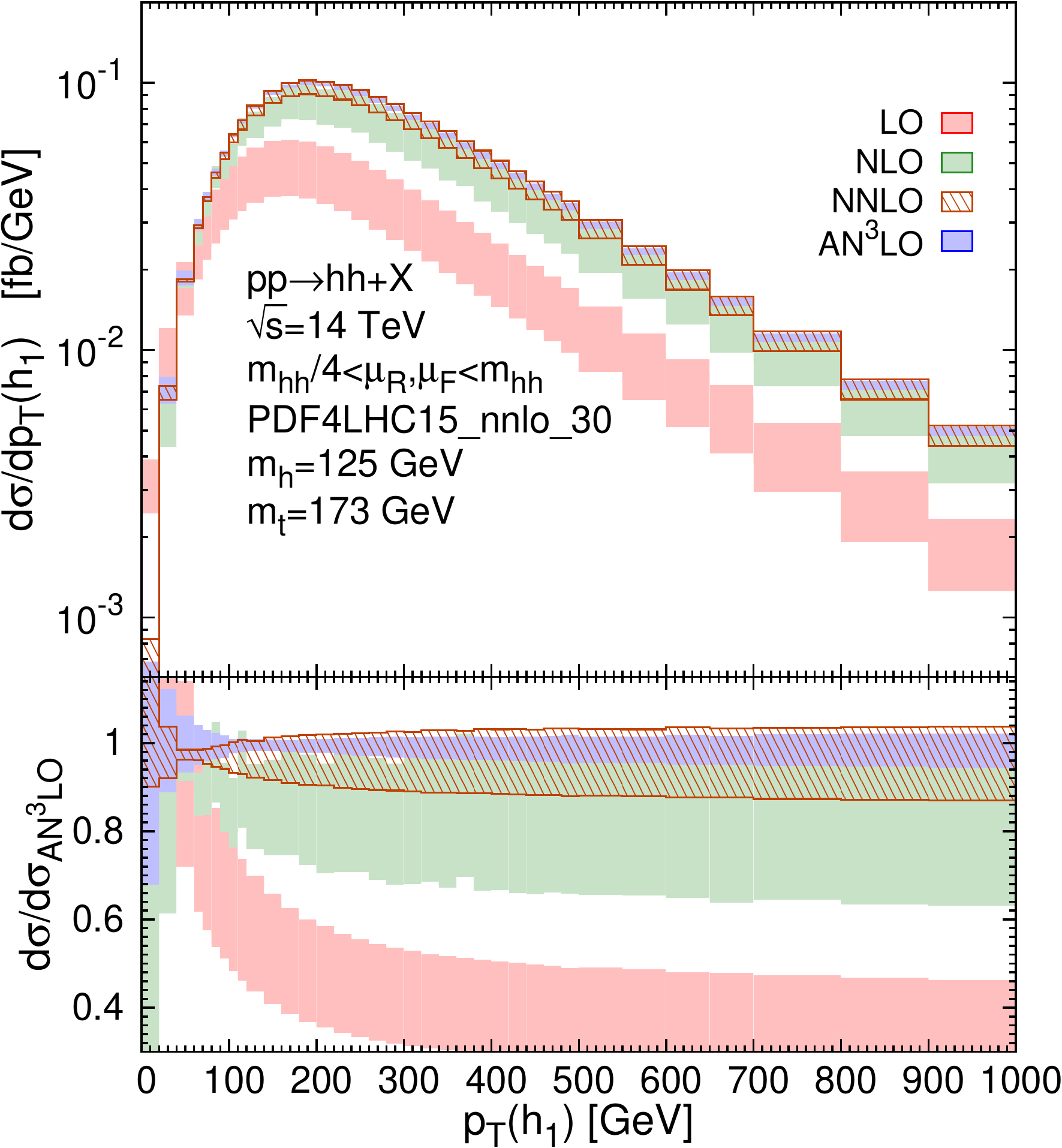}
    \includegraphics[scale=.33,draft=false,trim = 0mm 0mm 0mm 0mm,clip]{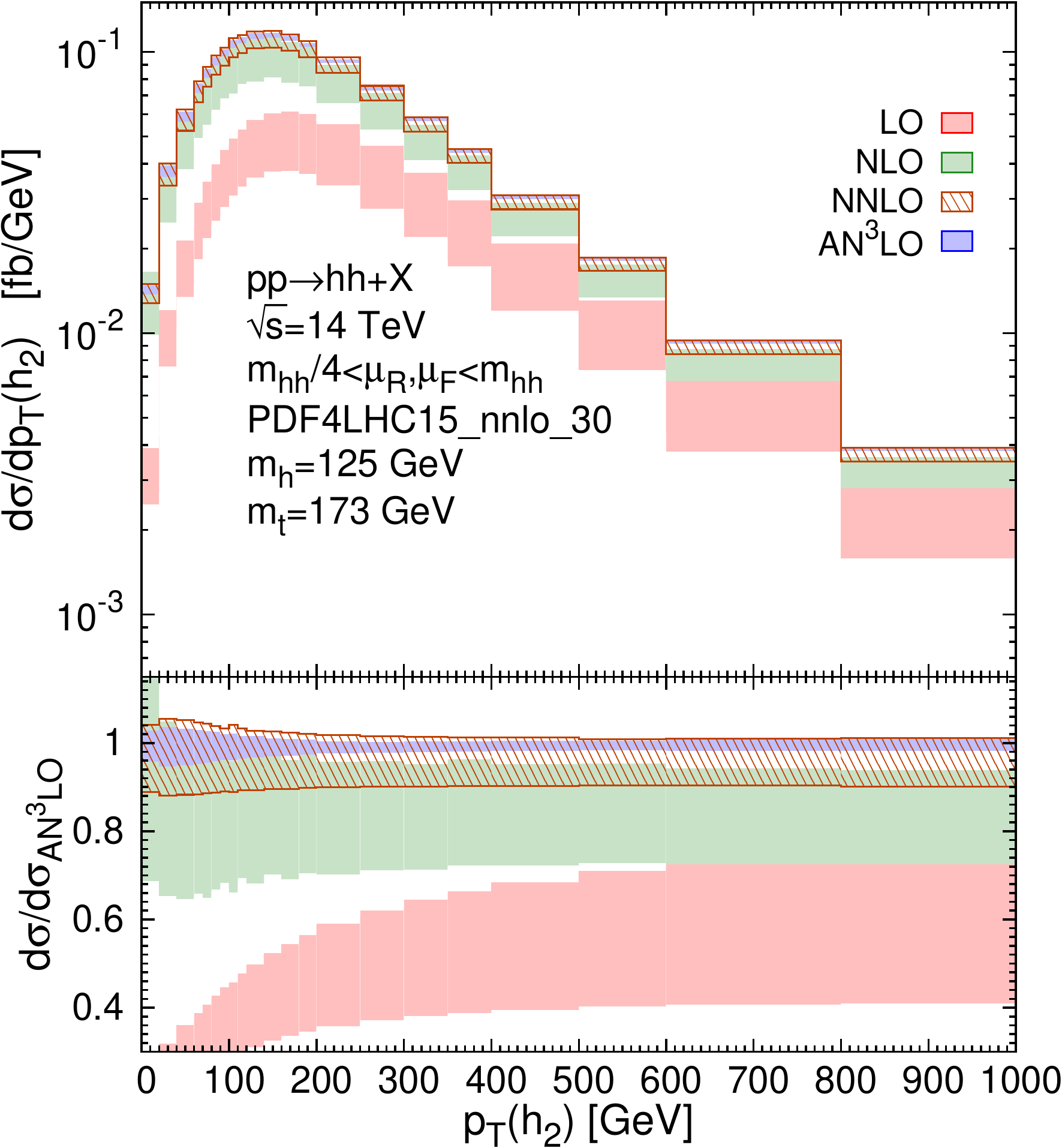}\\
    \includegraphics[scale=.33,draft=false,trim = 0mm 0mm 0mm 0mm,clip]{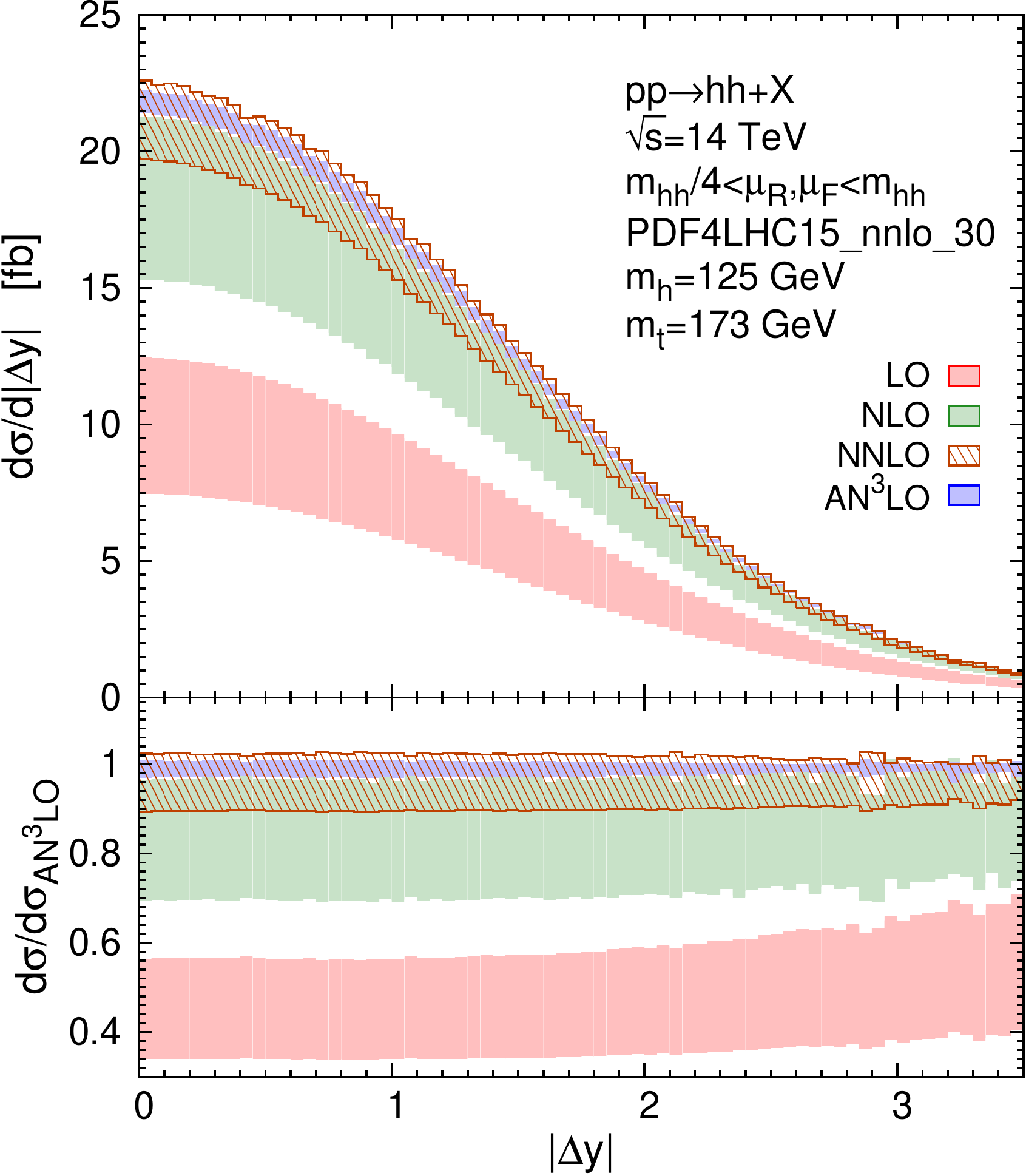}
    \includegraphics[scale=.33,draft=false,trim = 0mm 0mm 0mm 0mm,clip]{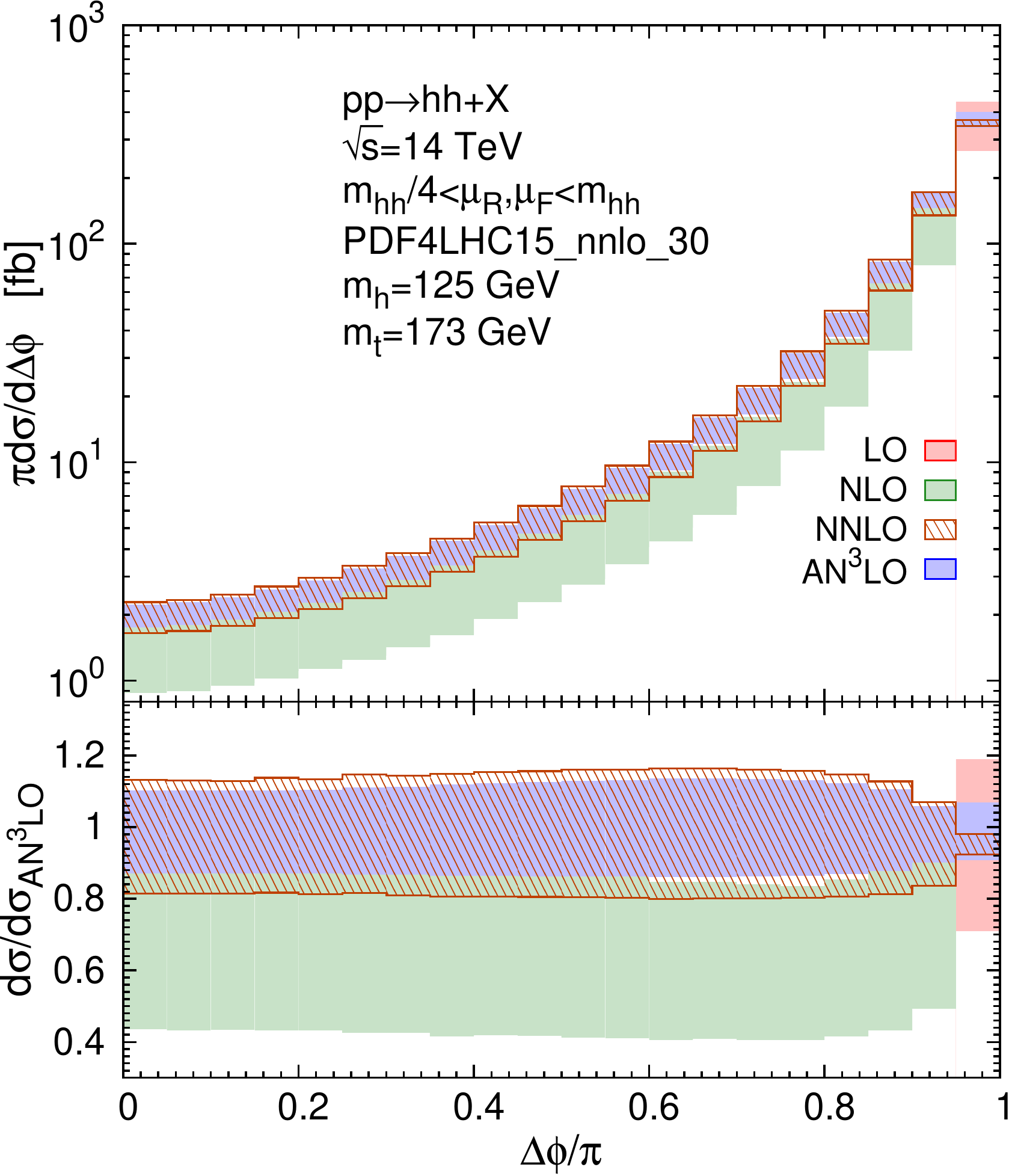}\\
    \vspace{0cm}
    \caption{Various distributions [$y_{hh}$ (up left), $y_h$ (up right), $p_T(h_1)$ (middle left), $p_T(h_2)$ (middle right), $\Delta y$ (low left), and $\Delta \phi$ (low right)] for the Higgs boson pair production in proton-proton collisions at $\sqrt{s}=14$ TeV.}
    \label{fig:othersLHC14}
\end{figure}

We have shown 6 differential distributions in figure~\ref{fig:othersLHC14} from LO to AN$^3$LO at $\sqrt{s}=14$ TeV, while the same distributions at other energies can be found in appendix~\ref{app:addplots}. They are the rapidity distribution of the Higgs boson pair ($O=y_{hh}$, up left), the rapidity distribution of a randomly selected Higgs boson~\footnote{Such a distribution is equivalent to the average of the two histograms, where each histogram represents a rapidity distribution of one labelled Higgs boson in the di-Higgs events.} ($O=y_h$, up right), the transverse momenta of leading-$p_T$ ($O=p_T(h_1)$, middle left) and subleading-$p_T$ ($O=p_T(h_2)$, middle right) of the two Higgs bosons, the absolute rapidity difference ($O=|\Delta y|$, low left) and the azimuthal angle difference ($O=\Delta \phi$, low right) between the Higgs pair. In all cases, AN$^3$LO corrections significantly reduce the scale uncertainties with respect to NNLO distributions, except $p_T(h_1)\rightarrow 0$ and $\Delta \phi \rightarrow \pi$. Like the dijet hadroproduction case~\cite{Frixione:1997ks,Frederix:2016ost}, the region  of $p_T(h_1)\rightarrow 0$ is largely populated by IR quanta radiations, which makes fixed-order perturbative calculations problematic. In addition, the $\Delta \phi$ distribution is quite special as all the LO events locate at $\Delta \phi=\pi$, i.e. back-to-back of the Higgs boson pair in the transverse plane. For all the $\Delta \phi\neq \pi$ bins, an N$^k$LO calculation only gives the N$^{k-1}$LO accuracy. On the contrast, the N$^k$LO accuracy can be achieved by a complete N$^k$LO calculation in the end point $\Delta \phi=\pi$. The region is however sensitive to the soft gluon emissions, which yields large logarithms to spoil the fixed-order perturbative calculations. Such a feature can be deduced from the fact that the scale uncertainty bands do not shrink from LO to AN$^3$LO. The pathological behaviour should be cured after performing the soft-gluon resummation in the region.

\section{N$^3$LO corrections with top-quark mass effects\label{sec:n3lomt}}

\subsection{Top-quark mass approximations at N$^3$LO }

It is well known that the top-quark mass effects are important in the Higgs boson pair production. Therefore, any relevant phenomenology studies should take into account these effects. However, the direct improvements of perturbative calculations with full top-quark mass dependence are technically very challenging because the lowest order is already loop-induced. The state-of-the-art calculation without performing $1/m_t^2$ expansion is NLO in $\alpha_s$. A standard way to improve the perturbative calculations is to combine the NLO full top-quark mass calculations (denote as {\bf NLO$_{m_t}$}) with the higher-order infinite top-quark mass calculations. The combination of the two different calculations is not unique, and therefore relies on various approximations.

There are several approximations to combine the differential cross sections in the infinite top-quark mass limit $d\sigma^{{\rm N}^k{\rm LO}}_{m_t\to\infty}$ and those with full top-quark mass dependence $d\sigma^{{\rm N}^l{\rm LO}}_{m_t}$ ($l<k$). In our case, we have $k=0,1,2,3$ and $l=0,1$. They are:
\begin{itemize}
\item {\bf N$^k$LO}$\oplus${\bf N$^l$LO$_{m_t}$}: this approximation simply improves the leading $m_t$ expansion term in $d\sigma^{{\rm N}^k{\rm LO}}_{m_t}-d\sigma^{{\rm N}^l{\rm LO}}_{m_t}$, i.e.
\begin{eqnarray}
d\sigma^{{\bf N}^k{\bf LO\oplus N}^l{\bf LO_{m_t}}}&=&d\sigma^{{\rm N}^l{\rm LO}}_{m_t}+\Delta \sigma_{m_t\to\infty}^{k,l},
\end{eqnarray}
where we have defined $\Delta \sigma_{m_t\to\infty}^{k,l}=d\sigma^{{\rm N}^k{\rm LO}}_{m_t\to\infty}-d\sigma^{{\rm N}^l{\rm LO}}_{m_t\to\infty}$.
\item {\bf N$^k$LO}$_{\bf B-i}\oplus${\bf N$^l$LO$_{m_t}$}: the correction part $\Delta\sigma^{k,l}_{m_t\to\infty}$ is simply improved by $\frac{d\sigma^{\rm LO}_{m_t}}{d\sigma^{\rm LO}_{m_t\to\infty}}$, i.e.
\begin{eqnarray}
d\sigma^{{\bf N}^k{\bf LO_{B-i}}\oplus {\bf N}^l{\bf LO_{m_t}}}&=&d\sigma^{{\rm N}^l{\rm LO}}_{m_t}+\Delta\sigma^{k,l}_{m_t\to\infty}\frac{d\sigma^{\rm LO}_{m_t}}{d\sigma^{\rm LO}_{m_t\to\infty}}.
\end{eqnarray}
\item {\bf N$^k$LO}$\otimes${\bf N$^l$LO$_{m_t}$}: this assumes that the QCD K factor $\frac{d\sigma^{{\rm N}^k{\rm LO}}_{m_t\to\infty}}{d\sigma^{{\rm N}^l{\rm LO}}_{m_t\to\infty}}$ in the infinite top-quark mass limit also applies to the other top-quark mass dependent terms. It is defined as
\begin{eqnarray}
d\sigma^{{\bf N}^k{\bf LO}\otimes{\bf N}^l{\bf LO_{m_t}}}&=&d\sigma^{{\rm N}^l{\rm LO}}_{m_t}\frac{d\sigma^{{\rm N}^k{\rm LO}}_{m_t\to\infty}}{d\sigma^{{\rm N}^l{\rm LO}}_{m_t\to\infty}}=d\sigma^{{\rm N}^l{\rm LO}}_{m_t}+\Delta \sigma_{m_t\to\infty}^{k,l}\frac{d\sigma_{m_t}^{{\rm N}^l{\rm LO}}}{d\sigma_{m_t\to\infty}^{{\rm N}^l{\rm LO}}}.
\end{eqnarray}
\end{itemize}
Other approximations are of course still possible (e.g. those introduced in refs.~\cite{Grazzini:2018bsd,Maltoni:2014eza}). However, they require the knowledge of the fully-differential  distributions, which is not known at N$^3$LO. In particular, the ``FT approximation''~\footnote{In the so-called FT approximation, the matrix elements in the infinite top-quark mass limit for each partonic subprocess are improved/reweighted by the ratios of the one-loop full top-quark mass squared amplitudes over the tree-level $m_t\to +\infty$ squared amplitudes.} introduced in refs.~\cite{Grazzini:2018bsd,Maltoni:2014eza} is considered as the most advanced predictions. We leave the FT approximation at N$^3$LO for a future study. Here, we decide to restrict ourselves with the above three approximations. Among them, {\bf N$^k$LO}$\otimes${\bf N$^l$LO$_{m_t}$} is expected to be the most accurate predictions, while {\bf N$^k$LO}$\oplus${\bf N$^l$LO$_{m_t}$} is the worst approximation because the finite top-quark mass effects are missing in the correction $\Delta\sigma^{k,l}_{m_t\to\infty}$. In the following, we will present the results under three approximations for comparison.

\subsection{Results}

With the same setup as described in the section \ref{sec:setup}, the full $m_t$-dependent NLO (differential) cross sections can be obtained by the public code~\cite{Heinrich:2017kxx,Heinrich:2019bkc} available in the {\sc\small Powheg-Box}~\cite{Nason:2004rx,Frixione:2007vw,Alioli:2010xd}. The scale uncertainties for each approximation in the present paper are estimated by taking the envelope of 9-point variations $\xi_R=\mu_R/\mu_0,\xi_F=\mu_F/\mu_0$ with $\mu_0=m_{hh}/2,\xi_R,\xi_F\in \{0.5,1,2\}$. The (differential) cross sections at each point are defined as
\begin{eqnarray}
d\sigma^{{\bf N}^k{\bf LO}\oplus{\bf N}^l{\bf LO_{m_t}}}(\xi_R,\xi_F)&=&d\sigma^{{\rm N}^l{\rm LO}}_{m_t}(\xi_R,\xi_F)+\Delta \sigma_{m_t\to\infty}^{k,l}(\xi_R,\xi_F),\nonumber\\
d\sigma^{{\bf N}^k{\bf LO_{B-i}}\oplus {\bf N}^l{\bf LO_{m_t}}}(\xi_R,\xi_F)&=&d\sigma^{{\rm N}^l{\rm LO}}_{m_t}(\xi_R,\xi_F)+\Delta\sigma^{k,l}_{m_t\to\infty}(\xi_R,\xi_F)\frac{d\sigma^{\rm LO}_{m_t}(1,1)}{d\sigma^{\rm LO}_{m_t\to\infty}(1,1)},\nonumber\\
d\sigma^{{\bf N}^k{\bf LO}\otimes{\bf N}^l{\bf LO_{m_t}}}(\xi_R,\xi_F)&=&d\sigma^{{\rm N}^k{\rm LO}}_{m_t\to\infty}(\xi_R,\xi_F)\frac{d\sigma^{{\rm N}^l{\rm LO}}_{m_t}(1,1)}{d\sigma^{{\rm N}^l{\rm LO}}_{m_t\to\infty}(1,1)}.\label{eq:mtscale}
\end{eqnarray}

\subsubsection{Inclusive total cross sections}

The  inclusive total cross sections after taking into account the top-quark mass effects are tabulated in table~\ref{tab:totxsmt}. The NLO cross section with full top-quark mass dependence (denoted by NLO$_{m_t}$) is 27.56 fb at $\sqrt{s}=13$ TeV,~\footnote{We have verified that the slightly offsets between our NLO$_{m_t}$ results and those in ref.~\cite{Grazzini:2018bsd} at $\sqrt{s}=13,14$ TeV can be attributed to the different PDFs. In our calculations, we always use the same NNLO PDF, while ref.~\cite{Grazzini:2018bsd} used a NLO PDF for the NLO calculations and a NNLO PDF in the NNLO calculations.} 
which is 6.8\% larger than the result in the infinite top-quark mass limit  (denoted by NLO)  shown in table~\ref{tab:totxs}.  
However, at 100 TeV,  the NLO$_{m_t}$ cross section~\footnote{A caveat for using the {\sc\small Powheg-Box} code to evaluate NLO$_{m_t}$ is the  presence of numerical errors because of the limitation of the two-loop numerical grid at large $m_{hh}$ and at high $p_T(h)$~\cite{Davies:2019dfy}. Such errors are negligible at 13 and 14 TeV and insignificant at 27 TeV, but may result in 1\% deviation at 100 TeV.} is more than 3 times smaller than the NLO result. 
This indicates that the large top-quark mass approximation is not valid any more at a very high energy collider. 

The  remaining scale uncertainties in NLO$_{m_t}$ cross sections are beyond 10\%. Such theoretical uncertainties are expected to be reduced by including higher-order QCD corrections. We evaluated the NNLO and N$^3$LO cross sections by using three approximations defined in the previous section based on the NLO$_{m_t}$ results. The central values as well as the scale uncertainties are presented in table~\ref{tab:totxsmt}. Because the finite $m_t$ corrections  in $\Delta \sigma^{k,1}_{m_t\to\infty},k=2,3$ are still missing, the N$^k$LO$\oplus$NLO$_{m_t}$ approximation is least accurate and even not reliable at 100 TeV, which is also implied in the shown pathological scale uncertainties. In contrast, both N$^k$LO$_{\rm B-i}\oplus$NLO$_{m_t}$ and N$^k$LO$\otimes$NLO$_{m_t}$ approximations have partially captured the finite top mass effects in the higher-order QCD correction pieces $\Delta \sigma^{k,1}_{m_t\to\infty}$. The differences between the two different approximations can be viewed as a way of estimating the remaining $1/m_t^2$ uncertainties, which are around 2-3\%.
In particular, we take N$^3$LO$\otimes$NLO$_{m_t}$ predictions as the state-of-the-art. The relative scale uncertainties in N$^k$LO$\otimes$NLO$_{m_t}$ are identical to those in N$^k$LO.

\begin{table}[h]
\centering
\begin{tabular}{|c|c|c|c|c|}
\hline
$\sqrt{s}$ & $13$ TeV & $14$ TeV & $27$ TeV & $100$ TeV\\\hline
NLO$_{m_t}$ & $27.56_{-13\%}^{+14\%}$ & $32.64^{+14\%}_{-12\%}$ & $126.2^{+12\%}_{-10\%}$  & $1119^{+13\%}_{-13\%}$ \\\hline
NNLO$\oplus$NLO$_{m_t}$ & $32.16_{-5.9\%}^{+5.9\%}$ & $38.29^{+5.6\%}_{-5.5\%}$ & $157.3^{+3.0\%}_{-4.7\%}$ & $1717^{+5.8\%}_{-12\%}$\\
NNLO$_{\rm B-i}\oplus$NLO$_{m_t}$ & $33.08^{+5.0\%}_{-4.9\%}$ & $39.16^{+4.9\%}_{-5.0\%}$  &  $150.8^{+4.6\%}_{-5.7\%}$ &  $1330^{+4.0\%}_{-7.2\%}$ \\
NNLO$\otimes$NLO$_{m_t}$ & $32.47^{+5.3\%}_{-7.8\%}$ & $38.42^{+5.2\%}_{-7.6\%}$ & $147.6^{+4.8\%}_{-6.7\%}$ & $1298^{+4.2\%}_{-5.3\%}$ \\\hline
N$^3$LO$\oplus$NLO$_{m_t}$ & $33.06^{+2.1\%}_{-2.9\%}$ & $39.40^{+1.7\%}_{-2.8\%}$  &  $163.3^{+4.0\%}_{-8.3\%}$ & $1833^{+14\%}_{-20\%}$ \\
N$^3$LO$_{\rm B-i}\oplus$NLO$_{m_t}$ & $34.17^{+1.9\%}_{-4.6\%}$ &  $40.44^{+1.9\%}_{-4.7\%}$ & $155.5^{+2.3\%}_{-5.0\%}$ & $1372^{+2.8\%}_{-5.0\%}$\\
N$^3$LO$\otimes$NLO$_{m_t}$ & $33.43^{+0.66\%}_{-2.8\%}$ & $39.56^{+0.64\%}_{-2.7\%}$  & $151.7^{+0.53\%}_{-2.4\%}$  & $1333^{+0.51\%}_{-1.8\%}$\\\hline
\end{tabular}
\caption{The inclusive total cross sections (in unit of fb) of Higgs boson pair production at different centre-of-mass energies $\sqrt{s}$ within the considered approximations.
  The quoted relative uncertainties are from the 9-point scale variations.}
\label{tab:totxsmt}
\end{table}

\subsubsection{Invariant mass distributions}

\begin{figure}[h]
    \centering
    \subfigure[$\sqrt{s}=13$ TeV]{\includegraphics[width=0.45\textwidth]{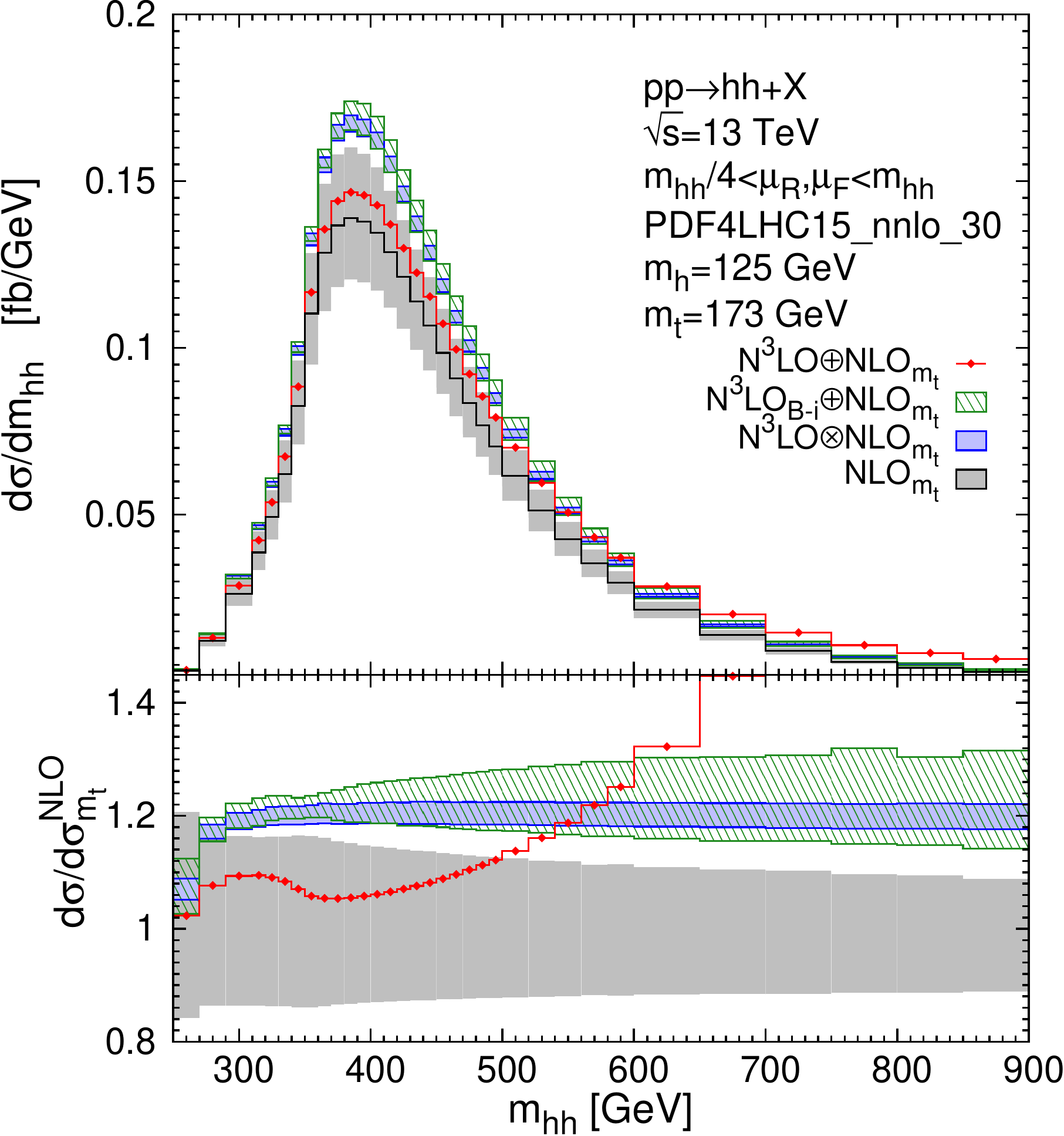}}
    \subfigure[$\sqrt{s}=14$ TeV]{\includegraphics[width=0.45\textwidth]{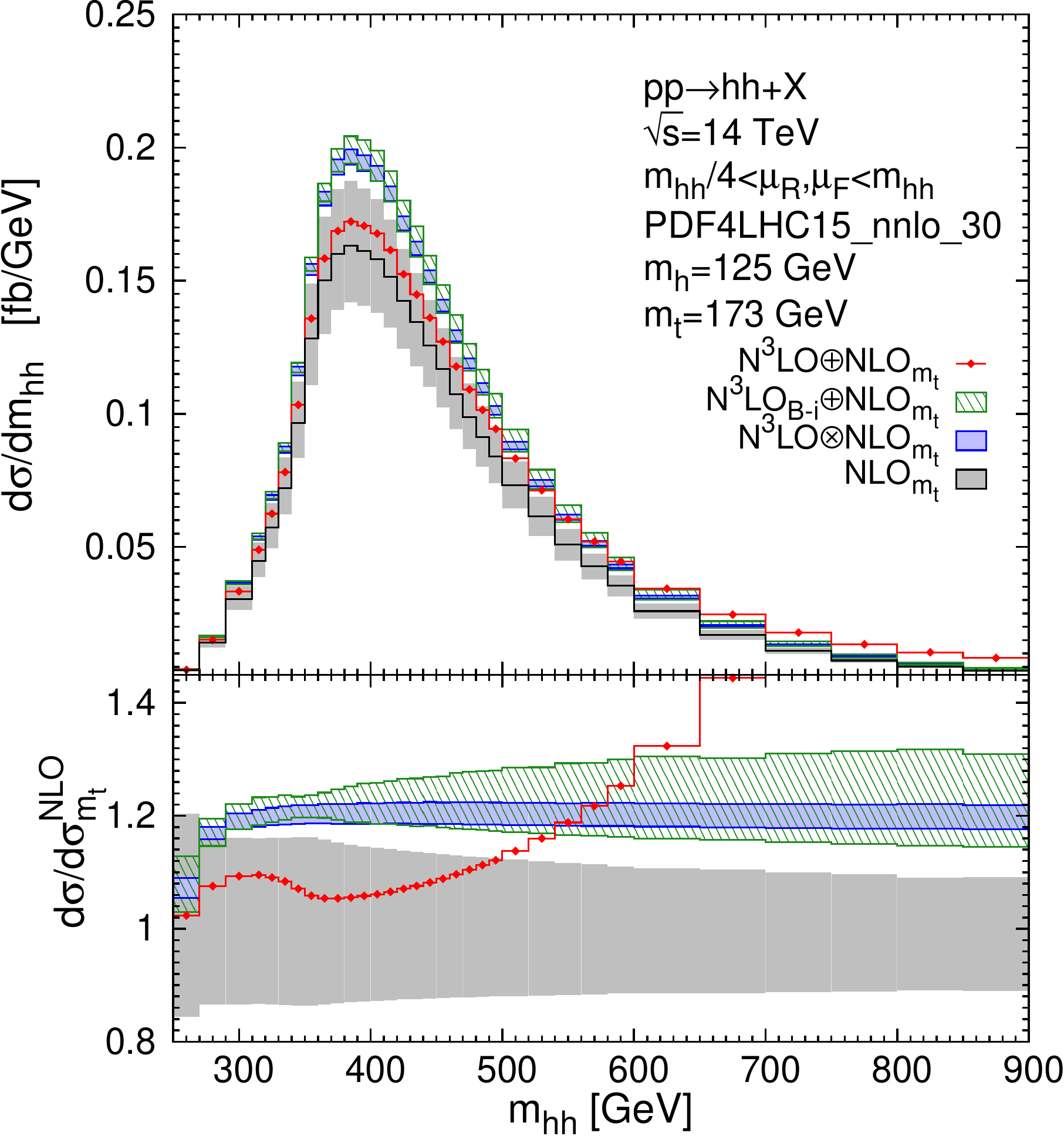}}\\
    \subfigure[$\sqrt{s}=27$ TeV]{\includegraphics[width=0.45\textwidth]{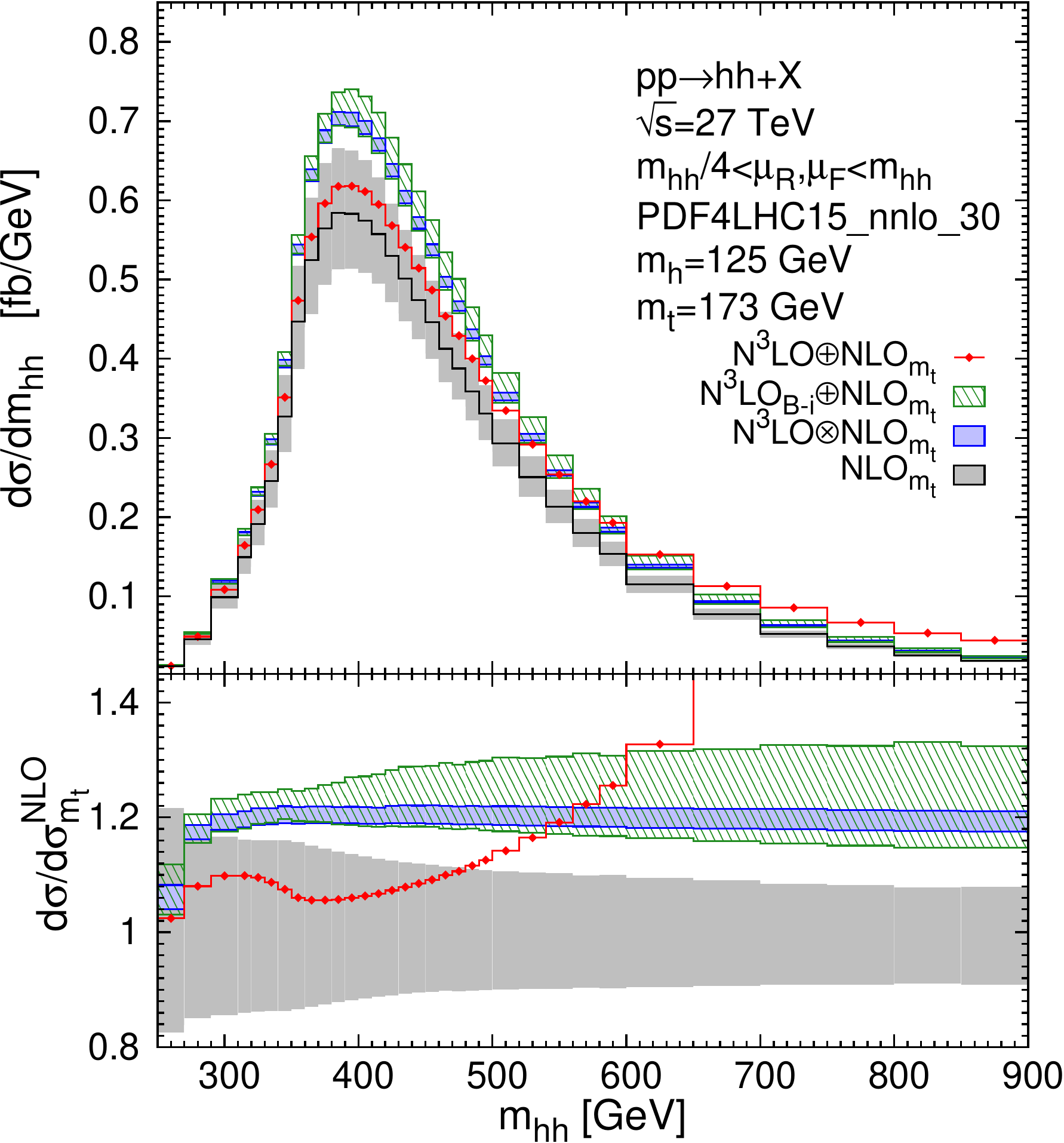}}
    \subfigure[$\sqrt{s}=100$ TeV]{\includegraphics[width=0.45\textwidth]{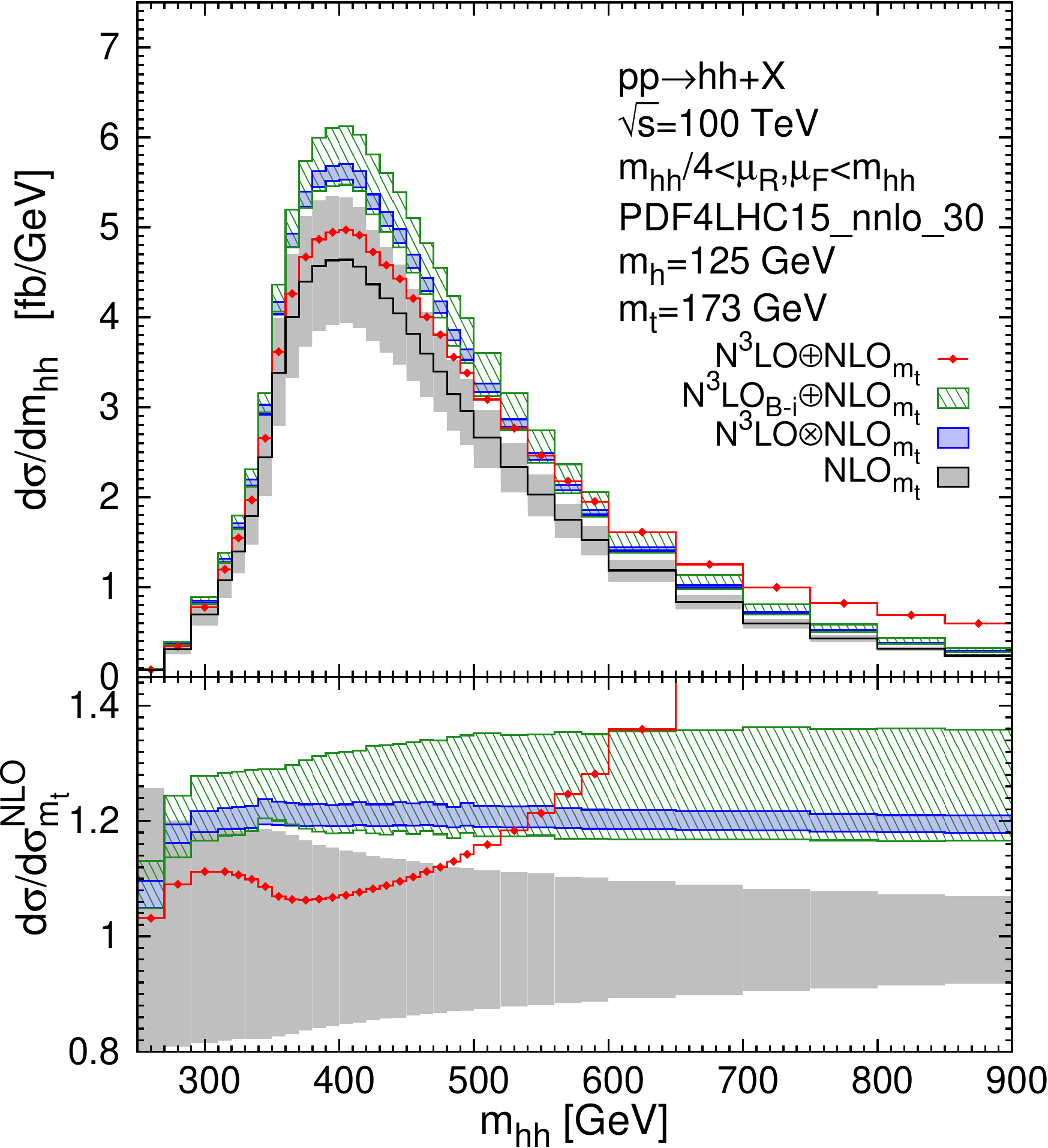}}
    \vspace{0cm}
    \caption{Invariant mass distributions of the Higgs boson pair under three top-quark mass approximations at $\sqrt{s}=13,14,27,100$ TeV. The bands represent the scale uncertainties.  The red, green, blue and black curves are the N$^3$LO$\oplus$NLO$_{m_t}$, N$^3$LO$_{\rm B-i}\oplus$NLO$_{m_t}$, N$^3$LO$\otimes$NLO$_{m_t}$ and NLO$_{m_t}$ predictions, respectively. The bottom panel shows the ratios to the NLO$_{m_t}$ distribution. }
    \label{fig:mhhmt1}
\end{figure}

\begin{figure}[h]
    \centering
    \subfigure[$\sqrt{s}=13$ TeV]{\includegraphics[width=0.45\textwidth]{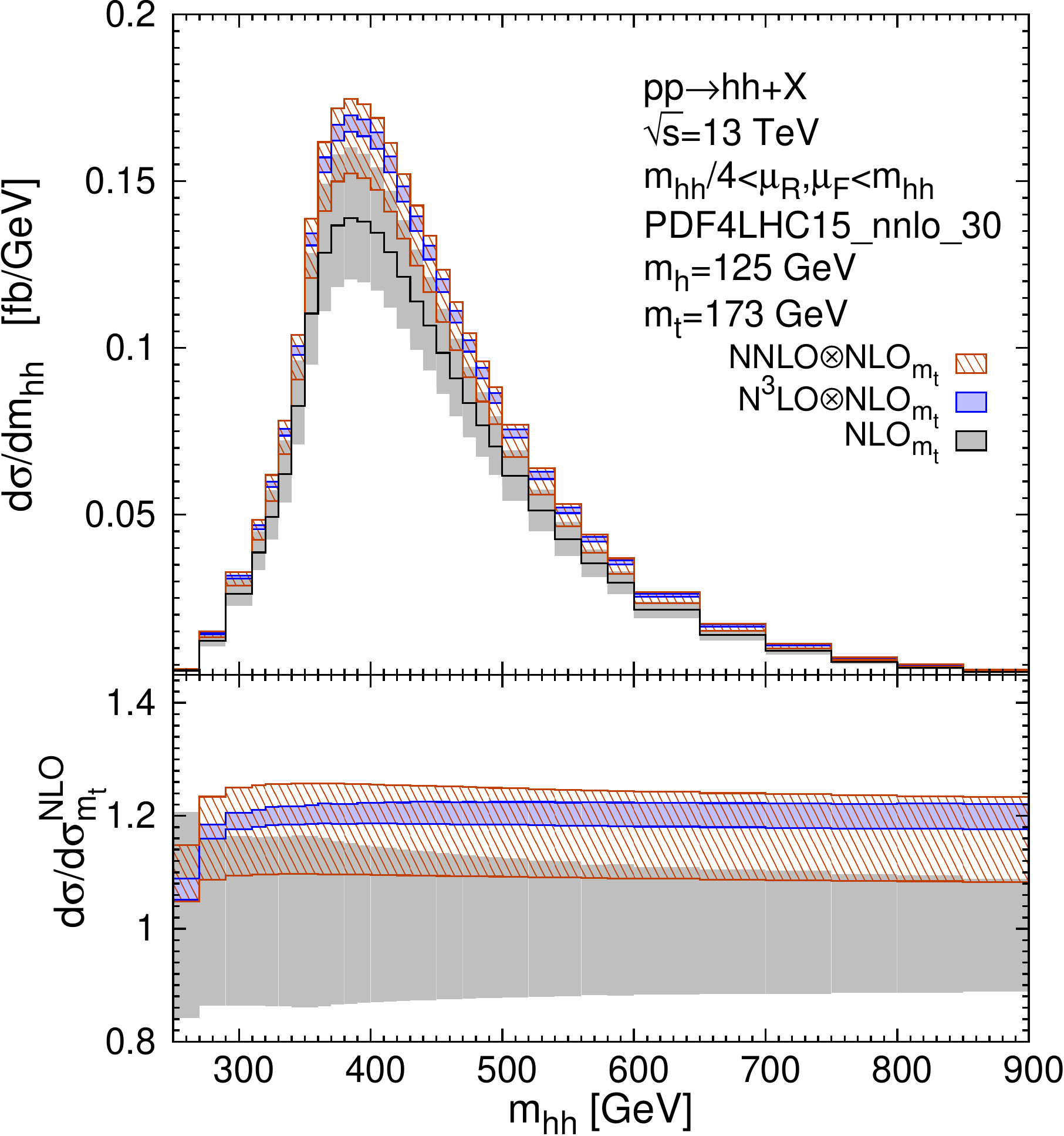}}
    \subfigure[$\sqrt{s}=14$ TeV]{\includegraphics[width=0.45\textwidth]{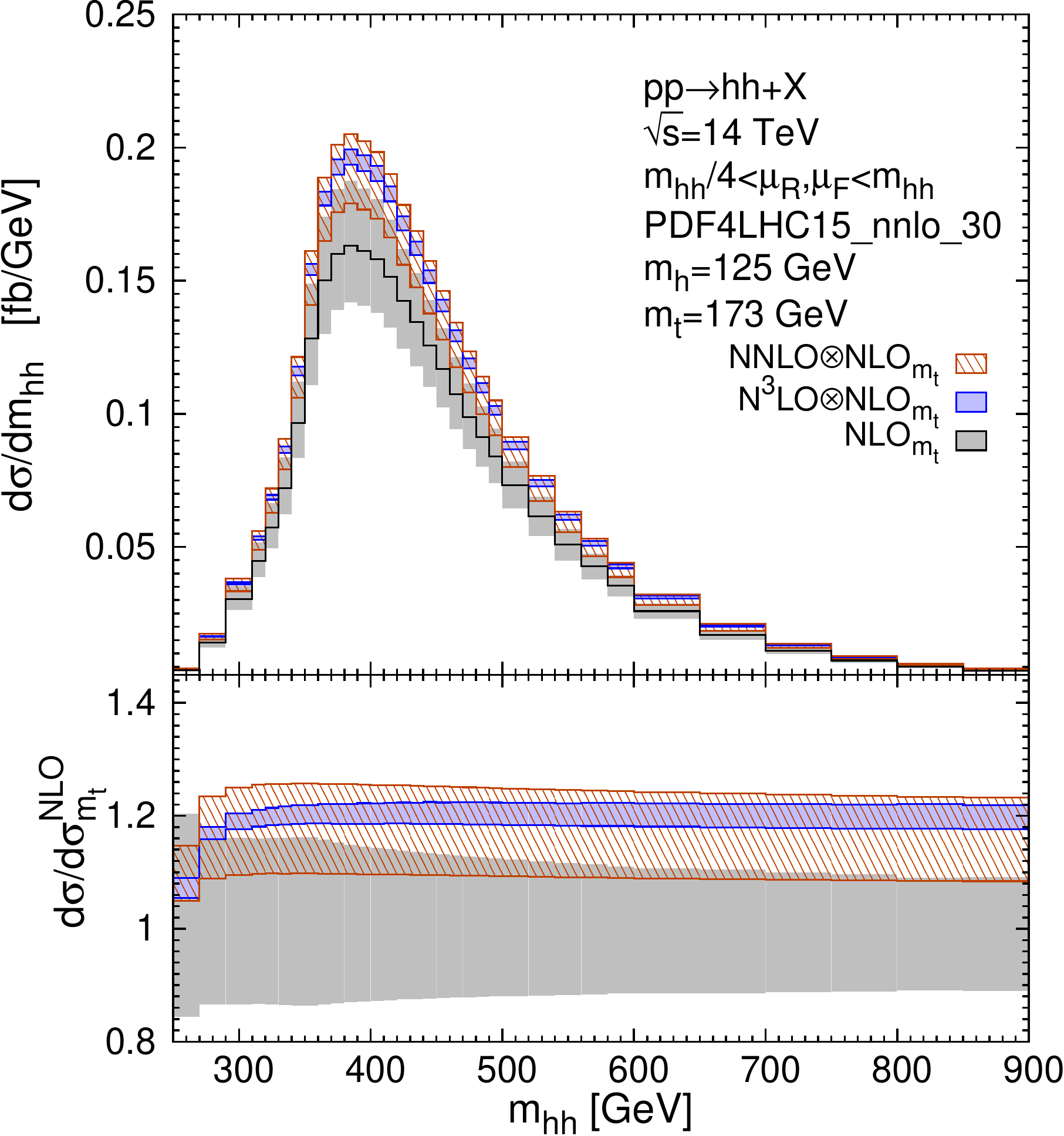}}\\
    \subfigure[$\sqrt{s}=27$ TeV]{\includegraphics[width=0.45\textwidth]{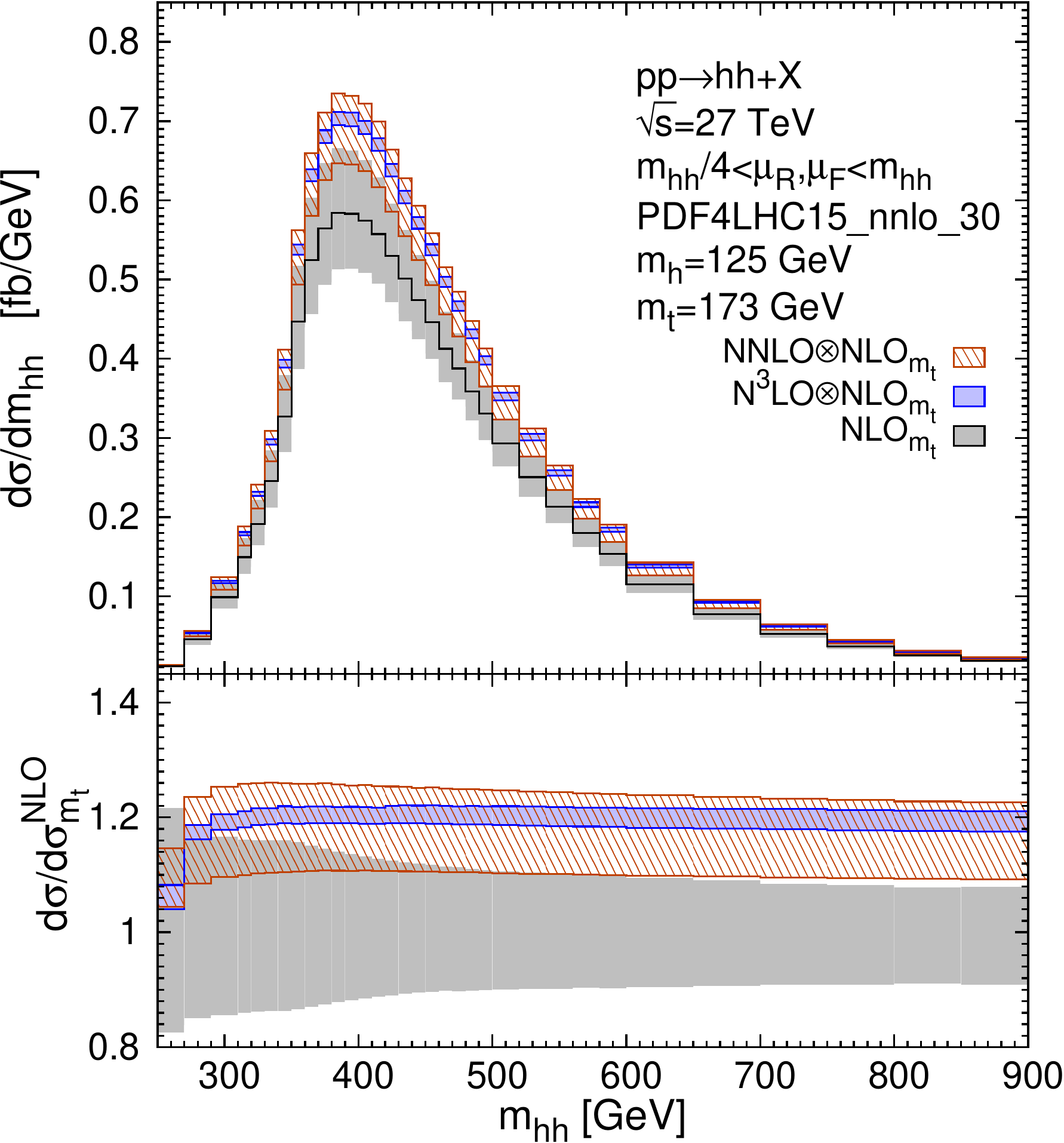}}
    \subfigure[$\sqrt{s}=100$ TeV]{\includegraphics[width=0.45\textwidth]{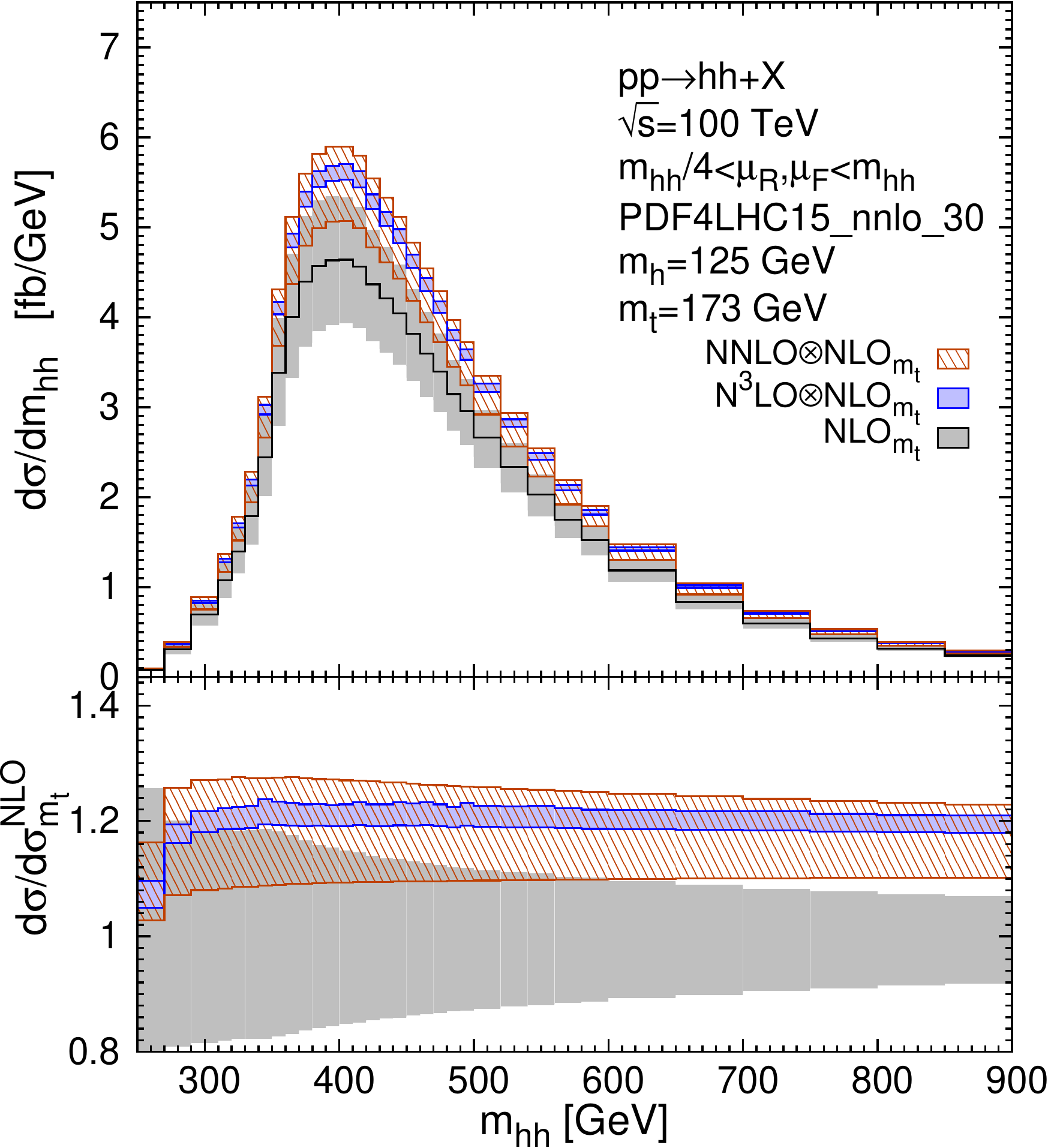}}
    \vspace{0cm}
    \caption{Comparisons of invariant mass distributions under N$^3$LO$\otimes$NLO$_{m_t}$ and NNLO$\otimes$NLO$_{m_t}$ approximations at $\sqrt{s}=13,14,27,100$ TeV. The bands represent the scale uncertainties.  The dark-orange, blue and black curves are the NNLO$\otimes$NLO$_{m_t}$, N$^3$LO$\otimes$NLO$_{m_t}$ and NLO$_{m_t}$ predictions, respectively. The bottom panel shows the ratios to the NLO$_{m_t}$ distribution. }
    \label{fig:mhhmt2}
\end{figure}

The invariant mass $m_{hh}$ distributions at 4 different energies $\sqrt{s}$ are shown in both figure~\ref{fig:mhhmt1} and figure~\ref{fig:mhhmt2}. In figure~\ref{fig:mhhmt1}, we have computed three different $m_t$ approximations at N$^3$LO. They are N$^3$LO$\oplus$NLO$_{m_t}$ (red lines), N$^3$LO$_{\rm B-i}\oplus$NLO$_{m_t}$ (green bands) and N$^3$LO$\otimes$NLO$_{m_t}$ (blue bands) together with the pure NLO$_{m_t}$ predictions (black bands). The N$^3$LO$\oplus$NLO$_{m_t}$ predictions significantly overshoot the other predictions when $m_{hh}>600$ GeV. Besides, the theoretical accuracy estimated via the scale variations in the N$^3$LO$_{\rm B-i}\oplus$NLO$_{m_t}$ predictions is degraded to NLO accuracy when $m_{hh}$ becomes larger than two times of the top-quark mass where the scale cancellations are not guaranteed. For N$^3$LO$\otimes$NLO$_{m_t}$, because of the manner of varying $\xi_R,\xi_F$ in differential cross sections eq.(\ref{eq:mtscale}), their relative scale uncertainties are exactly the same as N$^3$LO in section \ref{sec:mhhnomt}. Comparisons between NNLO$\otimes$NLO$_{m_t}$ and N$^3$LO$\otimes$NLO$_{m_t}$ predictions are given in figure~\ref{fig:mhhmt2}. Similar to what has been found at NNLO in ref.~\cite{Grazzini:2018bsd}, the higher-order QCD corrections are quite small near the threshold region $m_{hh}\simeq 2m_h$. The K factors $\frac{{\rm N^3LO\otimes NLO_{m_t}}}{{\rm NLO_{m_t}}}$ are almost  constants (around 1.2) at larger $m_{hh}$. A lesson from NNLO tells us that the NNLO$\otimes$NLO$_{m_t}$ predictions feature different shapes as the FT approximation. Therefore, it would be quite desirable to carry out the latter approximation at N$^3$LO, which is however beyond the scope of the present paper.

\subsubsection{Other differential distributions}

\begin{figure}[h]
    \centering
    \includegraphics[scale=.33,draft=false,trim = 0mm 0mm 0mm 0mm,clip]{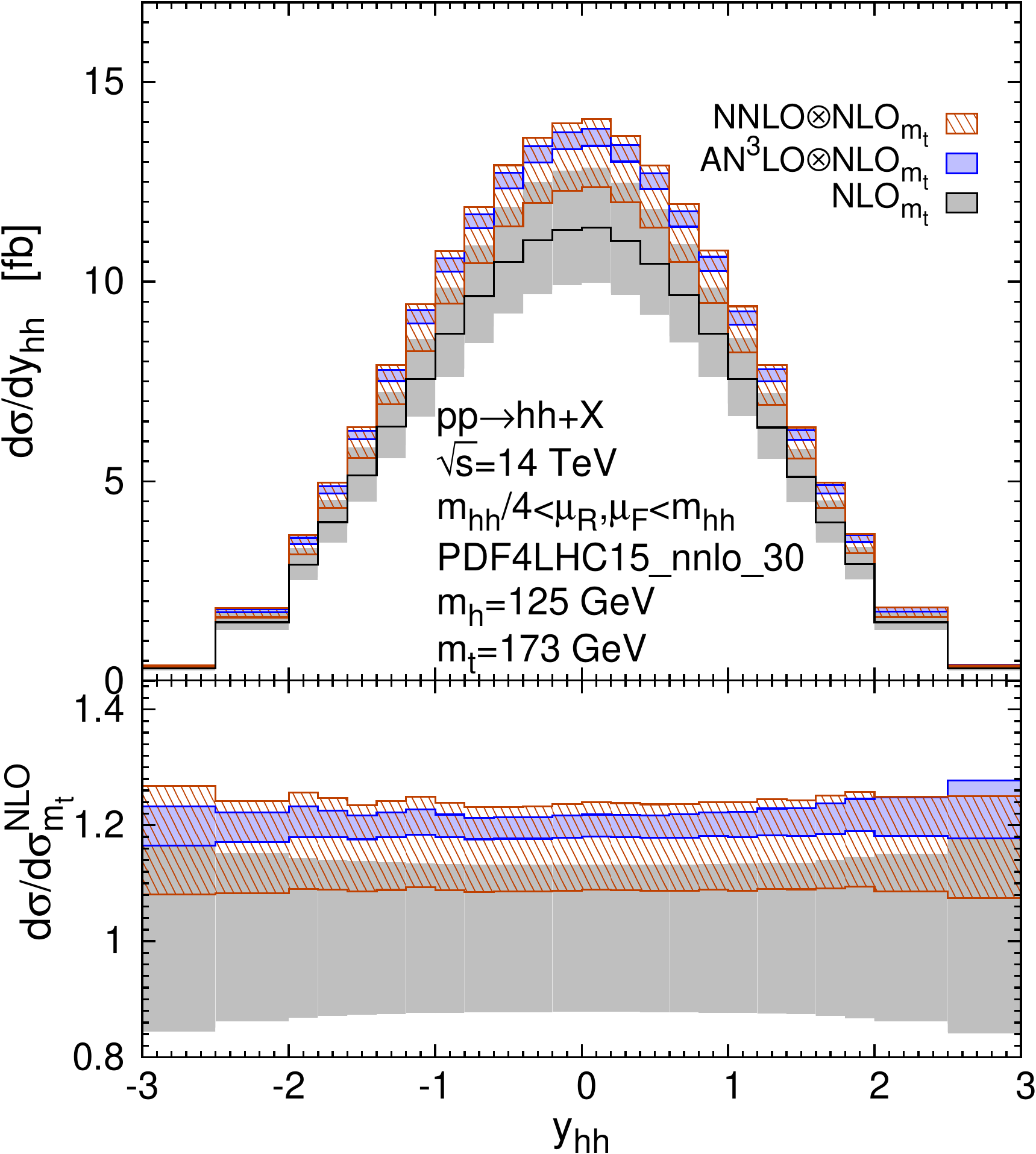}
    \includegraphics[scale=.33,draft=false,trim = 0mm 0mm 0mm 0mm,clip]{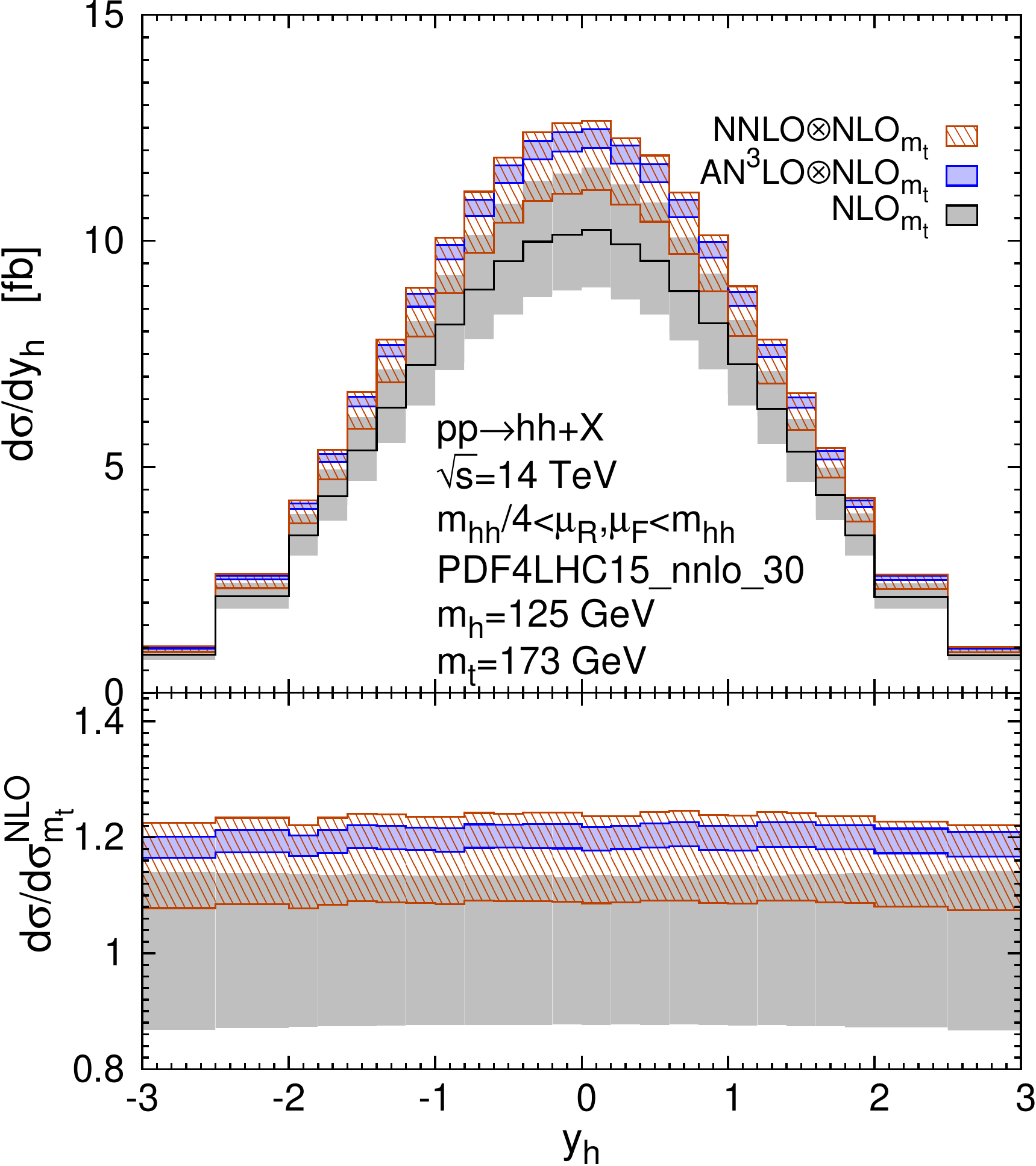}\\
    \includegraphics[scale=.33,draft=false,trim = 0mm 0mm 0mm 0mm,clip]{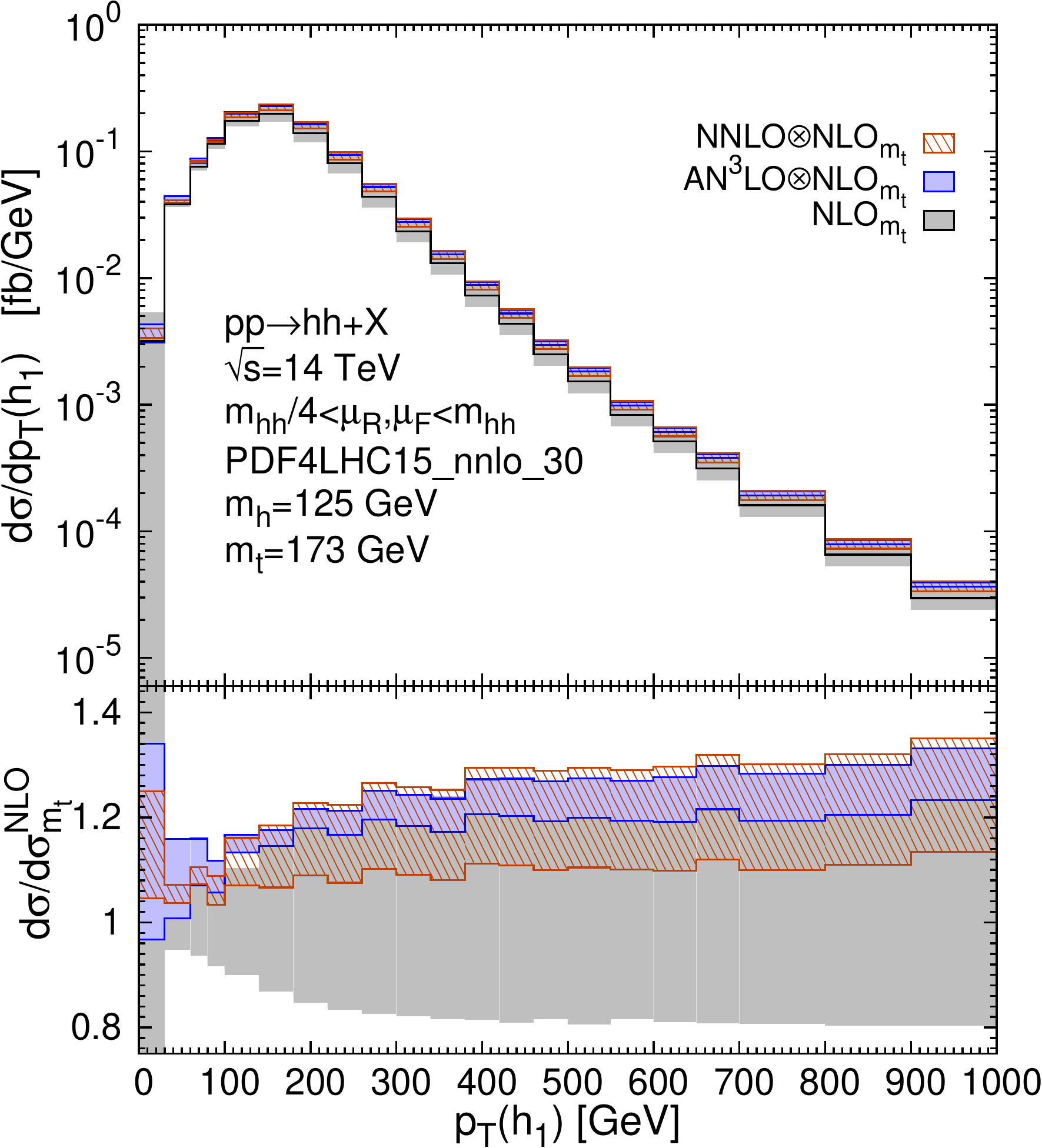}
    \includegraphics[scale=.33,draft=false,trim = 0mm 0mm 0mm 0mm,clip]{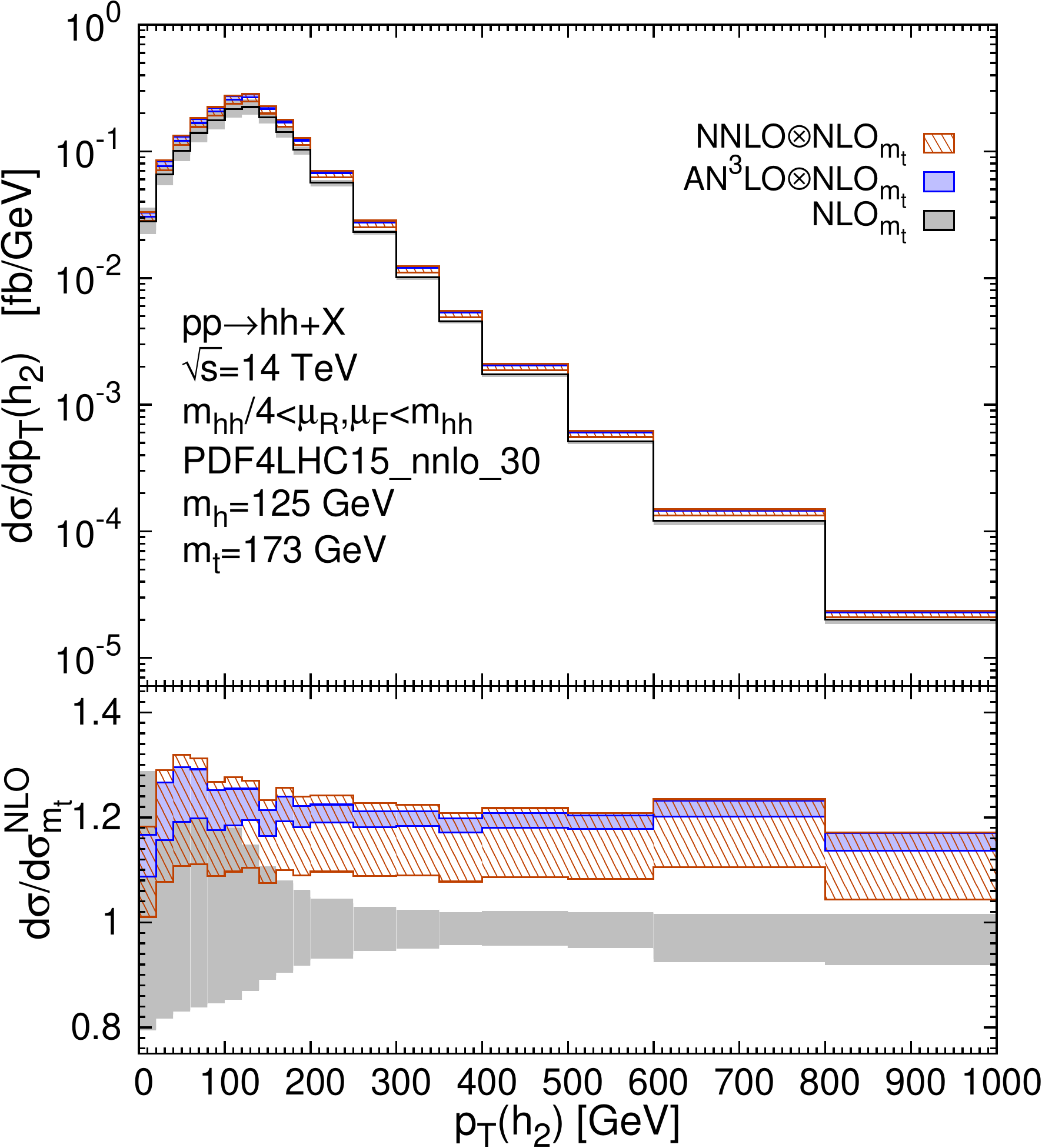}\\
    \includegraphics[scale=.33,draft=false,trim = 0mm 0mm 0mm 0mm,clip]{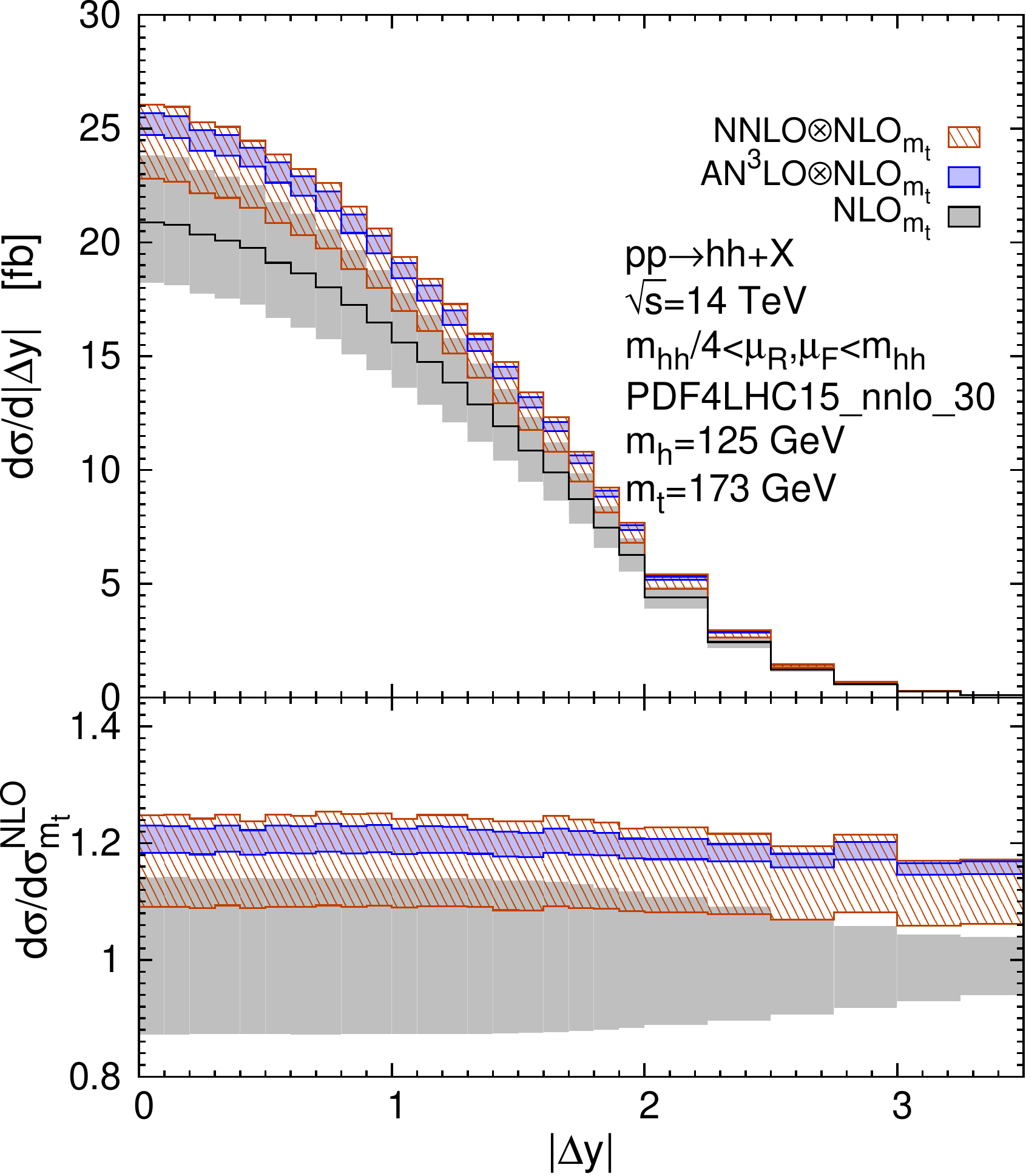}
    \includegraphics[scale=.33,draft=false,trim = 1.8mm 0mm 0mm 0mm,clip]{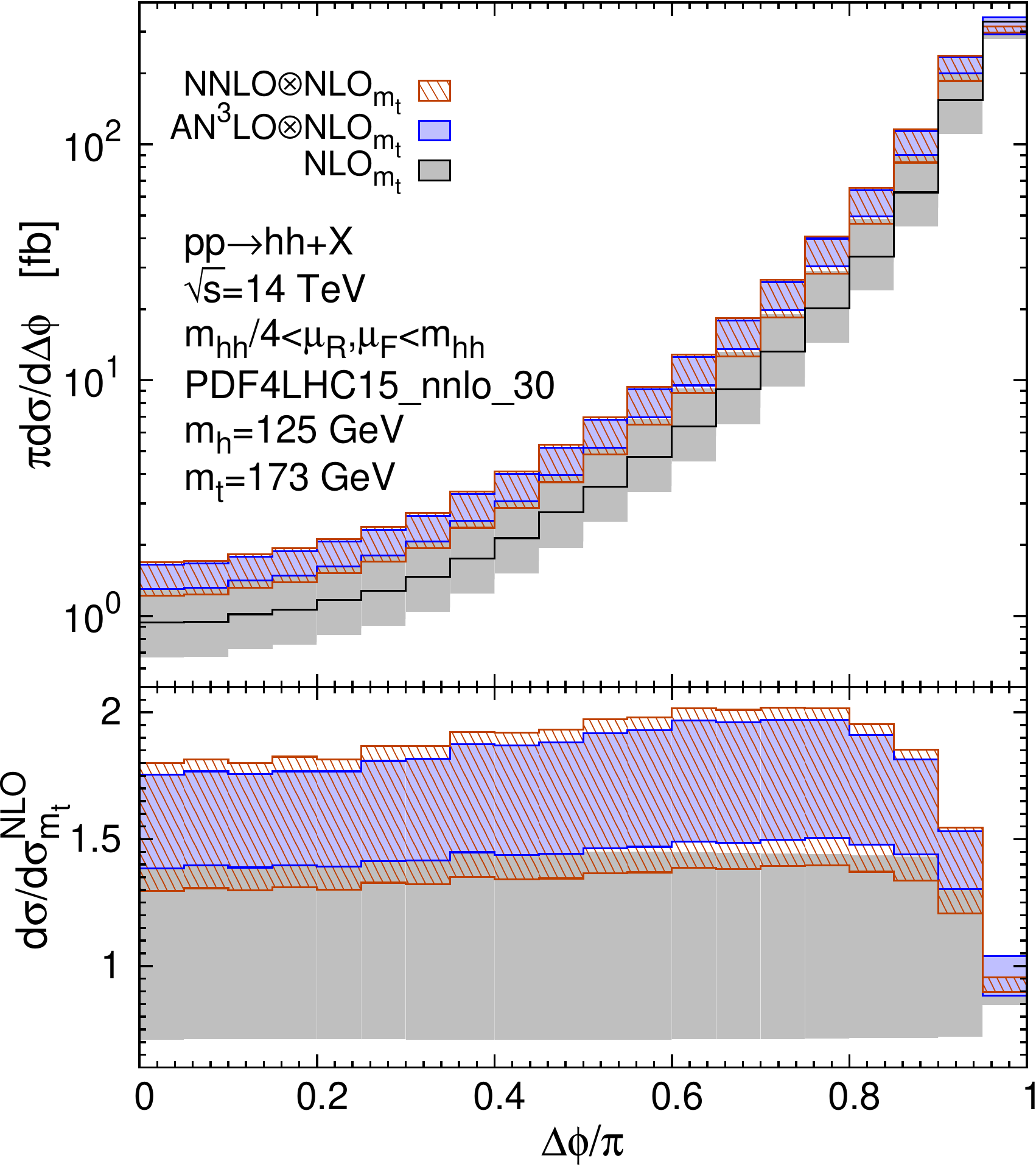}\\
    \vspace{0cm}
    \caption{Various distributions [$y_{hh}$ (up left), $y_h$ (up right), $p_T(h_1)$ (middle left), $p_T(h_2)$ (middle right), $|\Delta y|$ (low left), and $\Delta \phi$ (low right)] with top-quark mass effects for the Higgs boson pair production in proton-proton collisions at $\sqrt{s}=14$ TeV.}
    \label{fig:othersMTLHC14}
\end{figure}

With the approximation eq.(\ref{eq:N3LOapprox}) used at N$^3$LO in other observables, we are able to report our predictions for fully differential distributions of the Higgs boson pair production. We have shown 6 differential kinematic distributions at $\sqrt{s}=14$ TeV in figure~\ref{fig:othersMTLHC14} as our illustrative examples, while the same differential cross sections at $\sqrt{s}=13,27,100$ TeV can be found in appendix~\ref{app:addplots}. These kinematics are the rapidity of the Higgs pair (up left panel of figure~\ref{fig:othersMTLHC14}), the rapidity  of a random Higgs boson (up right panel of figure~\ref{fig:othersMTLHC14}), the transverse momenta $p_T$ of the harder (middle left panel of figure~\ref{fig:othersMTLHC14}) and the softer Higgs (middle right panel of figure~\ref{fig:othersMTLHC14}), the absolute rapidity difference $|\Delta y|$ (low left panel of figure~\ref{fig:othersMTLHC14}) and the azimuthal angle difference $\Delta \phi$  (low right panel of figure~\ref{fig:othersMTLHC14}) between the two Higgs particles. For the sake of clarity, we will only show the results of NLO$_{m_t}$ (black), NNLO$\otimes$NLO$_{m_t}$ (dark-orange) and AN$^3$LO$\otimes$NLO$_{m_t}$ (blue), where we have adopted the AN$^3$LO calculations to approximate the N$^3$LO differential cross sections.

The rapidity distribution of the Higgs boson pair 
reported in the up-left panel of figure~\ref{fig:othersMTLHC14} receives  approximately a uniform K factor $\frac{{\rm AN^3LO\otimes NLO_{m_t}}}{{\rm NLO_{m_t}}}\simeq 1.2$. The shape of the distribution is mainly driven by the partonic luminosity encoded in the PDF. The scale uncertainty band is reduced from NNLO$\otimes$NLO$_{m_t}$ to AN$^3$LO$\otimes$NLO$_{m_t}$ by a factor of four.

Because the rapidity distributions of the leading-$p_T$ and subleading-$p_T$ Higgs bosons are sensitive to soft-gluon radiations, i.e. not IR safe at fixed orders, we instead show the rapidity distribution of a random Higgs boson. The latter histogram is equivalent to the arithmetic mean of the former two histograms. Similar to the $y_{hh}$ distribution, the higher-order QCD corrections only change the shape slightly. The central region has a bit larger radiative corrections than the forward and backward regions. The difference is however quite insignificant, which is only at 1-2 percent level. The importance of the inclusion of $\mathcal{O}(\alpha_s^5)$ corrections is evident from the obvious reduction of theoretical uncertainties.

The differential cross sections in the transverse momenta of the leading-$p_T$ (harder) and the subleading-$p_T$ (softer) Higgs bosons can be found in the middle panels of figure~\ref{fig:othersMTLHC14}. These two transverse momenta are identical at LO. Beyond LO, due to the presence of extra real radiations, the difference between the two emerges. It is quite often that the Higgs boson will pick up a larger $p_T$ if it recoils against the hardest real radiation. For this reason, the real emission topologies are dominant in the tail of the $p_T(h_1)$ distribution, which results in the growth of the scale uncertainties in the high $p_T(h_1)$ bins. The AN$^3$LO$\otimes$NLO$_{m_t}$ scale uncertainty is $^{+2\%}_{-5\%}$ at the bin $p_T(h_1)\in [800,900]$ GeV, while those of NNLO$\otimes$NLO$_{m_t}$ and NLO$_{m_t}$ are $^{+7\%}_{-10\%}$ and $^{+25\%}_{-19\%}$ respectively in the same bin. At low $p_T(h_1)$, the QCD radiative corrections become perturbatively unstable~\footnote{It can be clearly seen from the fact that the scale uncertainties in the NLO$_{m_t}$ result blow up in the first bin.} due to the large logarithms of $\left(p_T(h_1)-p_T(h_2)\right)/\mu_0<p_T(h_1)/\mu_0\rightarrow 0$. The scale uncertainties of AN$^3$LO$\otimes$NLO$_{m_t}$ are larger than NNLO$\otimes$NLO$_{m_t}$ in the first three bins. Such a pathological behaviour reflects the fact that more large logarithms due to the soft-gluon radiations appear in the higher order $\alpha_s$ calculation.
On the other hand, the subleading $p_T$ distribution receives quite uniform K factors at NNLO and AN$^3$LO except the first bin, where the K factors are lower in the first bin than others. It has been shown in ref.~\cite{Borowka:2016ypz} that the NLO QCD corrections are vanishing in the tail of the $p_T(h_2)$ distribution. This makes the NLO$_{m_t}$ scale variation very small. However, we  do not have  an understanding in depth for such a behaviour at the moment. In the tail, we find that only AN$^3$LO$\otimes$NLO$_{m_t}$ has the comparable size of the scale variation with NLO$_{m_t}$.

Finally, we are in the position to discuss the two kinematic correlation distributions between the Higgs boson pair. They are the rapidity difference $\Delta y$ and the azimuthal angle difference $\Delta \phi$ in the low two panels of figure~\ref{fig:othersMTLHC14}. The significance of the higher-order QCD corrections is slowly reduced from the two near Higgs boson ($|\Delta y|\sim 0$) region to the region where the two Higgs particles are far away (i.e. $|\Delta y|$ is large). This is because a large $|\Delta y|$ usually corresponds to a large invariant mass of the Higgs pair $m_{hh}$, where the latter is proportional to our hard scale. In particular,  in Born kinematics, we have $m_{hh}=2\sqrt{m_h^2+p_T^2}\cosh{\frac{\Delta y}{2}}$, where $p_T$ is the transverse momentum of an arbitrary Higgs boson.

The radiative corrections are dramatic in the $\Delta \phi$ distribution. 
All the Born-like $2\rightarrow 2$ events locate at $\Delta \phi=\pi$, as the two Higgs bosons are always in the back-to-back configuration in the transverse plane. All the contributions to the $\Delta \phi<\pi$ regime must be from the events with at least one additional jet in the final states. 
In the bins of $\Delta \phi <\pi$, NLO$_{m_t}$, NNLO$\otimes$NLO$_{m_t}$ and AN$^3$LO$\otimes$NLO$_{m_t}$ results correspond to the true LO, NLO and NNLO accuracy in $\alpha_s$.  
The K factor $\frac{{\rm AN^3LO\otimes NLO_{m_t}}}{{\rm NLO_{m_t}}}$ increases slightly from $\Delta \phi=0$ to $\Delta \phi = 0.8\pi$ and then drops quickly from  $\Delta \phi = 0.8\pi$ to $\Delta \phi = \pi$. 
The 9-point scale variations shift their central values by $^{+45\%}_{-29\%}$, $^{+15\%}_{-17\%}$ and $^{+10\%}_{-13\%}$ respectively in the first bin $\Delta \phi \in [0,0.05]\pi$. The uncertainty reduction from NNLO to AN$^3$LO is not as immense as in other cases. Since a small kick by soft gluon radiations will make the two Higgs bosons not being  back-to-back anymore, a reliable prediction for the region $\Delta \phi\sim \pi$ can only be achieved after performing a resummation calculation.

\subsubsection{Assessment of the top-quark mass approximations}

Before we close the section, we will discuss how good are our top-quark mass approximations. Since the full NNLO and N$^3$LO calculations with the full $m_t$ dependence are  absent, the way of estimating the remaining $m_t$ uncertainties is not unique. 

One obvious way is to assess the missing $m_t$ uncertainties by trying different approximations. This has been taken at NNLO in ref.~\cite{Grazzini:2018bsd} even with the most advanced one -- the FT approximation. In the inclusive cross sections, the FT approximation gives smaller predictions than other approximations, including the NNLO$\otimes$NLO$_{m_t}$ approximation, because of the additional $m_t$ contributions in the former. The difference is amplified a bit with the  increasing of $\sqrt{s}$. At NNLO, the difference between the FT approximation and the NNLO$\otimes$NLO$_{m_t}$ approximation is 5\% at 13 TeV to 9\% at 100 TeV. This is not surprising since the $m_t$ corrections become more important at larger energies. Given that the $m_t$ corrections are more-or-less orthogonal to the $\alpha_s$ corrections, we expect the N$^3$LO$\otimes$NLO$_{m_t}$ numbers in table~\ref{tab:totxsmt} should be lowered by a similar amount after we applied the FT approximation at N$^3$LO. Besides this normalisation, the shapes of NNLO$\otimes$NLO$_{m_t}$ and the FT approximation at NNLO are very close for $y_{hh}, p_T(h_1),p_T(h_2)$ and $\Delta \phi$ distributions, while those for $m_{hh}$ are quite distinct. The deviation between the N$^3$LO FT approximation and the N$^3$LO$\otimes$NLO$_{m_t}$ scheme can be viewed as a way to assign the theoretical uncertainties from the missing top mass corrections. Such a difference is expected to be similar to what has been found at NNLO~\cite{Grazzini:2018bsd}.

We can also follow the NLO discussions in ref.~\cite{Borowka:2016ypz} to assess the goodness of our top-quark mass approximations in the differential distributions, where both the NLO$_{m_t}$ (the results with the notation ``NLO'' in ref.~\cite{Borowka:2016ypz})  and the NLO$\otimes$LO$_{m_t}$ (those with the notation ``B-i, NLO HEFT'' in ref.~\cite{Borowka:2016ypz}) cross sections were computed. However, since we  have already used the full NLO$_{m_t}$ in our calculations, the remaining $m_t$ uncertainties are expected to be at least $\alpha_s$ suppressed with respect to the estimations from NLO versus NLO$\otimes$LO$_{m_t}$. The total cross sections are lowered by 14\% (24\%) at $\sqrt{s}=14$ (100) TeV from the NLO$\otimes$LO$_{m_t}$ approximation to the complete NLO$_{m_t}$ calculations. Both the FT approximation at NLO and the NLO$\otimes$LO$_{m_t}$ results overestimate the true NLO QCD corrections at large $m_{hh},p_T(h_1),p_T(h_2)$. On the other hand, the shapes of the rapidity distributions are quite similar between  NLO$\otimes$LO$_{m_t}$ and NLO$_{m_t}$. In the former cases, the missing $m_t$ correction uncertainties will be underestimated by using the first approach described in the previous paragraph. Instead, a better way to assess the top mass corrections in N$^3$LO$\otimes$NLO$_{m_t}$ is to multiply ($d\sigma^{\rm NLO}_{m_t}-d\sigma^{{\rm NLO}\otimes {\rm LO}_{m_t}}$) with $\alpha_s$.

\section{Summary\label{sec:summary}}

In the paper, we first carried out the N$^3$LO QCD corrections to the Higgs boson pair production via ggF at high-energy hadron colliders in the infinite top-quark mass limit. We have shown that the corrections at this order are essential and quite remarkable due to the huge reduction of the scale uncertainties, which amount to a factor of four with respect to the known NNLO results. It paves the way for the precision theoretical studies of the Higgs potential at the percent level. Besides the total cross sections, we are also able to predict the various differential distributions at N$^3$LO,  where an approximation is used in the distributions other than the Higgs pair invariant mass distributions. In general, we have shown very good perturbative convergences in all distributions, and the scale uncertainties are in good control. Besides the SM case, we have also studied the N$^3$LO impacts on the (differential) cross sections by varying the trilinear Higgs coupling $\lambda_{hhh}$ alone. The shapes are again found to be stable at N$^3$LO with respect to those at NNLO.

Based on these N$^3$LO results, we include the important top-quark mass effects at $\mathcal{O}(\alpha_s^4)$ and $\mathcal{O}(\alpha_s^5)$ via three different approximations, where the full $m_t$-dependent NLO calculations are taken from the public code~\cite{Heinrich:2017kxx,Heinrich:2019bkc}. The $m_t$ effects are indispensable for the realistic phenomenological applications. We take the (A)N$^3$LO$\otimes$NLO$_{m_t}$ approximation as our best predictions in this paper. The most advanced FT approximation for the process, requiring the full differential knowledge, will be left for our future studies. The theoretical uncertainties are further improved by the inclusion of both the N$^3$LO corrections and the finite $m_t$ corrections. The missing $m_t$ corrections are larger than the remaining scale uncertainties. Besides, there are several other additional uncertainty sources worthwhile being considered in order to improve the theoretical predictions further. They are the top-quark mass scheme dependence~\cite{Baglio:2018lrj}, electroweak corrections, bottom quark effects and the parametric uncertainties (e.g. $m_t$, $\alpha_s$ and PDF).
 
As a follow-up paper of our previous short letter ref.~\cite{Chen:2019lzz}, we have the opportunity to document all the technical details and validation materials here. In particular, we write down the analytic expressions of the one-loop amplitude and the new $R_2$ Feynman rules in the appendices. The NLO UFO model ready to be used in \mgshort\ is publically available and can be downloaded from \url{http://feynrules.irmp.ucl.ac.be/wiki/HEFT_DH}.

\section*{Acknowledgements} 
We thank Ding Yu Shao for collaborations at the early stage of this work. We are also grateful to Jonas Lindert for the clarifications of the results in ref.~\cite{Grazzini:2018bsd} and to Gudrun Heinrich about the two-loop grid used in {\sc\small Powheg-Box}.
LBC is supported by the National Natural Science Foundation of China(NSFC) under the grants 11747051 and 11805042.
HTL is supported by the Los Alamos National Laboratory LDRD program.
The work of HSS is supported by the ILP Labex (ANR-11-IDEX-0004-02, ANR-10-LABX-63). 
The work of JW has  been supported   by the program for Taishan scholars.

\newpage

\appendix

\section{Hard functions \label{app:hard}} 

The amplitudes for the Higgs boson pair production in the effective theory, $g(p_1)+g(p_2)\to h(p_3) + h(p_4)$, can be decomposed into two topologically distinct classes: Class-A with one effective vertex and Class-B with two effective vertices~\footnote{See figure 1 and figure 2 of ref.~\cite{Banerjee:2018lfq}.}, i.e.
\begin{align}
    \mathcal{M}_{ab}(gg\to hh)=\frac{i}{v^2}\epsilon^{\mu}(p_1)\epsilon^{\nu}(p_2)\left(\mathcal{M}_{ab}^{A ,\mu\nu}+\mathcal{M}_{ab}^{B ,\mu\nu}\right),
\end{align}
where $\epsilon^{\mu}(p_1)$ and $\epsilon^{\mu}(p_2)$ are the polarisation vectors of the two initial gluons. The prefactor $\frac{i}{v^2}$ in the above equation is chosen in order to recycle the same notations used in ref.~\cite{Banerjee:2018lfq}.
The amplitudes for Class-A and Class-B can be decomposed into two Lorentz covariant and gauge invariant terms~\cite{Glover:1987nx} 
\begin{align}
    \mathcal{M}_{ab}^{A/B ,\mu\nu} = \delta_{ab}\left(\mathcal{T}_{1}^{\mu\nu } \mathcal{M}_1^{A/B}+ \mathcal{T}_{2}^{\mu\nu } \mathcal{M}_2^{A/B}\right)
\end{align}
where the tensors are given by
\begin{eqnarray}
	\mathcal{T}_1^{\mu\nu} & = & g^{\mu \nu} - {1 \over p_1\cdot p_2} 
	p_1^\nu p_2^\mu  \,,
	  \\
	\mathcal{T}_2^{\mu\nu} & = & g^{\mu \nu} + {1 \over p_1\cdot p_2~ p_T^2} \Big(
	m_h^2~ p_2^\mu p_1^\nu - 2 p_1 \cdot p_3~ p_2^\mu p_3^\nu -2 p_2\cdot p_3~ p_3^\mu p_1^\nu
	+2 p_1\cdot p_2~ p_3^\mu p_3^\nu \Big)\,.\nonumber
\end{eqnarray}
with $p_T^2 =  (\hat{t}\hat{u}-m_h^4)/\hat{s}$.  The Mandelstam variables are defined as
\begin{align}
    \hat{s}=(p_1+p_2)^2, \qquad \hat{t}=(p_1-p_3)^2, \qquad \hat{u}=(p_2-p_3)^2.
\end{align}

For Class-A, after performing renormalisation in the $\overline{\rm MS}$ scheme, we have
\begin{align}
    \mathcal{M}^{A}_1 =& i\frac{\hat{s}}{2} \left( C_{hh} - C_h \frac{6\lambda_{hhh} v^2}{\hat{s}-m_h^2}\right) C_g \,,
    \nonumber \\ 
     \mathcal{M}^{A}_2 =& 0\,,
\end{align}
where $C_g$ is the gluon structure function which has been calculated up to three loops~\cite{Baikov:2009bg,Gehrmann:2010ue}.

For N$^3$LO QCD corrections,  we need the two-loop virtual correction to Class-B amplitudes. 
This was computed in~\cite{Banerjee:2018lfq}, where the finite two-loop four-point amplitudes are obtained by subtracting the IR divergences following the method in ref.~\cite{Catani:1998bh}.  
In our framework, a different subtraction method, namely the $\overline{\rm MS}$ subtraction, is applied, and thus
we have reconstructed the full amplitudes with IR poles in Class-B and then performed the renormalisation procedure according to the method in refs.~\cite{Becher:2009qa,Becher:2009cu}. 
As a result, we obtain the finite part
\begin{align}
     \mathcal{M}_i^B&=
     \mathcal{M}_i^{B,(0)} 
    +\frac{\alpha_s}{4\pi}\left[  \mathcal{M}_i^{B,(1), fin} +   \mathcal{M}_i^{B,(0)} \left( -3 L_s^2 -\frac{23}{3} L_s +\frac{\pi^2}{2}\right) \right]
    \nonumber \\ & 
    + \left( \frac{\alpha_s}{4\pi} \right)^2 \left[  \mathcal{M}_i^{B,(2), fin} +   \mathcal{M}_i^{B,(1)} \left( -3 L_s^2 -\frac{23}{3} L_s +\frac{\pi^2}{2}\right)  + \mathcal{M}_i^{B,(0)}  \left( \frac{9}{2}L_s^4  +\frac{46}{3}L_s^3
    \nonumber \right.\right. \\ & \left. \left.
  + \left(\frac{3 \pi ^2}{2}-\frac{151}{3}\right) L_s^2 + \left(18 \zeta_3+\frac{23 \pi ^2}{6} -\frac{1316}{9}\right)L_s -\frac{23 \zeta_3}{2}-\frac{19 \pi ^2}{54}\right)  
    \right]\, + \mathcal{O}(\alpha_s^5),
\end{align}
with $L_s=\ln( -\frac{\mu_R^2}{\hat{s}+i 0})$.
The Born amplitudes are given by
\begin{align}
    {\cal M}_{1}^{B,(0)} &=  i \frac{\alpha_s^2}{18 \pi^2} \hat{s}, \nonumber\\
    {\cal M}_{2}^{B,(0)} &=  i \frac{\alpha_s^2}{36 \pi^2} \frac{(\hat{t}+\hat{u})(\hat{t}\hat{u}-m_h^4)}{\hat{t}\hat{u}}.
\end{align}
$\mathcal{M}_i^{B,(j), fin}$ is the finite $j$-loop amplitude defined in eq.~(2.24) of ref.~\cite{Banerjee:2018lfq}.  
The one-loop amplitudes $  \mathcal{M}_i^{B,(1), fin}$  including the real and imaginary contributions are needed in this paper.  However, the explicit analytical results can not be found  in the literature. In this work we calculated the one-loop amplitudes using {\sc\small FeynArts}~\cite{Hahn:2000kx} and {\sc\small FIRE}~\cite{Smirnov:2014hma} packages, and the results read
\begin{align}
	\frac{{\cal M}_{1}^{B,(1),fin}}{{\cal M}_{1}^{B,(0)}} & =  -C_A\left(1+2\frac{m_h^4}{\hat{s}^2}\right)\Bigg[\text{Li}_2\left(1-\frac{m_h^4}{\hat{t}\hat{u}}\right)
	+2\text{Li}_2\left(\frac{m_h^2}{\hat{t}}\right)+2\text{Li}_2\left(\frac{m_h^2}{\hat{u}}\right)
	\nn \\ &
	-\frac{1}{2}\ln^2\left(\frac{\hat{t}}{\hat{u}}\right)-\frac{2\pi^2}{3} 
	+2\ln\left(1-\frac{m_h^2}{\hat{t}}\right)\ln\left(-\frac{m_h^2}{\hat{t}}\right)
	+2\ln\left(1-\frac{m_h^2}{\hat{u}}\right)\ln\left(-\frac{m_h^2}{\hat{u}}\right)
	\nn \\ & 
	-2i \pi \ln\left(\frac{(m_h^2-\hat{t})(m_h^2-\hat{u})}{\hat{t} \hat{u}-m_h^4}\right)\Bigg]
    +C_A\left(\frac{2m_h^2}{\hat{s}}+\frac{58}{9}\right)-\frac{10 }{9}n_f
        \nn \\ & 
    -\frac{11C_A-2n_f}{6}\left(\ln\left(\frac{\hat{t} \hat{u} \hat{s}^2}{\mu_R^8}\right)-2i\pi\right)
    +2C_h^{(1)}\,, 
\\
   \frac{{\cal M}_{2}^{B,(1),fin}}{{\cal M}_{2}^{B,(0)}} & =  -C_A\frac{\hat{t} \hat{u} (\hat{t}^2+\hat{u}^2-2m_h^4)((\hat{t}+\hat{u})^2-2m_h^4)}{(\hat{t}+\hat{u})(\hat{t} \hat{u} -m_h^4)^2\sqrt{\hat{s}(\hat{s}-4m_h^2)}}\left(4\text{Li}_2(y)+\ln^2(-y)+\frac{\pi^2}{3}\right)
   \nn\\&
   -2\pi^2 C_A\frac{\hat{t} \hat{u}(\hat{t}^2+\hat{u}^2)-2m_h^4 \hat{t} \hat{u}+2m_h^8}{3(\hat{t} \hat{u}-m_h^4)^2}+\frac{67}{9}C_A-\frac{10}{9}n_f
   \nn\\&
   +\frac{11C_A-2n_f}{3}\left(-\ln\left(\frac{\hat{s}}{\mu_R^2}\right)+i \pi-\frac{\hat{t}\ln\left(-\frac{\hat{u}}{\mu_R^2}\right)+\hat{u}\ln\left(-\frac{\hat{t}}{\mu_R^2}\right)}{\hat{t}+\hat{u}}\right) 
   \nn\\& 
   + C_A \bigg[
   \frac{\hat{u}(\hat{t}^4+\hat{t}^2\hat{u}^2-2m_h^4 \hat{t} \hat{u}+2m_h^8)}
   {(\hat{t}+\hat{u})(\hat{t} \hat{u} -m_h^4)^2}
   \left(-4\text{Li}_2\left(\frac{m_h^2}{\hat{t}}\right)
   -2\ln\left(-\frac{\hat{t}}{m_h^2}\right)\ln\left(\frac{\hat{s}}{m_h^2}\right)
      \nn \right. \\& \left.
   +\ln^2\left(-\frac{\hat{t}}{m_h^2}\right)
   +4\ln\left(1-\frac{m_h^2}{\hat{t}}\right)\ln\left(-\frac{\hat{t}}{m_h^2}\right)+\frac{\pi^2}{3}+ 2 i \pi \left(\ln\left(-\frac{\hat{t}}{m_h^2}\right)
      \nn \right. \right. \\& \left. \left.  
   +2\ln\left(1-\frac{m_h^2}{\hat{t}}\right)-\ln\left(\frac{\hat{s}}{m_h^2}\right) \right)\right) 
   + \hat{t} \leftrightarrow \hat{u} \bigg]+2C_h^{(1)}.
\end{align}
The dimensionless parameter $y$ is defined as $y=-\frac{\sqrt{\hat{s}}-\sqrt{\hat{s}-4m_h^2}}{\sqrt{\hat{s}}+\sqrt{\hat{s}-4m_h^2}}$. These analytical expressions have been cross-checked against {\sc\small MadLoop}~\cite{Hirschi:2011pa,Alwall:2014hca} and the scale-dependent terms in $  \mathcal{M}_i^{B,(2), fin}$~\cite{Banerjee:2018lfq}. 
The analytic results of the two-loop amplitudes have been obtained in ref.~\cite{Banerjee:2018lfq} and are expressed in terms of  the  multiple  polylogarithms,
 which can be evaluated numerically by the public {\sc\small Mathematica} package {\sc\small PolyLogTools}~\cite{Duhr:2019tlz}. 

The hard functions of class-$a$, -$b$ and -$c$ are given by
\begin{align}
    H^a =& \frac{1}{32v^4}|\mathcal{M}_1^A|^2\,,
    \nonumber \\
    H^b =& \frac{1}{16v^4}\Re[ \mathcal{M}_1^A  \mathcal{M}_1^{B*}]\,,
    \nonumber \\
    H^c =& \frac{1}{32v^4}\left(|\mathcal{M}_1^{B}|^2+| \mathcal{M}_2^{B}|^2\right)\,,
\end{align}
where we have averaged over the spins and colours of the two initial gluons and taken into account the symmetry factor $\frac{1}{2}$ for the two identical Higgs bosons. 
Note that the renormalisation is performed at the amplitude level and there is no interference between the two Lorentz structures. 

\section{The NLO model and  Feynman rules for the rational $R_2$ terms \label{app:model}}

The NLO simulations in the \mgshort\ framework require the derivations of two necessary ingredients from the effective Lagrangian $\mathcal{L}_{\rm eff}$ in eq.(\ref{eq:effL}) and the SM Lagrangian $\mathcal{L}_{\rm SM}$ on top of the information provided in a LO UFO model~\cite{Degrande:2011ua}. They are the UV counterterms to perform the one-loop renormalisation and the rational $R_2$ terms~\cite{Ossola:2008xq} originating from the integration of the $(d-4)$ parts of the loop integrands after decomposing their numerators into $4$-dimensional and $(d-4)$-dimensional pieces, where $d$ is the dimension of the spacetime in the dimensional regularisation. 

The QCD UV renormalisation counterterms in the theory can be related to the renormalisations of the strong coupling $\alpha_s$ and the wavefunctions of gluons and massless quarks. They are however identical to the QCD theory in the SM. Therefore, we will refrain from presenting them in the paper.

Similarly to the UV renormalisation, the computations of $R_2$ are also equivalent to those of tree-level amplitudes with a universal set of theory-dependent Feynman rules (see refs.~\cite{Draggiotis:2009yb,Garzelli:2009is,Garzelli:2010qm,Shao:2011tg,Pittau:2011qp} for QCD and electroweak corrections in the SM and refs.~\cite{Shao:2012ja,Page:2013xla} for the beyond the SM cases~\footnote{A collection of NLO-ready UFO models can be found at \url{http://feynrules.irmp.ucl.ac.be/wiki/NLOModels}.}). They can be derived once and for all (for each model) by just considering the one-particle irreducible one-loop Feynman diagrams. For di-Higgs production in the theory $\mathcal{L}_{\rm eff}+\mathcal{L}_{\rm SM}$, we have rederived the analytical expressions of $R_2$ Feynman rules for zero Higgs and one Higgs vertices by using an in-house {\sc\small Mathematica} programme with the aid of {\sc\small FeynRules}~\cite{Alloul:2013bka} and {\sc\small FeynArts}~\cite{Hahn:2000kx} packages. They have been successfully validated against those in the literature~\cite{Draggiotis:2009yb,Page:2013xla}. Besides, the results for vertices involving two Higgs bosons are new. The nonzero $R_2$ vertices involving two Higgs bosons are:\\
\begin{center}
\vspace{-0.7cm}
\framebox{$hh$~~~~$hhgg$~~~~$hhggg$~~~~$hhgggg$~~~~$hhq\bar{q}$~~~~$hhq\bar{q}g$}
\vspace{-0.2cm}
\end{center}
They read
\begin{eqnarray*}
\plaat{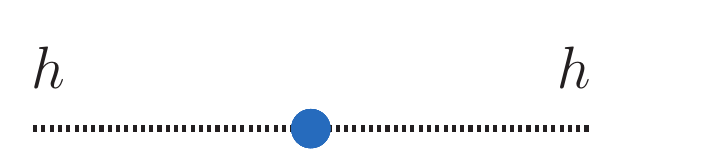}{0.15}{-1}&=&R_2(hh),~~\plaat{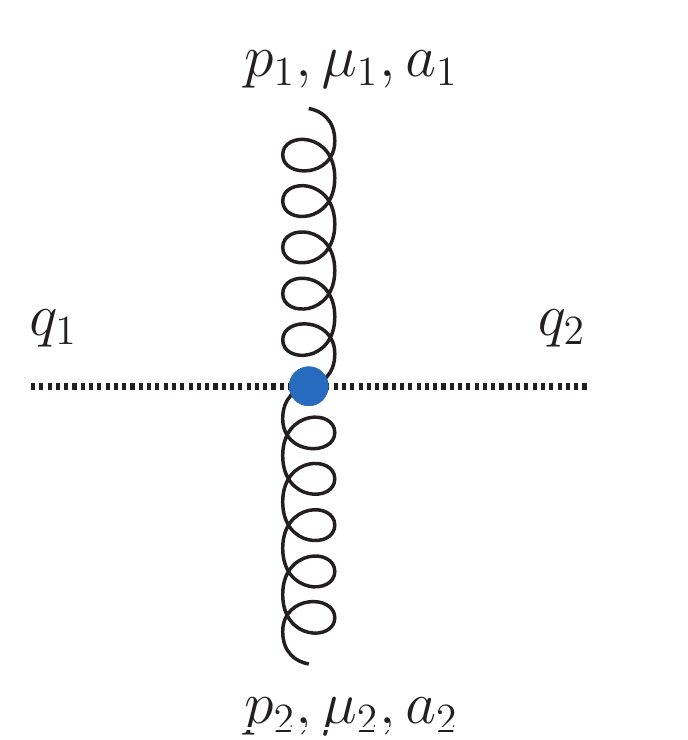}{0.15}{-31}=R_2(hhgg),~~\plaat{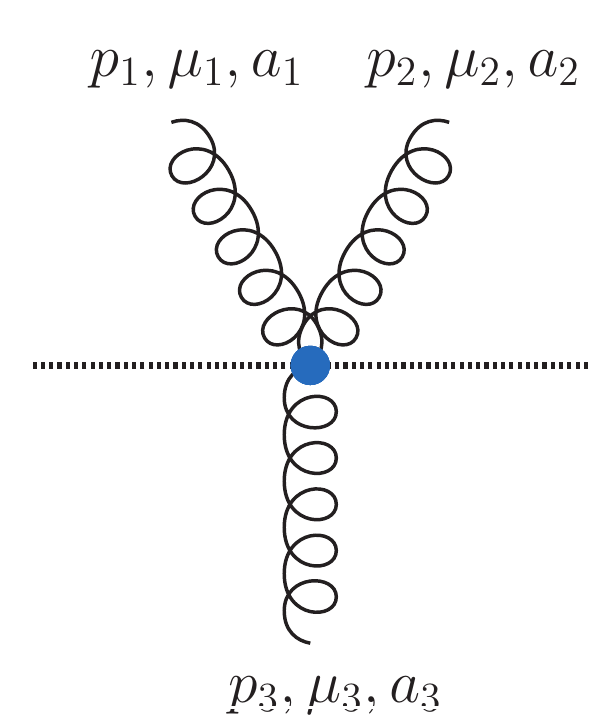}{0.15}{-35}=R_2(hhggg),\nonumber\\
\plaat{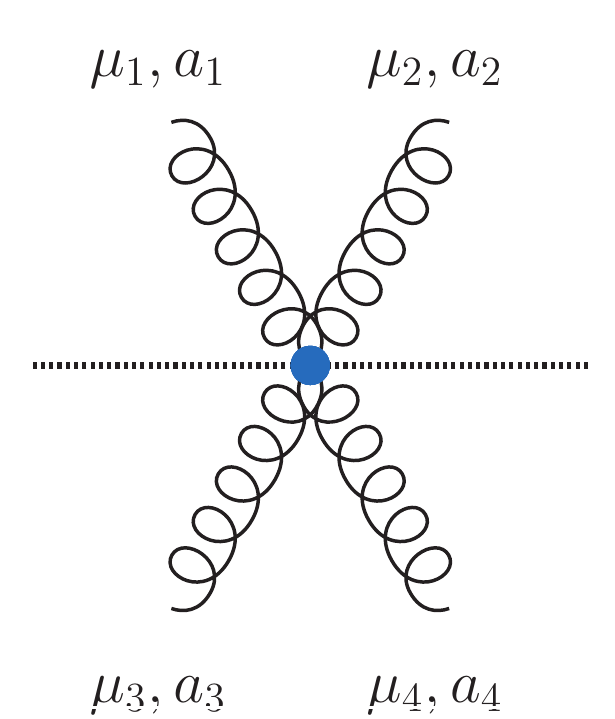}{0.15}{-35}&=&R_2(hhgggg),~~\plaat{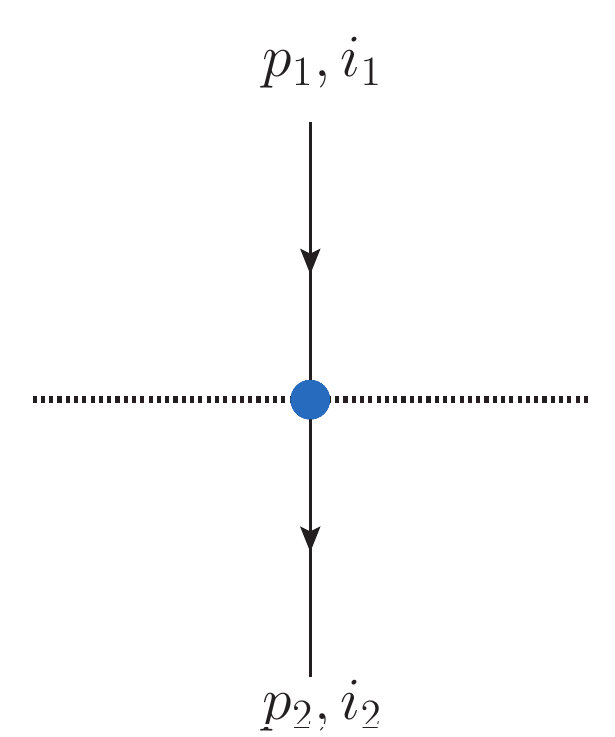}{0.15}{-33}=R_2(hhq\bar{q}),~~\plaat{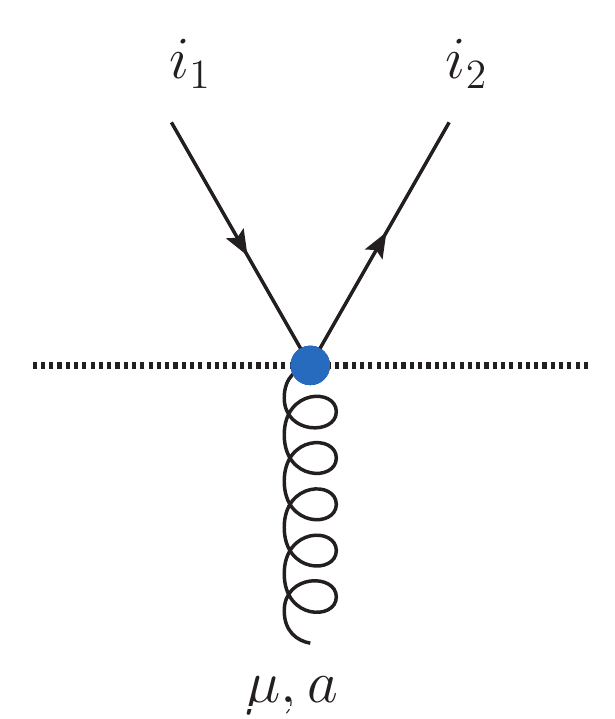}{0.15}{-35}=R_2(hhq\bar{q}g)
\end{eqnarray*}
with the expressions:
\begin{eqnarray}
R_2(hh)&=&-\frac{iC_h^2}{1920\pi^2v^2}\left(N_c^2-1\right)\left(30\lambda_{\rm HV}+17\right)\left(q^2\right)^2,\\
R_2(hhgg)&=&-\frac{iC_{hh}g_s^2}{384\pi^2v^2}N_c\delta_{a_1a_2}\left[p_1^{\mu_1}p_2^{\mu_2}+89p_1^{\mu_2}p_2^{\mu_1}
+14\left(p_1^{\mu_1}p_1^{\mu_2}+p_2^{\mu_1}p_2^{\mu_2}\right)\right.\\
&&~~~~~~~~~~~~~~~~~~~~~~~~\left.-g^{\mu_1\mu_2}\left(17p_1^2+17p_2^2+93p_1\cdot p_2\right)\right]\nonumber\\
&&-\frac{iC_h^2g_s^2}{3840\pi^2v^2}N_c\delta_{a_1a_2}\left[12p_1^{\mu_1}p_2^{\mu_2}+1152p_1^{\mu_2}p_2^{\mu_1}
+266\left(p_1^{\mu_1}p_1^{\mu_2}+p_2^{\mu_1}p_2^{\mu_2}\right)\right.\nonumber\\
&&\left.-g^{\mu_1\mu_2}\left(305p_1^2+305p_2^2+1200p_1\cdot p_2\right)-44\left(q_1^{\mu_1}q_2^{\mu_2}+q_1^{\mu_2}q_2^{\mu_1}\right)+70g^{\mu_1\mu_2}q_1\cdot q_2\right],\nonumber\\
R_2(hhggg)&=&-\left(\frac{15C_{hh}g_s^3}{128\pi^2v^2}+\frac{151C_h^2g_s^3}{1280\pi^2v^2}\right)N_cf_{a_1a_2a_3}V^{\mu_1\mu_2\mu_3}\left(p_1,p_2,p_3\right),\\
R_2(hhgggg)&=&-\frac{iC_{hh}g_s^4}{128\pi^2v^2}X_{a_1 a_2 a_3 a_4}^{\mu_1\mu_2\mu_3\mu_4}-\frac{iC_h^2g_s^4}{1920\pi^2v^2}Y^{\mu_1\mu_2\mu_3\mu_4}_{a_1a_2a_3a_4},\\
R_2(hhq\bar{q})&=&-\left[\frac{iC_{hh}g_s^2}{32\pi^2v^2}\lambda_{\rm HV}+\frac{iC_h^2g_s^2}{128\pi^2v^2}\left(8\lambda_{\rm HV}+1\right)\right]C_F\delta_{i_1i_2}\left({p\mkern-7.5mu/}_1-{p\mkern-7.5mu/}_2\right),\\
R_2(hhq\bar{q}g)&=&\frac{iC_{hh}g_s^3}{64\pi^2v^2}\gamma^{\mu}t^a_{i_2i_1}\left[\frac{2\lambda_{\rm HV}+1}{N_c}-N_c\left(2\lambda_{\rm HV}+3\right)\right]\\
&&+\frac{iC_{h}^2g_s^3}{32\pi^2v^2}\gamma^{\mu}t^a_{i_2i_1}\left[\frac{2\lambda_{\rm HV}+1}{N_c}-N_c\left(2\lambda_{\rm HV}+4\right)\right].\nonumber
\end{eqnarray}
We have used the colour factors $N_c=3,C_F=\frac{N_c^2-1}{2N_c}=\frac{4}{3}$, the Gell-Mann matrices $t^a$ in the fundamental representation of SU$(N_c)$ group, the asymmetric structure constants $f_{a_1a_2a_3}$ of SU$(N_c)$, the colour charge $g_s=\sqrt{4\pi\alpha_s}$, the parameter $\lambda_{\rm HV}=1 (0)$ corresponding to dimensional regularisation (reduction) and the shorthand functions
\begin{eqnarray}
V^{\mu_1\mu_2\mu_3}\left(p_1,p_2,p_3\right)&=&g^{\mu_1\mu_2}\left(p_2-p_1\right)^{\mu_3}+g^{\mu_2\mu_3}\left(p_3-p_2\right)^{\mu_1}+g^{\mu_3\mu_1}\left(p_1-p_3\right)^{\mu_2},\\
X_{a_1 a_2 a_3 a_4}^{\mu_1\mu_2\mu_3\mu_4}&=&{\rm Tr}\left(T^{a_1}T^{a_2}T^{a_3}T^{a_4}\right)\left(+21g^{\mu_1\mu_2}g^{\mu_3\mu_4}-41g^{\mu_1\mu_3}g^{\mu_2\mu_4}+21g^{\mu_1\mu_4}g^{\mu_2\mu_3}\right)\nonumber\\
&+&{\rm Tr}\left(T^{a_1}T^{a_2}T^{a_4}T^{a_3}\right)\left(+21g^{\mu_1\mu_2}g^{\mu_3\mu_4}+21g^{\mu_1\mu_3}g^{\mu_2\mu_4}-41g^{\mu_1\mu_4}g^{\mu_2\mu_3}\right)\nonumber\\
&+&{\rm Tr}\left(T^{a_1}T^{a_3}T^{a_2}T^{a_4}\right)\left(-41g^{\mu_1\mu_2}g^{\mu_3\mu_4}+21g^{\mu_1\mu_3}g^{\mu_2\mu_4}+21g^{\mu_1\mu_4}g^{\mu_2\mu_3}\right),\nonumber\\
Y_{a_1 a_2 a_3 a_4}^{\mu_1\mu_2\mu_3\mu_4}&=&{\rm Tr}\left(T^{a_1}T^{a_2}T^{a_3}T^{a_4}\right)\left(+323g^{\mu_1\mu_2}g^{\mu_3\mu_4}-625g^{\mu_1\mu_3}g^{\mu_2\mu_4}+323g^{\mu_1\mu_4}g^{\mu_2\mu_3}\right)\nonumber\\
&+&{\rm Tr}\left(T^{a_1}T^{a_2}T^{a_4}T^{a_3}\right)\left(+323g^{\mu_1\mu_2}g^{\mu_3\mu_4}+323g^{\mu_1\mu_3}g^{\mu_2\mu_4}-625g^{\mu_1\mu_4}g^{\mu_2\mu_3}\right)\nonumber\\
&+&{\rm Tr}\left(T^{a_1}T^{a_3}T^{a_2}T^{a_4}\right)\left(-625g^{\mu_1\mu_2}g^{\mu_3\mu_4}+323g^{\mu_1\mu_3}g^{\mu_2\mu_4}+323g^{\mu_1\mu_4}g^{\mu_2\mu_3}\right),\nonumber
\end{eqnarray}
where $T^a$ is the colour matrix in the adjoint representation with its elements $\left(T^a\right)_{bc}=-if_{abc}$ and the trace of $T^a$ into the trace of the Gell-Mann matrices are
\begin{eqnarray}
{\rm Tr}\left(T^aT^bT^cT^d\right)=N_c\left({\rm Tr}\left(t^at^bt^ct^d\right)+{\rm Tr}\left(t^dt^ct^bt^a\right)\right)+\frac{1}{2}\left(\delta_{ab}\delta_{cd}+\delta_{ac}\delta_{bd}+\delta_{ad}\delta_{bc}\right).
\end{eqnarray}
All the momenta $p_i,q_j$ are treated as incoming vectors. 

Our NLO UFO model can be downloaded from\\ $~~~~~~~~~$\url{http://feynrules.irmp.ucl.ac.be/wiki/HEFT_DH}.

\section{Renormalisation scale dependence \label{app:scale}}
In the framework of SCET, the typical scales are hard, jet and soft scales in addition to a factorisation scale.
In order to reproduce the fixed-order results from the resummation formula, all these scales are usually set to be equal (to the factorisation scale),
i.e., there is only one scale in the expanded result.
Since we want to investigate the scale uncertainties by varying factorisation and renormalisation scales independently,
we must reconstruct the individual $\mu_R$ and $\mu_F$ dependence separately.
In this appendix, 
we present details about the method we used to obtain the $\mu_R$ dependence in the expanded results from 
transverse momentum resummation formula.
As a by-product, we find a close relation between the contributions from class-$a$ and class-$b$.

Given that the N$^k$LO cross section is scale $(\mu_R=\mu_F=\mu)$ invariant, we have
\begin{align}
    \frac{d}{d\ln \mu} \sigma_{hh}^{{\rm N}^k{\rm LO}}(\mu,\mu) &= \left(\frac{\partial}{\partial\ln \mu_R} \sigma_{hh}^{{\rm N}^k{\rm LO}}(\mu_R, \mu_F)  + \frac{\partial}{\partial\ln \mu_F} \sigma_{hh}^{{\rm N}^k{\rm LO}}(\mu_R, \mu_F) \right)\bigg|_{\mu_R=\mu_F=\mu} 
    \nonumber \\ &
    = 0 + \mathcal{O}(\as^{3+k}).
\end{align}
The individual renormalisation and factorisation scale dependence is rebuilt from the evolution equation
\begin{align}
    \sigma_{hh}^{{\rm N}^k{\rm LO}}(\mu_R, \mu_F)= \sigma_{hh}^{{\rm N}^k{\rm LO}}(\mu_F, \mu_F) + \int_{\mu_F}^{\mu_R} d\bar{\mu} \left(\frac{\partial}{\partial \bar{\mu}} \sigma^{{\rm N}^k{\rm LO}}_{hh}(\bar{\mu}, \mu_F)\right) \,,
\end{align}
where the first term on the right hand is derived by expanding the transverse momentum resummation formula in the framework of SCET
and the second term is given below.
Since we use $q_T$-subtraction to calculate the NNLO correction to the class-$b$ diagrams, we focus on the scale dependence in this class.

Firstly, we know that the total N$^3$LO cross section  is  independent of the renormalisation scale at each fixed order, i.e.,
\begin{align}
    \frac{\partial}{\partial \ln\mu_R} \sigma_{hh}^{\rm N^3LO} (\mu_R, \mu_F) =&  \frac{\partial}{\partial\ln\mu_R} \sigma_{hh}^{a,{\rm N^3LO}}(\mu_R, \mu_F)
    + \frac{\partial}{\partial \ln\mu_R} \sigma_{hh}^{b,{\rm NNLO}}(\mu_R, \mu_F)
    \nonumber \\ &
    + \frac{\partial}{\partial \ln\mu_R} \sigma_{hh}^{c,{\rm NLO}} (\mu_R, \mu_F) = 0 + \mathcal{O}(\as^6)\;.
\end{align}
We have only calculated explicitly the results up to $\mathcal{O}(\as^5)$, so we omit higher-order terms.
The first contribution on the right hand is known,
\begin{multline}
    \frac{\partial}{\partial\ln \mu_R}\sigma_{hh}^{a,{\rm N^3LO}} (\mu_R, \mu_F) = 
    \\
    \int dm_{hh} f_{h\to hh} \bigg[\sigma_{h}^{\rm N^3LO} (\mu_R, \mu_F)\bigg|_{m_h\to m_{hh}}\bigg]  \times \frac{d}{d\ln \mu_R}  \bigg(\frac{C_{hh}(\mu_R)}{C_h(\mu_R)} - \frac{6 \lambda_{hhh} v^2}{m_{hh}^2-m_h^2} \bigg)^2 \;.
\end{multline}
where $\sigma_{h}$ has the expansion $\sigma_{h}^{\rm N^3LO}=\sum_{i=0}^{3}{\sigma_{h}^{(i)}}$ with $\sigma_{h}^{(i)}\propto \as^{2+i}$.
The  class-$c$ cross section up to NLO QCD  is  scale invariant,
\begin{align}
    \frac{\partial}{\partial \ln \mu_R}\sigma_{hh}^{c,{\rm NLO}} (\mu_R, \mu_F) = 0+\mathcal{O}(\as^6)\;.
\end{align}
As a consequence, the renormalisation group equation for class-$b$ is derived,
\begin{align} \label{eq:RGb}
     \frac{\partial}{\partial \ln \mu_R} \sigma^{b,{\rm NNLO}}_{hh}(\mu_R, \mu_F) =& -2\int dm_{hh} f_{h\to hh} \bigg[ \sigma_{h}^{\rm N^3LO} (\mu_R, \mu_F)\bigg|_{m_h\to m_{hh}}\bigg]
     \nonumber \\ &
     \times \bigg(\frac{C_{hh}(\mu_R)}{C_h(\mu_R)} - \frac{6 \lambda_{hhh} v^2}{m_{hh}^2-m_h^2} \bigg)  \left( \frac{d}{d\ln \mu_R}  \frac{C_{hh}(\mu_R)}{C_h(\mu_R)}\right) \,.
\end{align}
The ratio of $C_{hh}(\mu_R)$ over $C_h(\mu_R)$ can be expanded in terms of  $a_s\equiv\as(\mu_R)/4\pi$,
\begin{align}
    \frac{C_{hh}(\mu_R)}{C_{h}(\mu_R)} = 1+ \delta_2 a_s^2 + \delta_3(\mu_R) a_s^3 + \mathcal{O}(a_s^4) \;
\end{align}
with the coefficient $\delta_2=\frac{2}{3}\left(16n_f+35\right)$ being scale independent and
\begin{align}
    \delta_3(\mu_R)=\frac{2}{27}\left[L_t\left(192n_f^2-2292 n_f-10602\right)-77n_f^2-578n_f+799\right]\;.
\end{align}
Therefore, we have
\begin{align}
    \frac{d}{d\ln \mu_R}  \frac{C_{hh}(\mu_R)}{C_h(\mu_R)} =& \left( \frac{da_s}{d\ln\mu_R}\frac{\partial }{\partial a_s } + \frac{\partial }{\partial \ln \mu_R} \right) \frac{C_{hh}(\mu_R)}{C_h(\mu_R)}
    \nonumber \\ 
    =& -4 \beta_0 \delta_2 a_s^3 +  a_s^3\frac{d \delta_3(\mu_R)}{d\ln\mu_R}+ \mathcal{O}(a_s^4)\equiv a_s^3 \chi +  \mathcal{O}(a_s^4)
\end{align}
with $\beta_0=(11 C_A-2 n_f)/3$ and 
\begin{align}
\chi=\frac{16}{9}\left(32n_f^2-420n_f-1461\right)\;.    
\end{align} 
Then, eq.~(\ref{eq:RGb}) turns out to be 
\begin{align} \label{eq:RGb2}
     \frac{\partial}{\partial \ln \mu_R} &\sigma^b_{hh}(\mu_R, \mu_F) = 
     \nonumber \\ &
     -2 a_s^3 \chi\; \int dm_{hh} f_{h\to hh} \bigg[ \sigma_{h}^{(0)} (\mu_R, \mu_F)\bigg|_{m_h\to m_{hh}}\bigg]\bigg(1- \frac{6 \lambda_{hhh} v^2}{m_{hh}^2-m_h^2} \bigg)+\mathcal{O}(a_s^6)
     \nonumber \\ 
     =& -\frac{3}{4}a_s^2\; \chi \;  \sigma^{b (1)}_{hh}(\mu_R, \mu_F) +\mathcal{O}(a_s^6)\;.
\end{align}
In the above equation, we have decomposed the class-$b$ cross section as $\sigma^b_{hh} =\sum_{i=1}{\sigma^{b (i)}_{hh}}$ with $\sigma^{b (i)}_{hh} \propto a_s^{2+i}$ and 
\begin{align}
    \sigma^{b (1)}_{hh}(\mu_R, \mu_F) =  \frac{8}{3} a_s \int dm_{hh} f_{h\to hh} \bigg[ \sigma_{h}^{(0)} (\mu_R, \mu_F)\bigg|_{m_h\to m_{hh}}\bigg]\bigg(1 - \frac{6 \lambda_{hhh} v^2}{m_{hh}^2-m_h^2} \bigg).
\end{align}
Notice that $\sigma^{b (1)}_{hh}$ is the LO class-$b$ cross section but has a close relation with the class-$a$ cross section; see eq.(\ref{eq:RGb}).
So eq.~(\ref{eq:RGb2}) indicates that the class-$b$ cross section has a non-vanishing dependence on $\mu_R$ only from two-loops. This is actually a consequence of the operator mixing as studied in ref.~\cite{Zoller:2016iam}.

\section{Additional plots\label{app:addplots}}

\begin{figure}[hbt!]
    \centering
    \includegraphics[scale=.38,draft=false,trim = 0mm 0mm 0mm 0mm,clip]{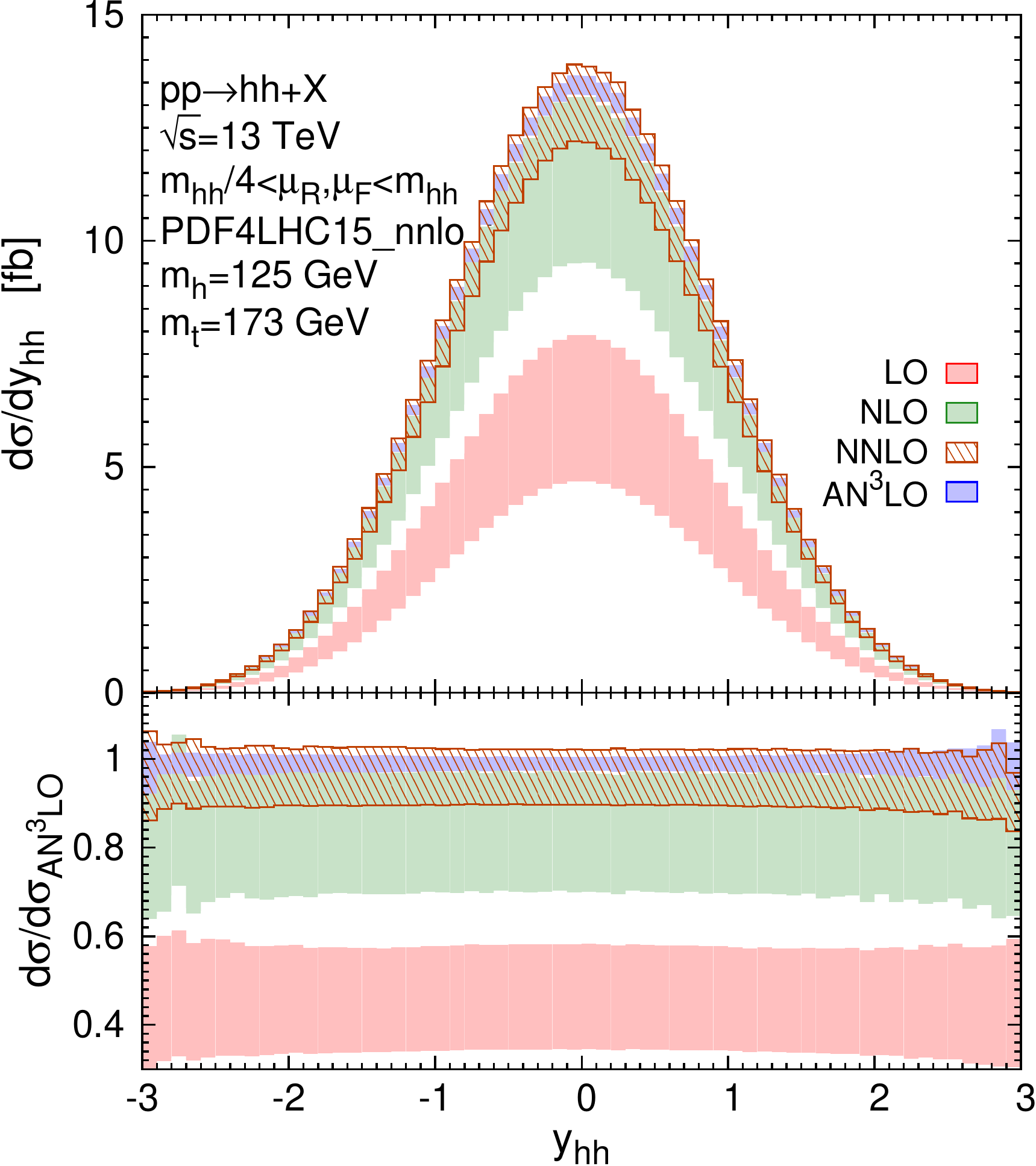}
    \includegraphics[scale=.38,draft=false,trim = 0mm 0mm 0mm 0mm,clip]{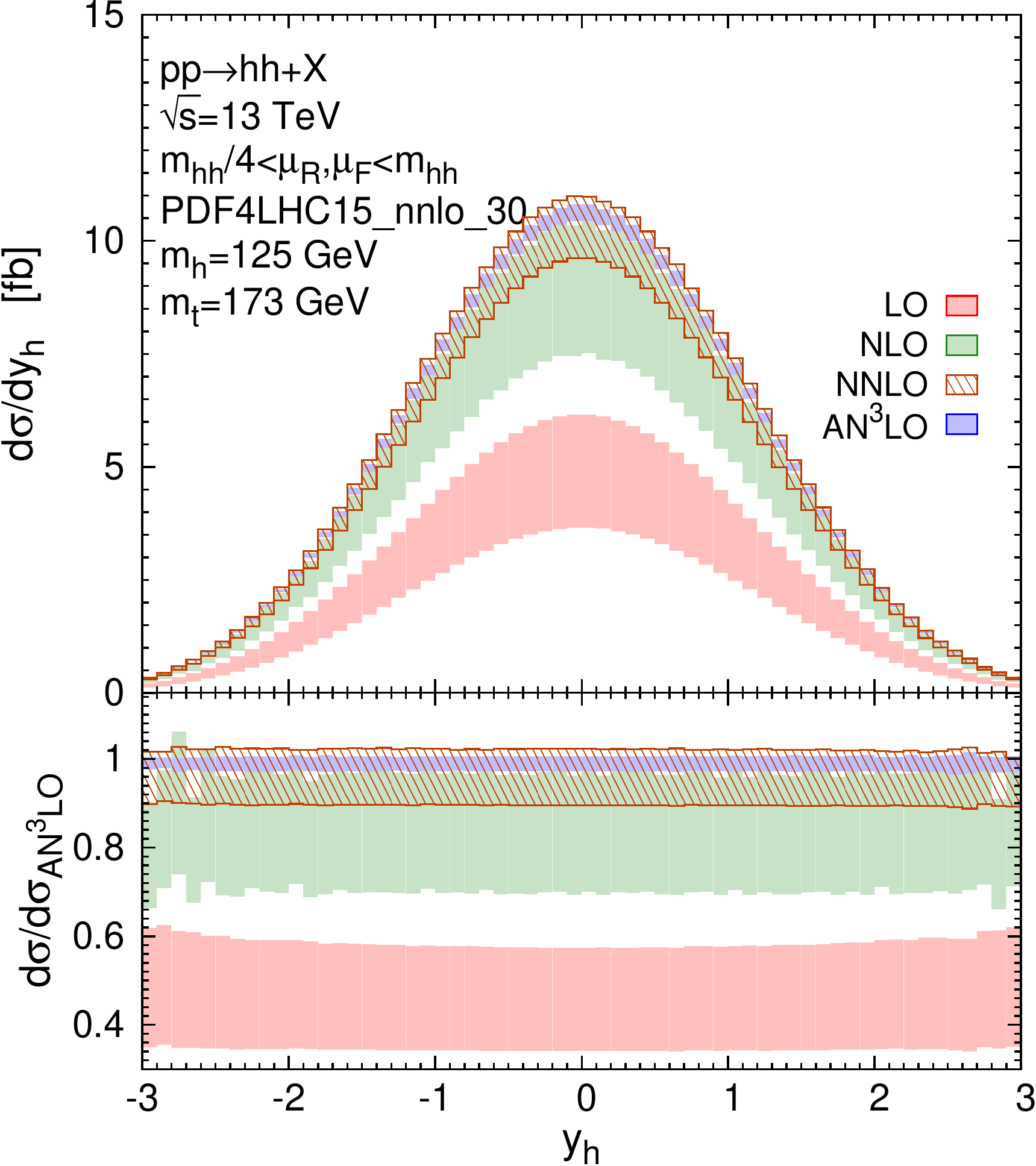}\\
    \includegraphics[scale=.38,draft=false,trim = 0mm 0mm 0mm 0mm,clip]{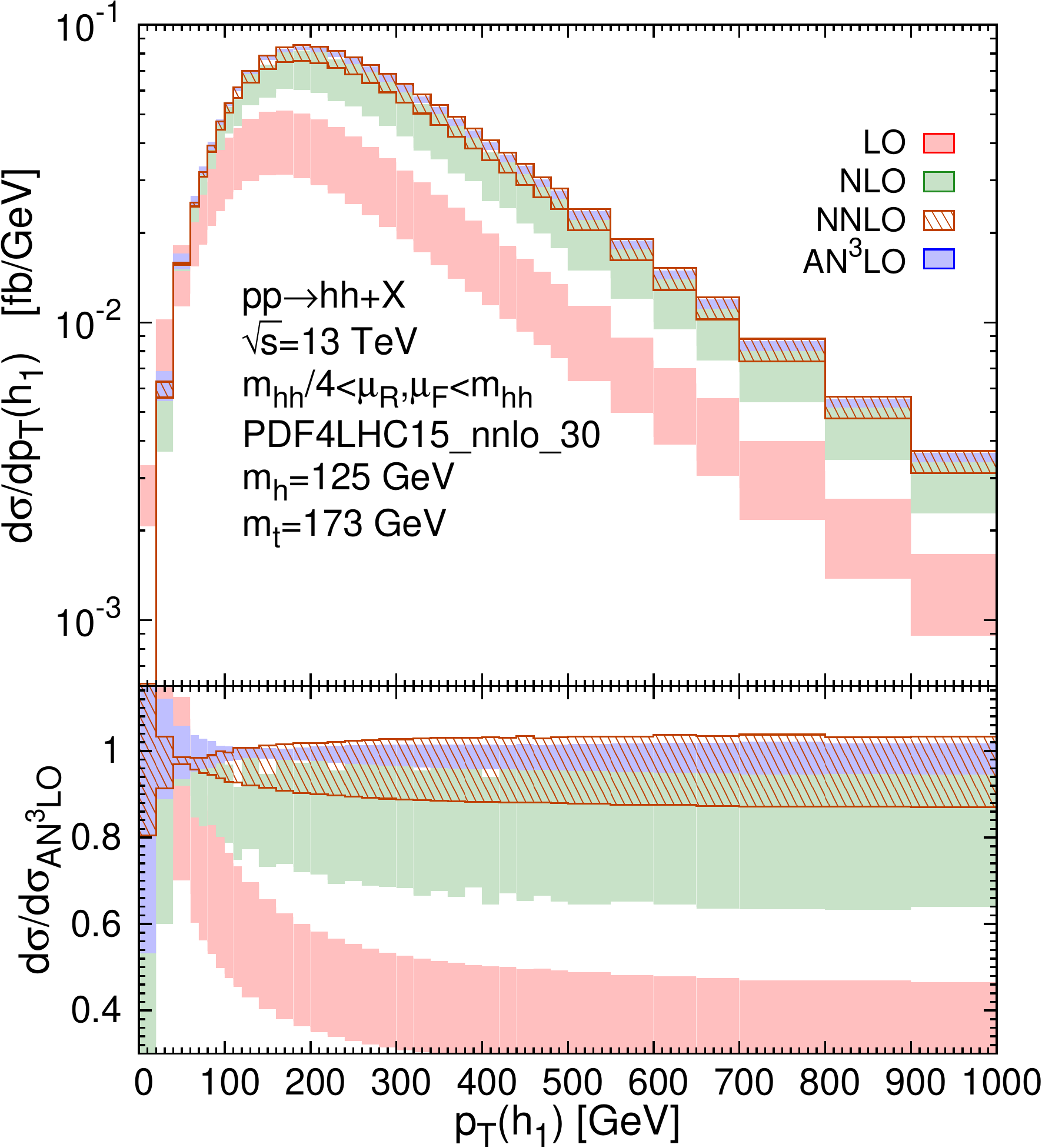}
    \includegraphics[scale=.38,draft=false,trim = 0mm 0mm 0mm 0mm,clip]{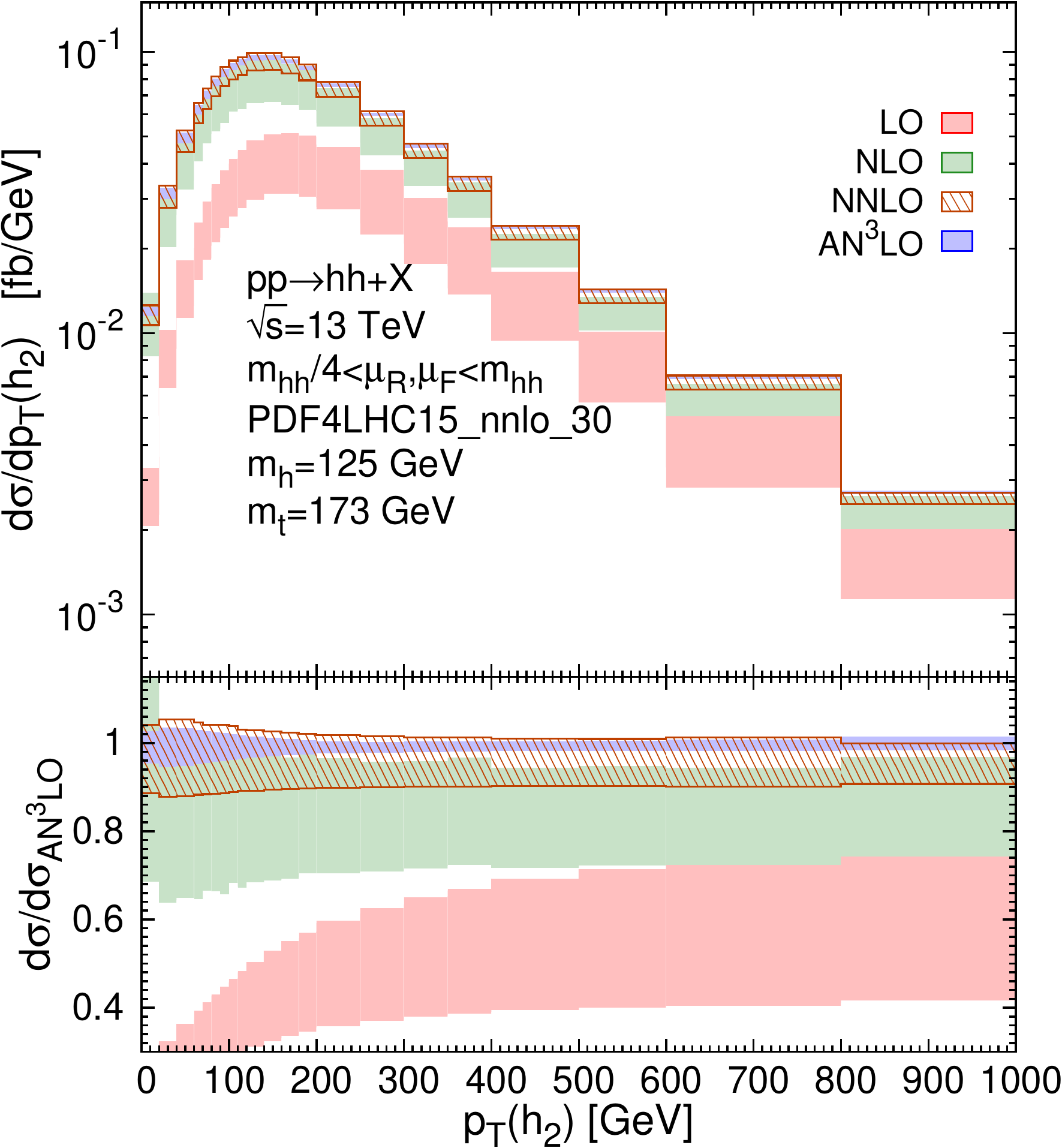}\\
    \includegraphics[scale=.38,draft=false,trim = 0mm 0mm 0mm 0mm,clip]{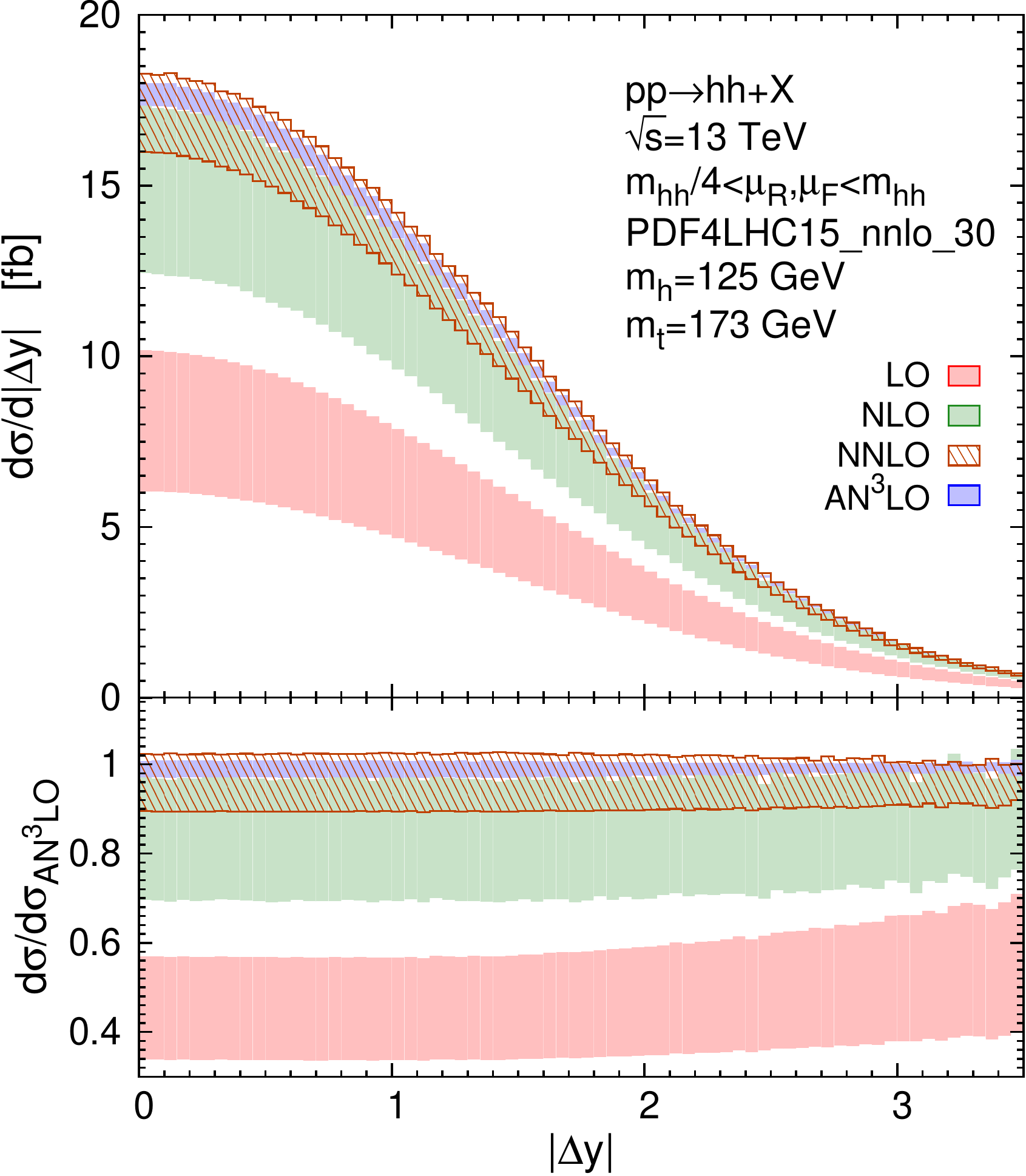}
    \includegraphics[scale=.38,draft=false,trim = 0mm 0mm 0mm 0mm,clip]{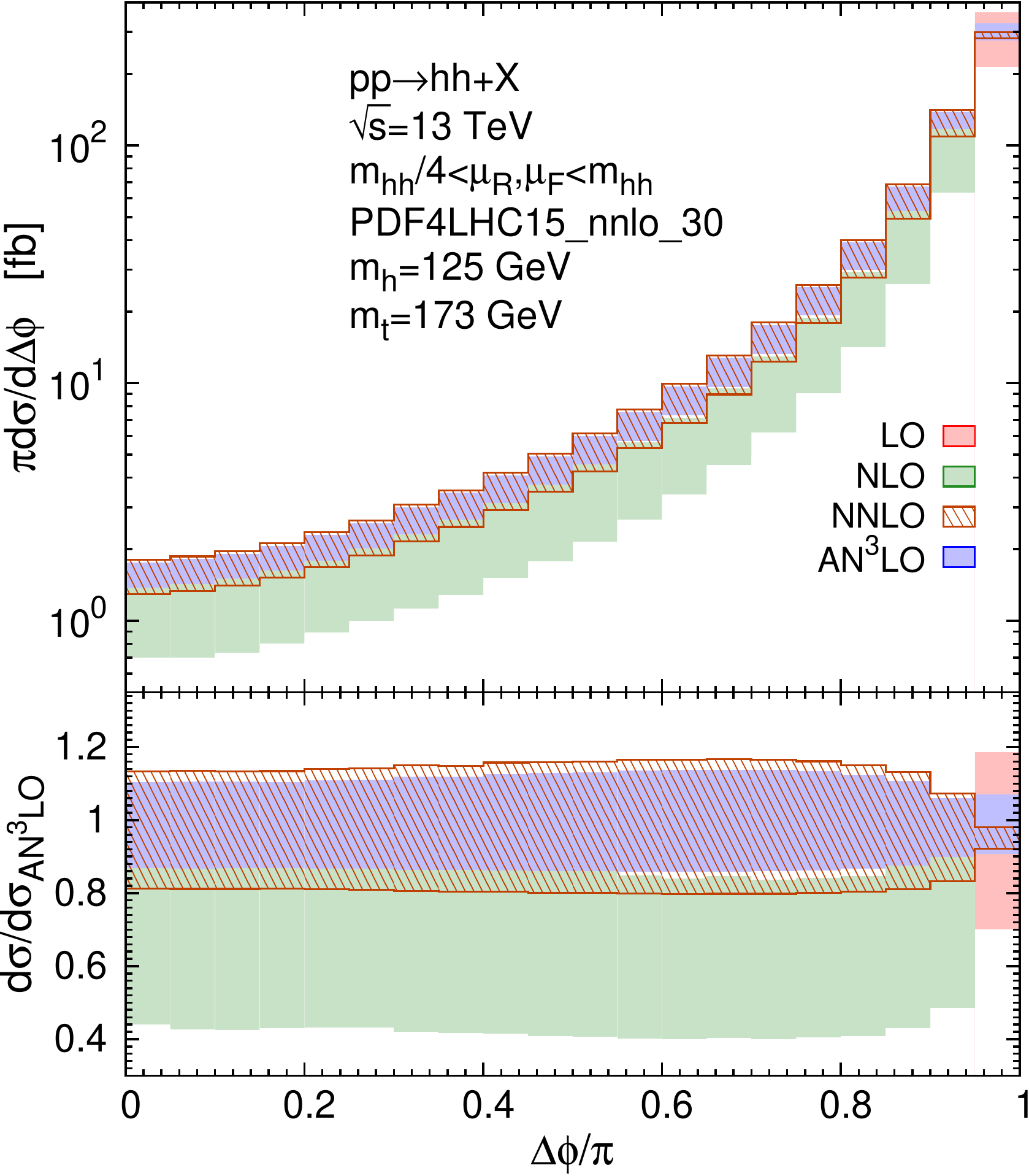}\\
    \vspace{0cm}
    \caption{Same as in figure~\ref{fig:othersLHC14} but at $\sqrt{s}=13$ TeV.}
    \label{fig:othersLHC13}
\end{figure}

\begin{figure}[hbt!]
    \centering
    \includegraphics[scale=.38,draft=false,trim = 0mm 0mm 0mm 0mm,clip]{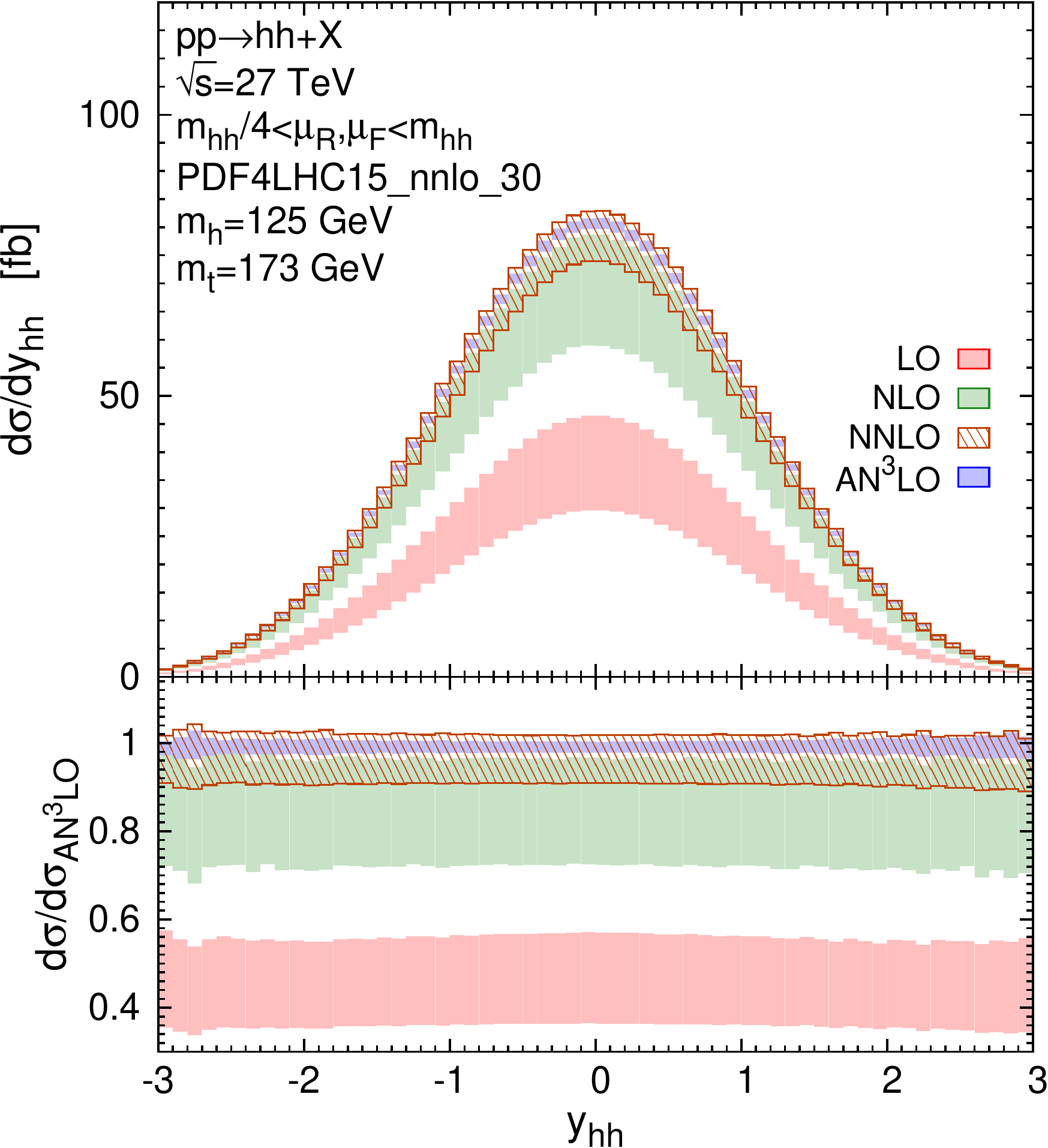}
    \includegraphics[scale=.38,draft=false,trim = 0mm 0mm 0mm 0mm,clip]{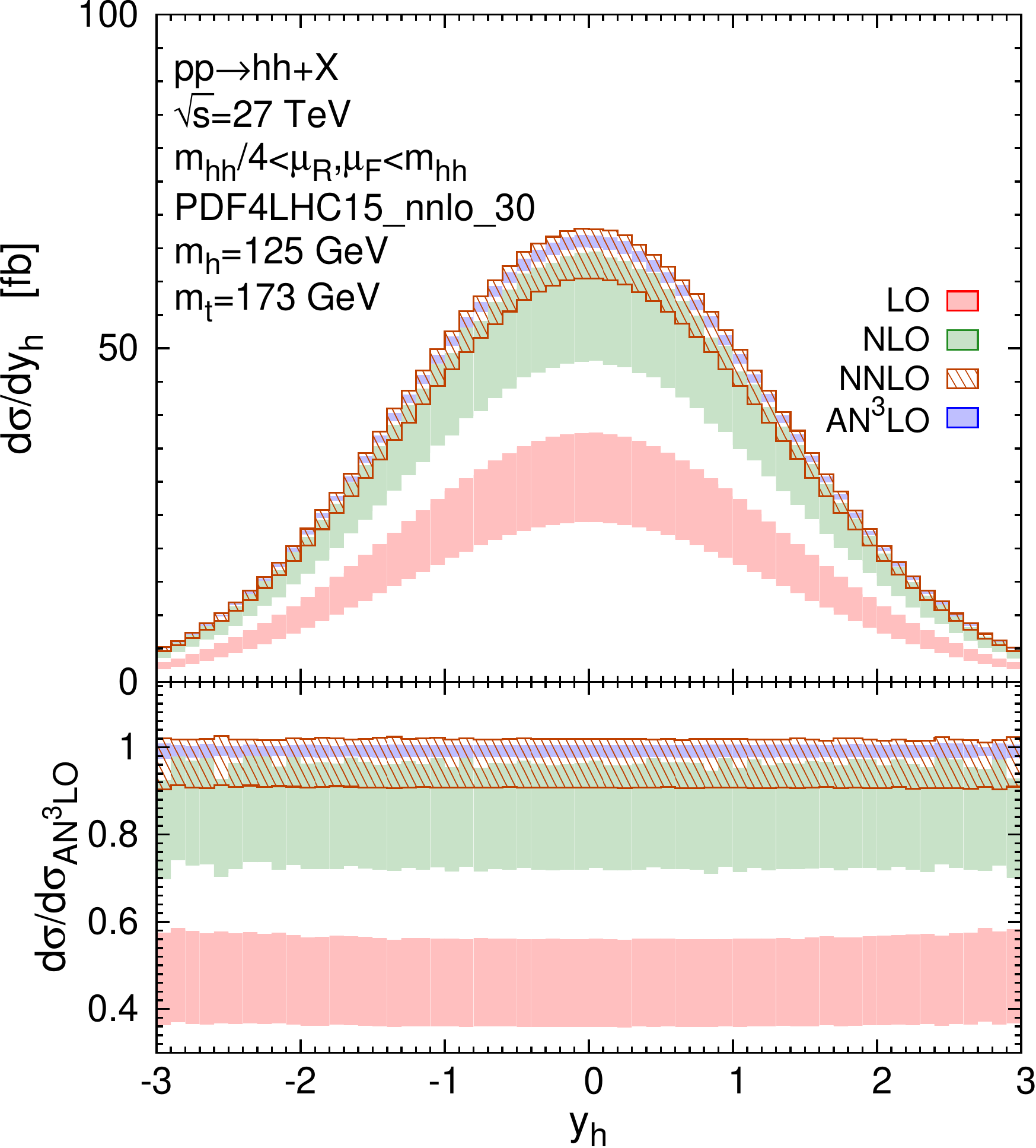}\\
    \includegraphics[scale=.38,draft=false,trim = 0mm 0mm 0mm 0mm,clip]{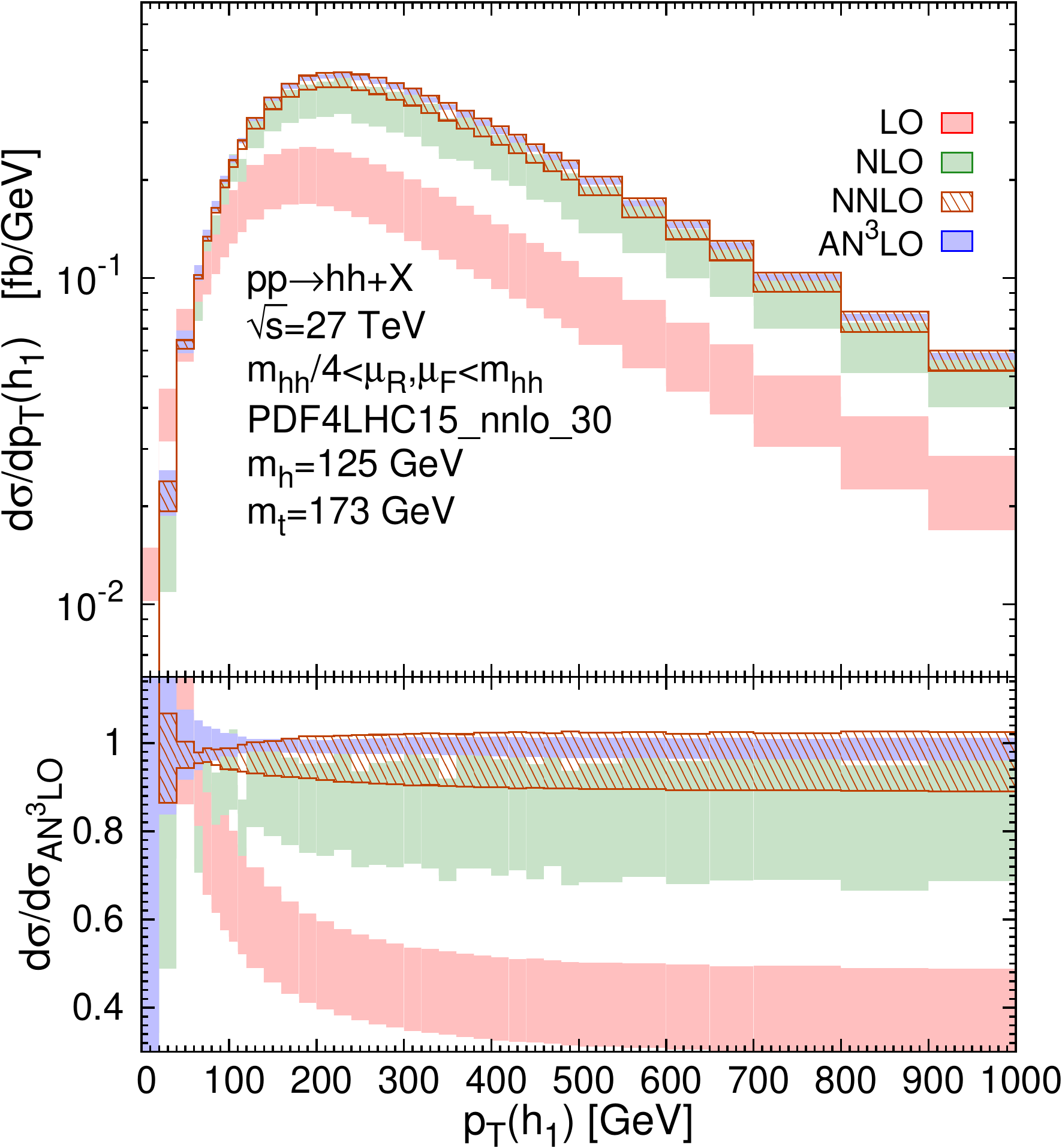}
    \includegraphics[scale=.38,draft=false,trim = 0mm 0mm 0mm 0mm,clip]{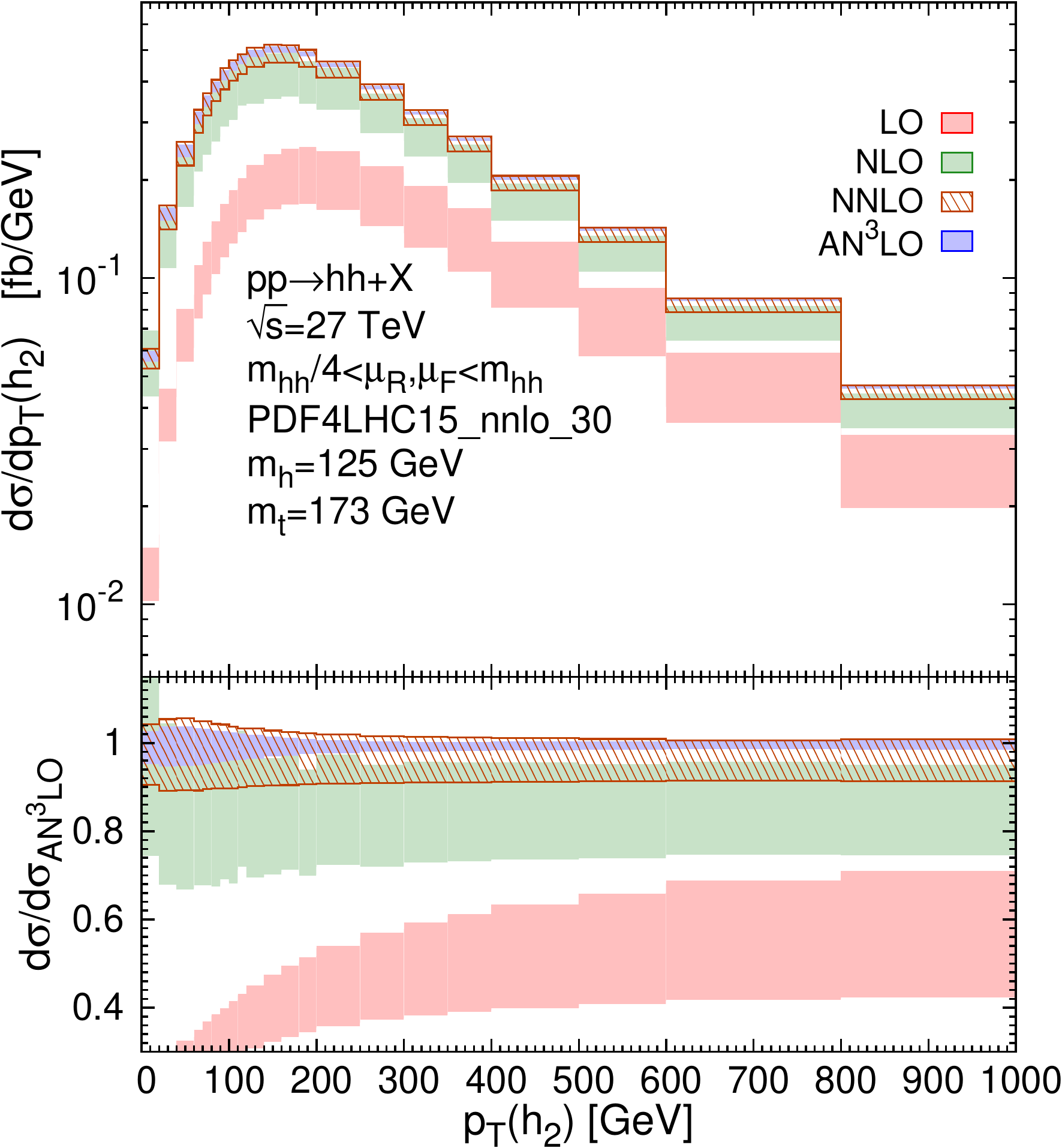}\\
    \includegraphics[scale=.38,draft=false,trim = 0mm 0mm 0mm 0mm,clip]{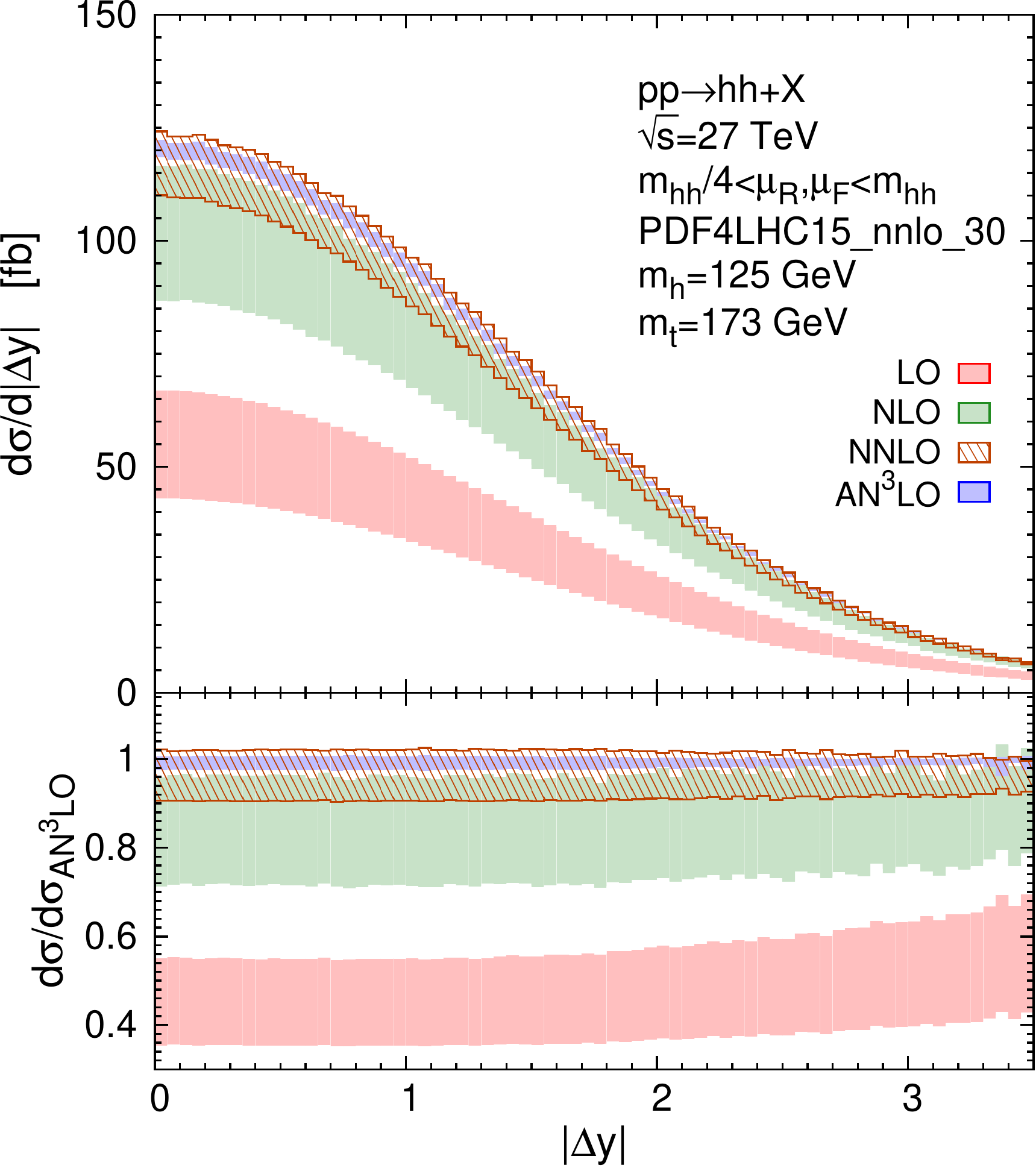}
    \includegraphics[scale=.38,draft=false,trim = 0mm 0mm 0mm 0mm,clip]{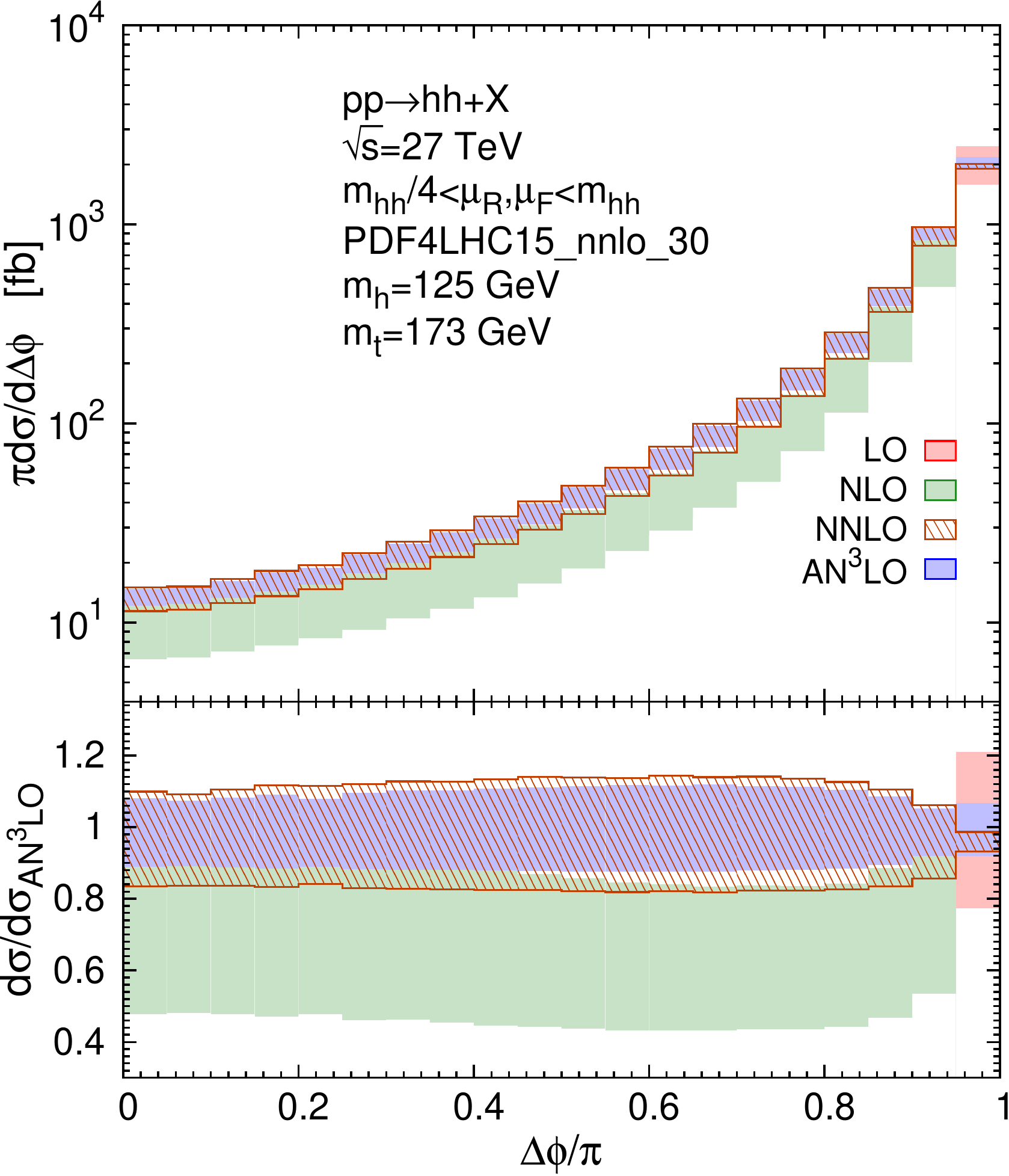}\\
    \vspace{0cm}
    \caption{Same as in figure~\ref{fig:othersLHC14} but at $\sqrt{s}=27$ TeV.}
    \label{fig:othersLHC27}
\end{figure}

\begin{figure}[hbt!]
    \centering
    \includegraphics[scale=.38,draft=false,trim = 0mm 0mm 0mm 0mm,clip]{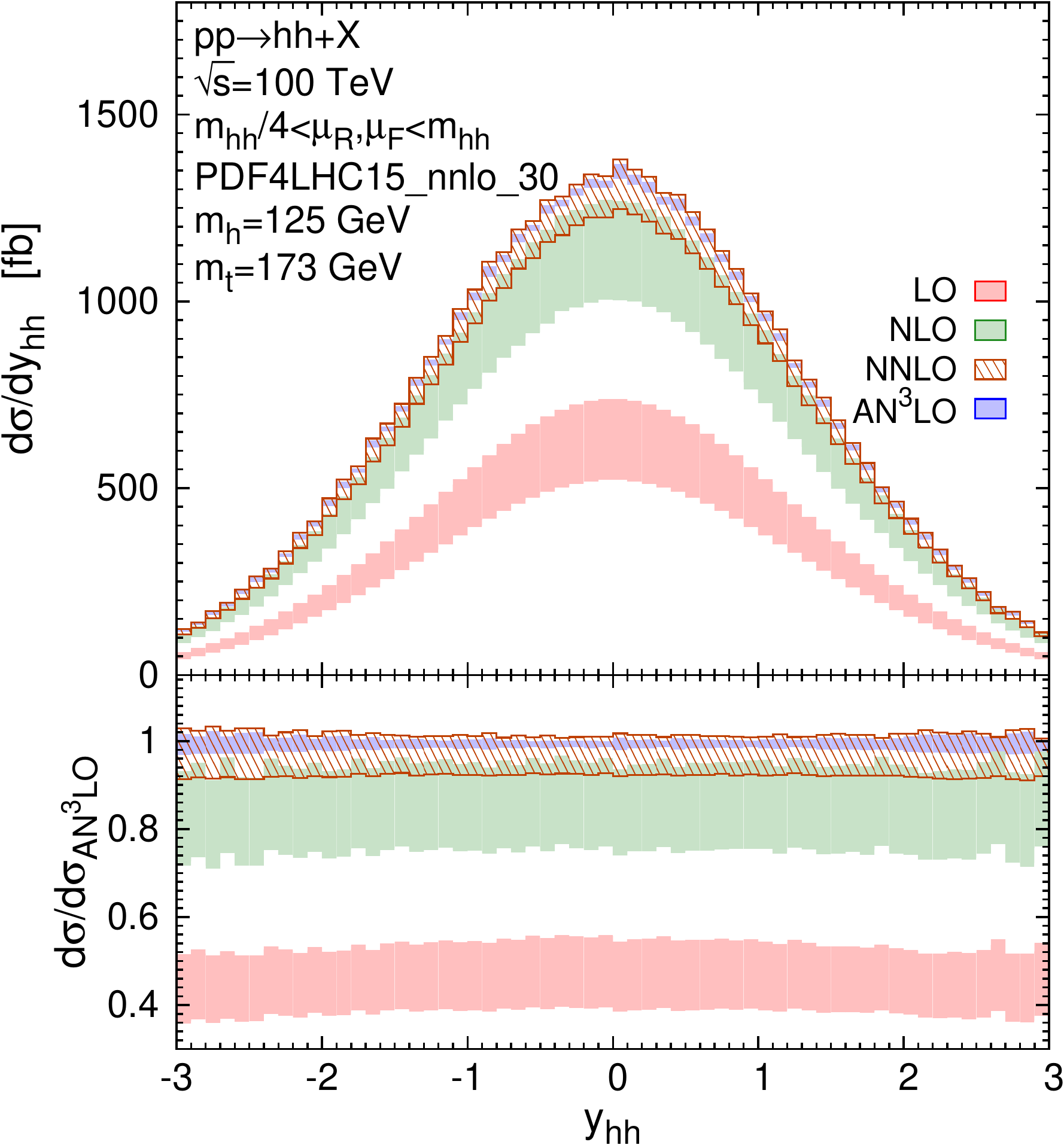}
    \includegraphics[scale=.38,draft=false,trim = 0mm 0mm 0mm 0mm,clip]{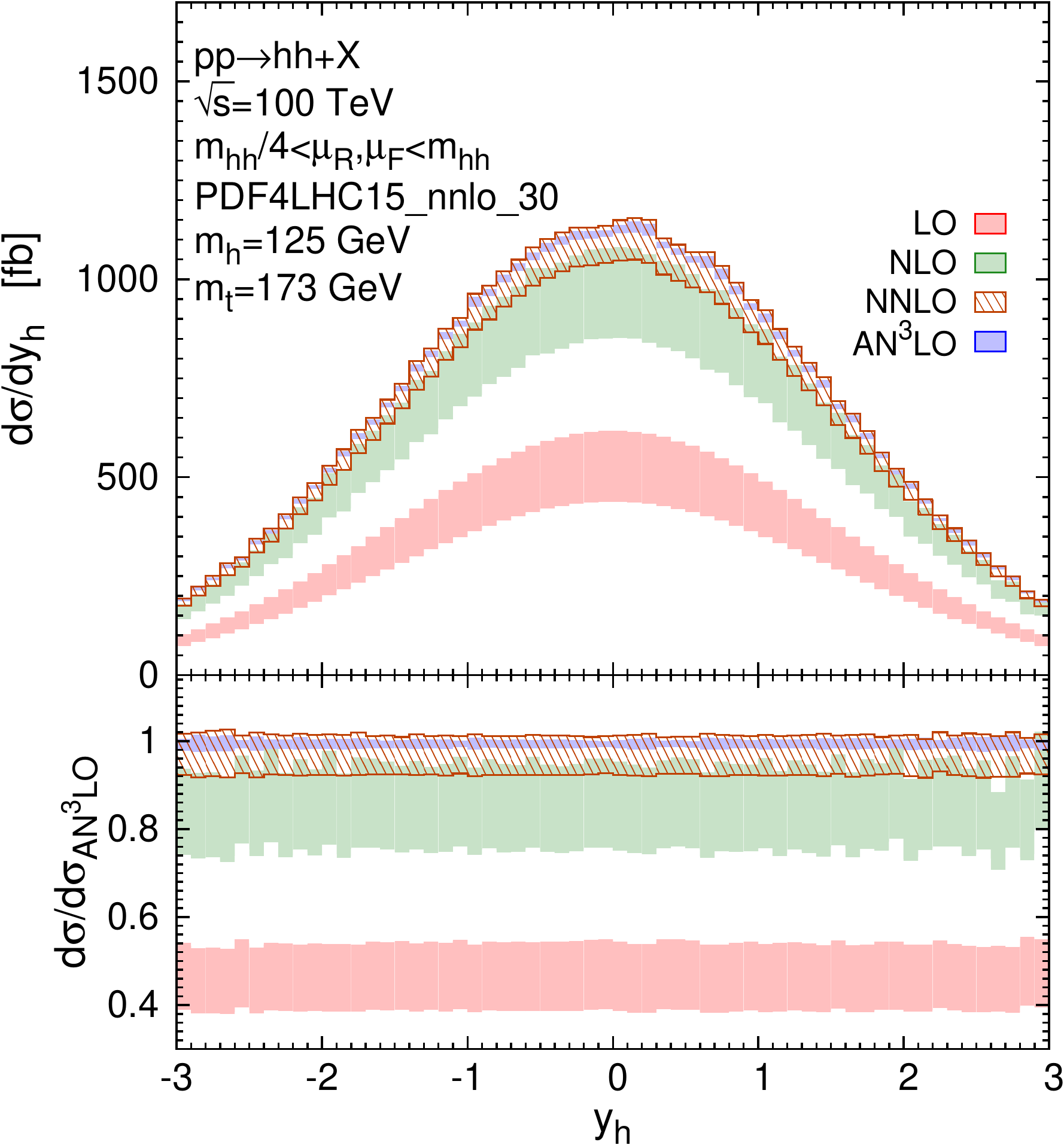}\\
    \includegraphics[scale=.38,draft=false,trim = 0mm 0mm 0mm 0mm,clip]{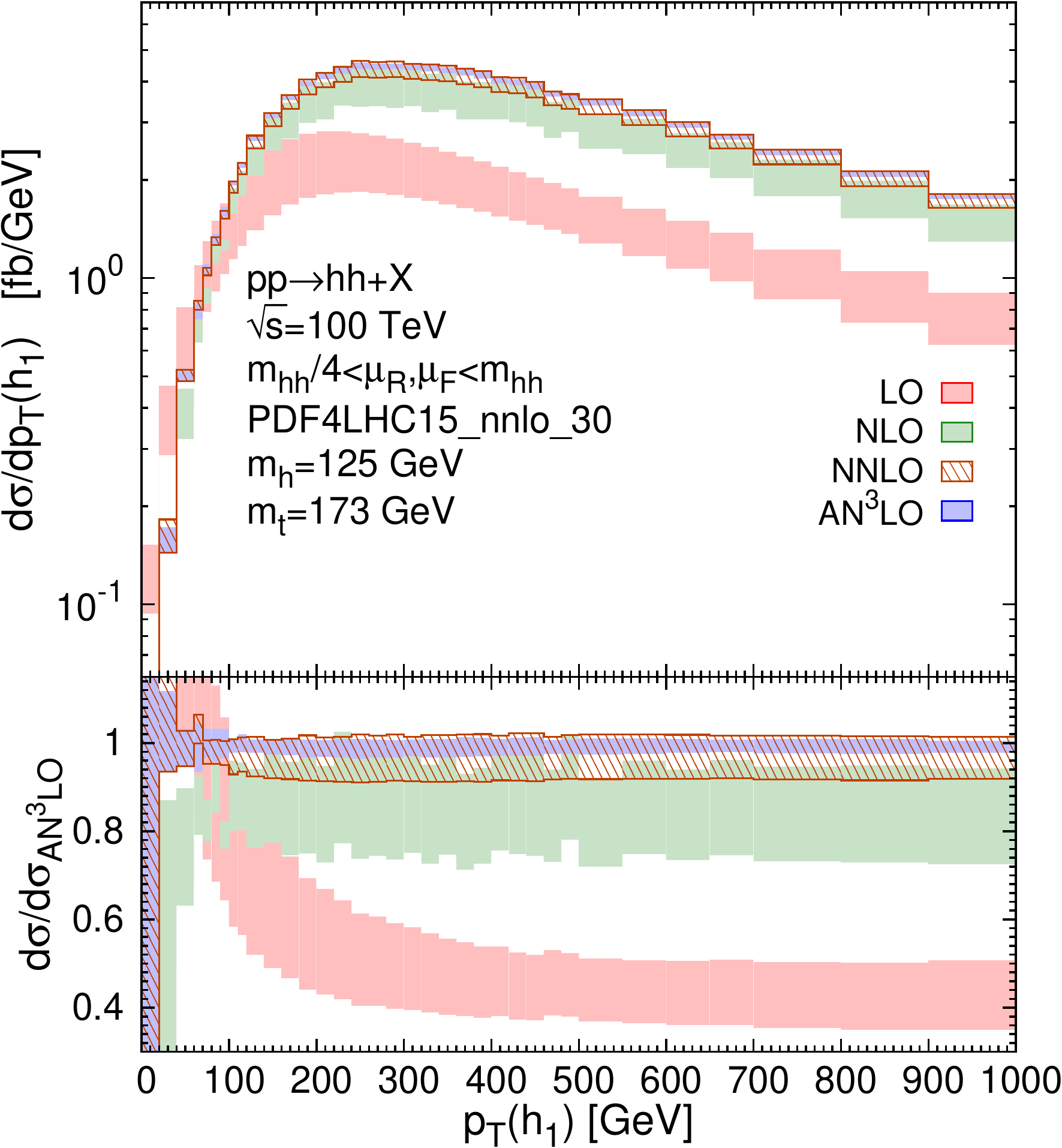}
    \includegraphics[scale=.38,draft=false,trim = 0mm 0mm 0mm 0mm,clip]{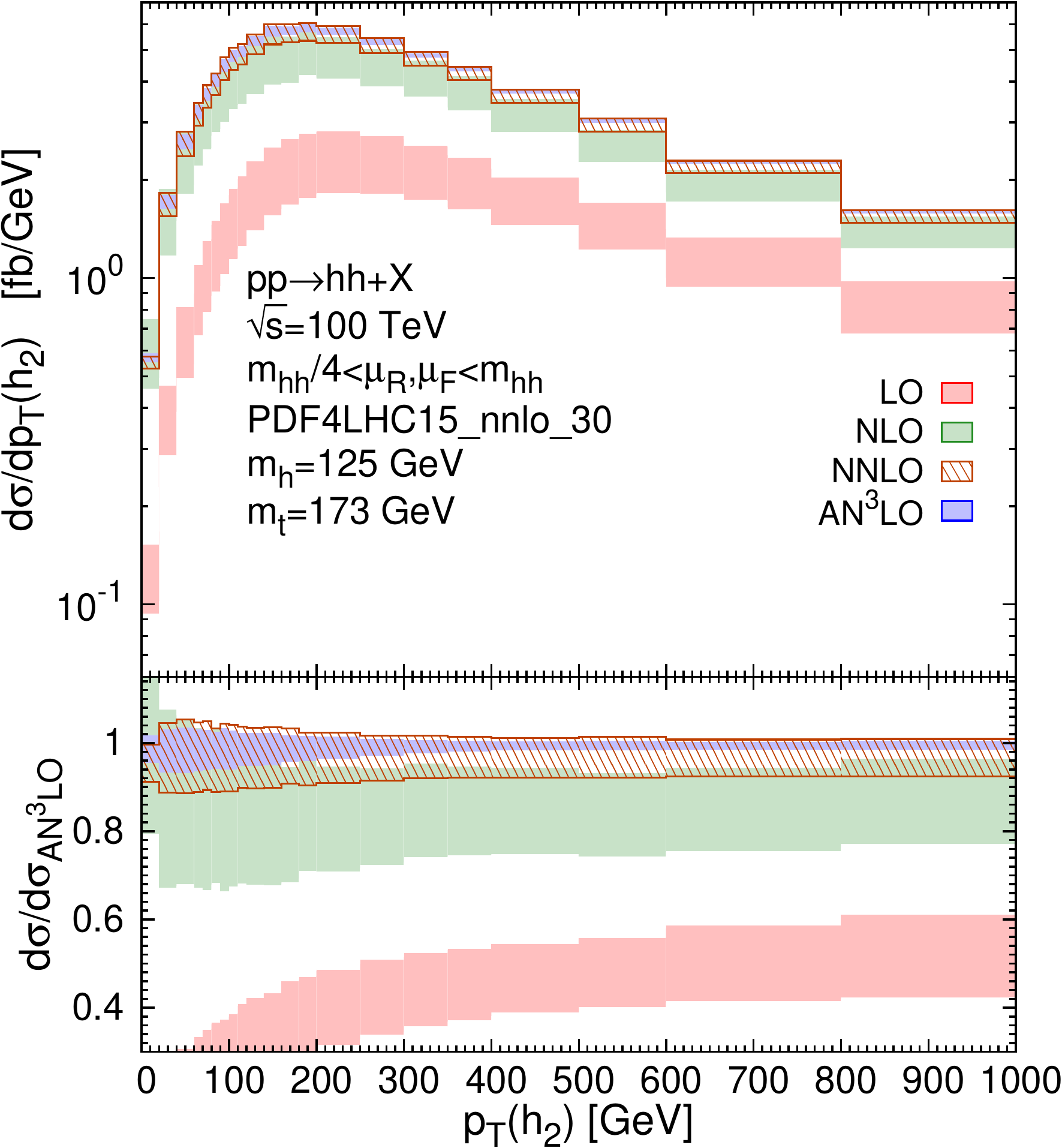}\\
    \includegraphics[scale=.38,draft=false,trim = 0mm 0mm 0mm 0mm,clip]{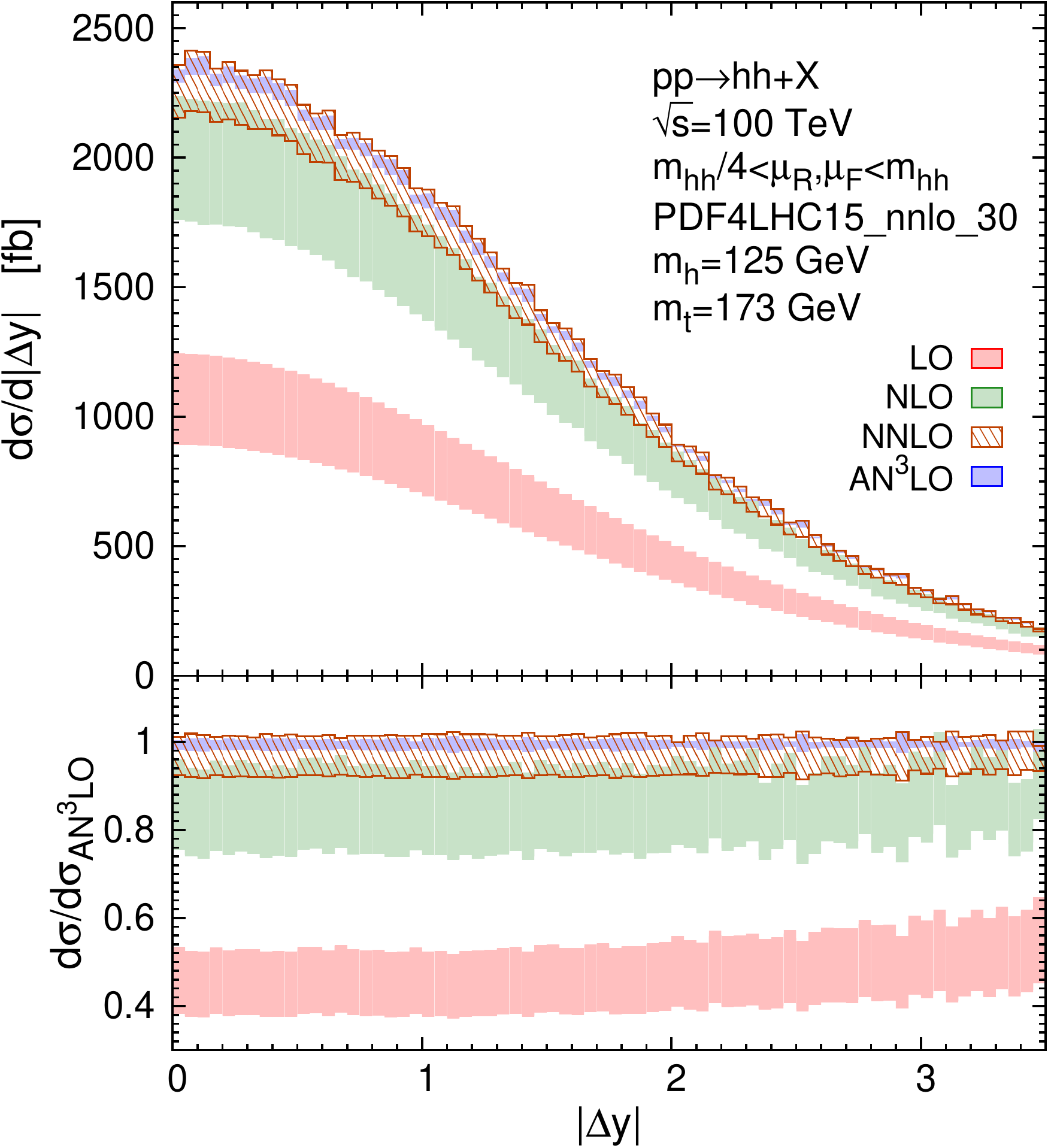}
    \includegraphics[scale=.38,draft=false,trim = 0mm 0mm 0mm 0mm,clip]{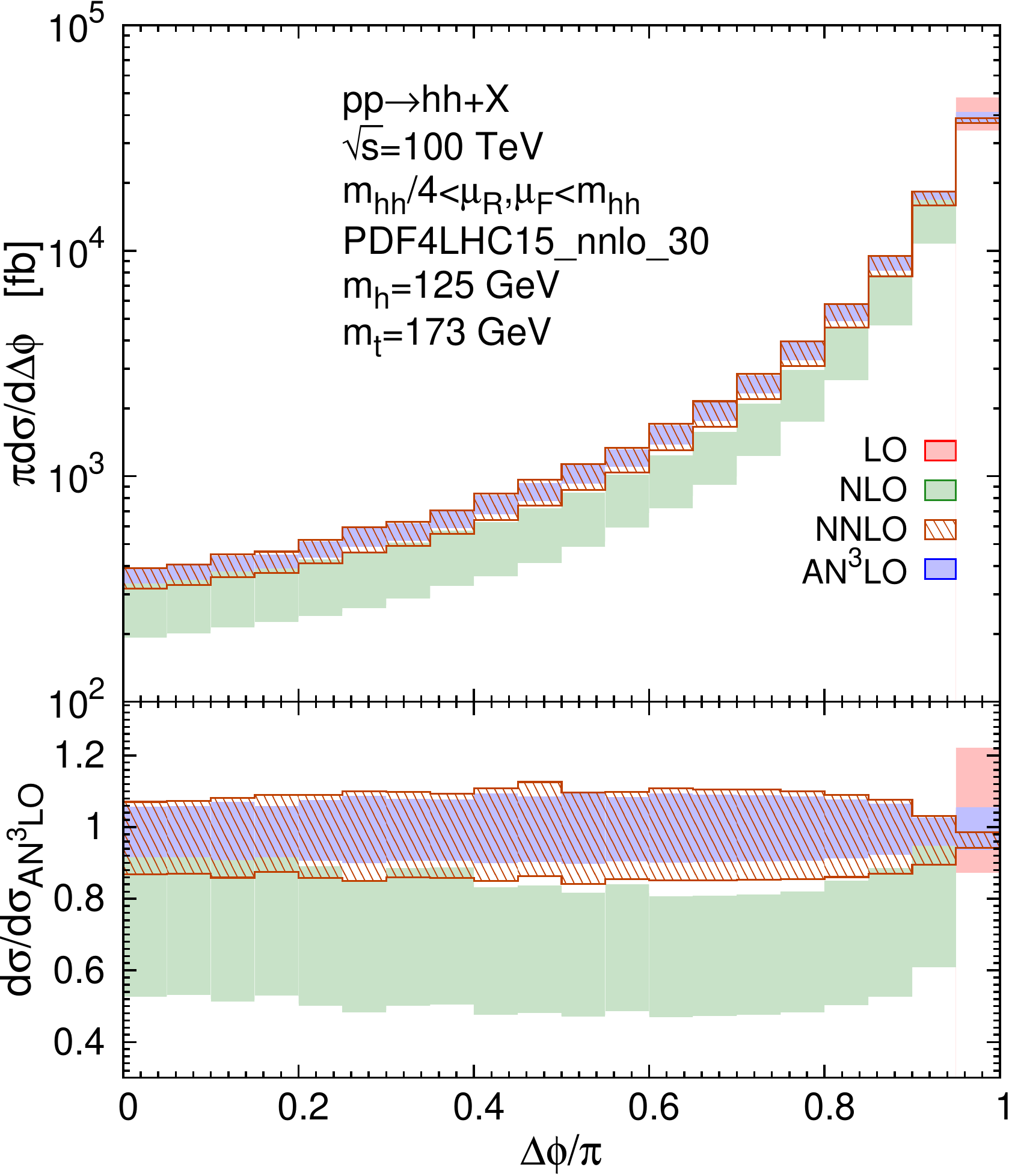}\\
    \vspace{0cm}
    \caption{Same as in figure~\ref{fig:othersLHC14} but at $\sqrt{s}=100$ TeV.}
    \label{fig:othersLHC100}
\end{figure}

\begin{figure}[h]
    \centering
    \includegraphics[scale=.38,draft=false,trim = 0mm 0mm 0mm 0mm,clip]{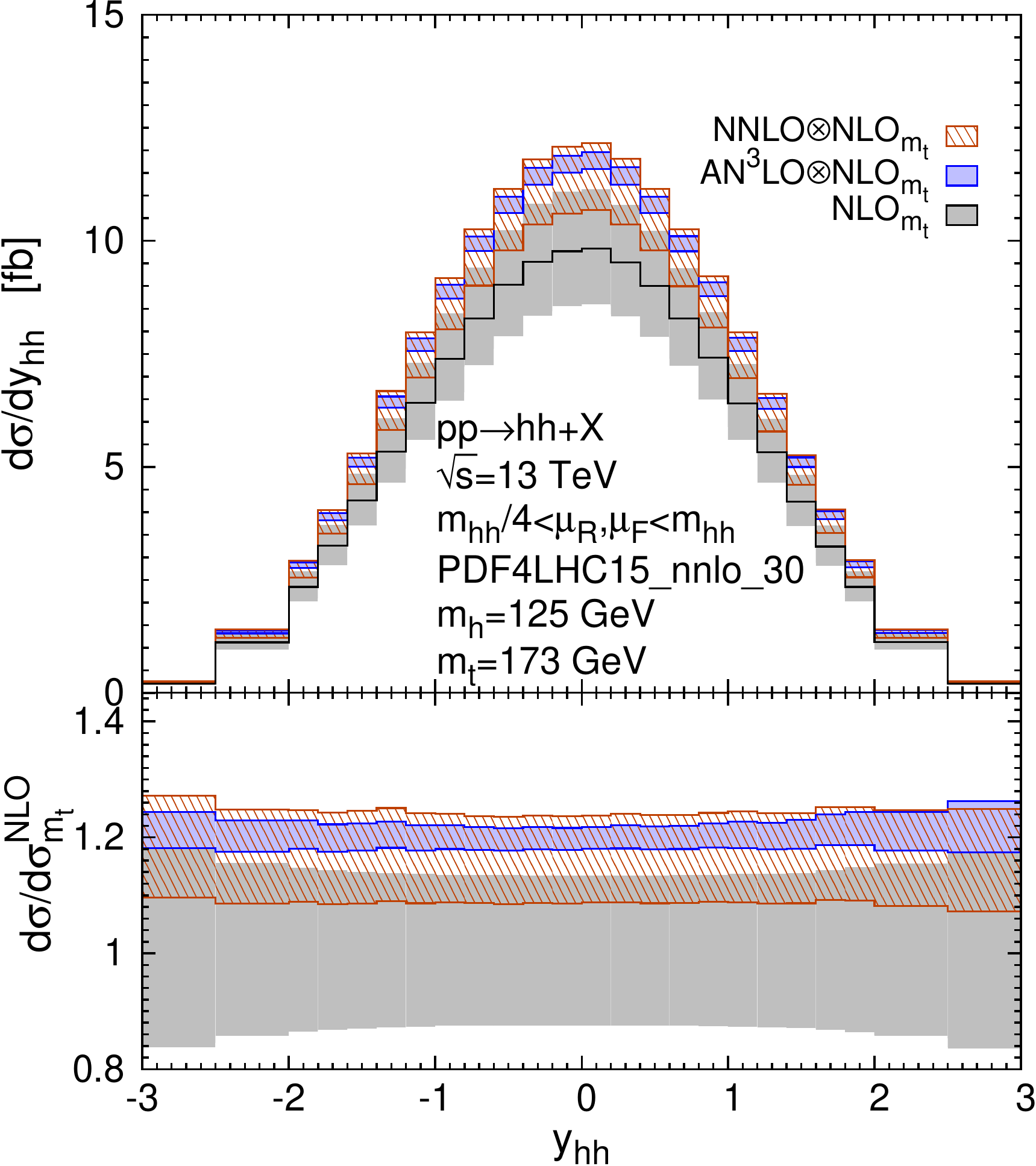}
    \includegraphics[scale=.38,draft=false,trim = 0mm 0mm 0mm 0mm,clip]{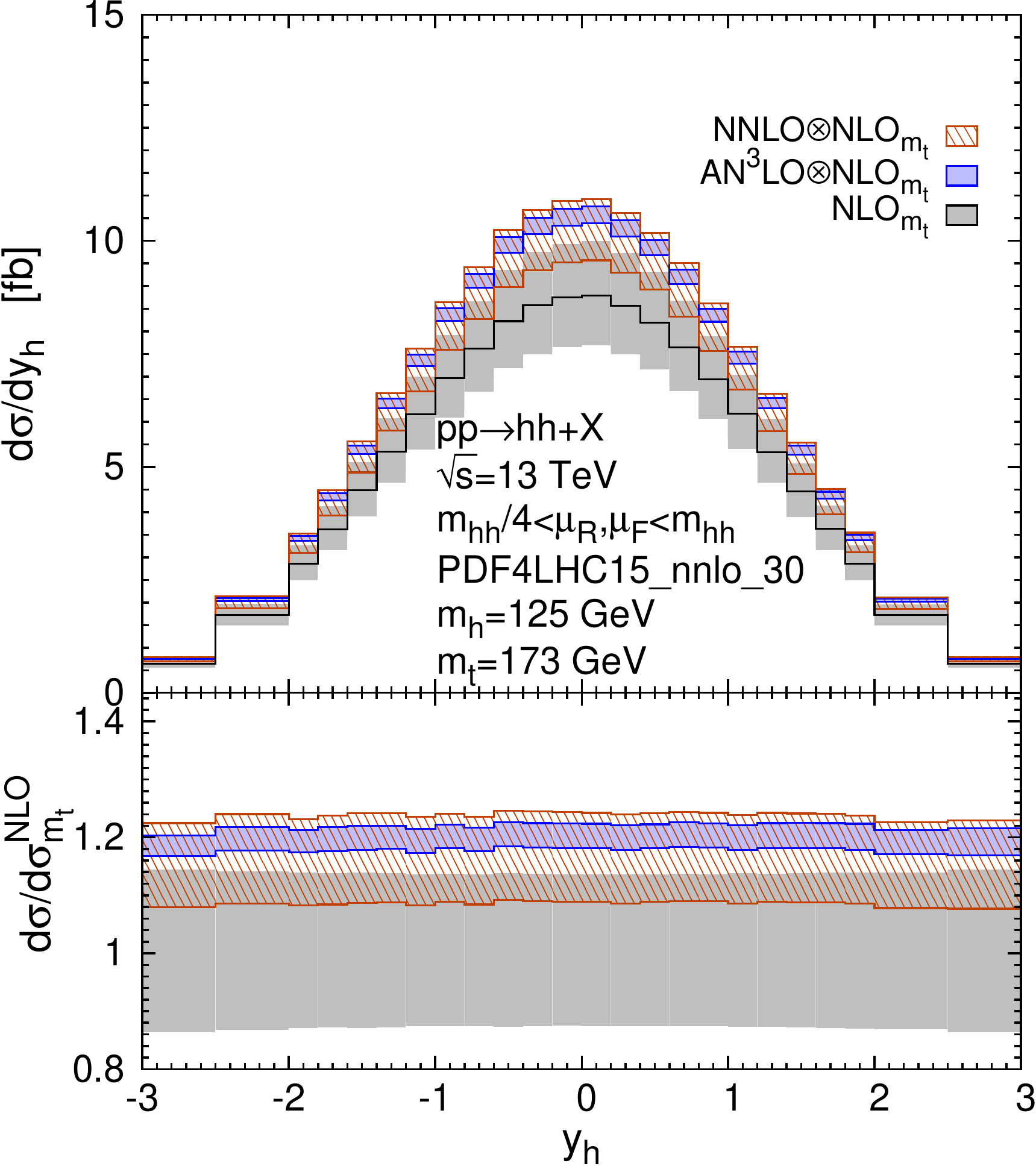}\\
    \includegraphics[scale=.38,draft=false,trim = 0mm 0mm 0mm 0mm,clip]{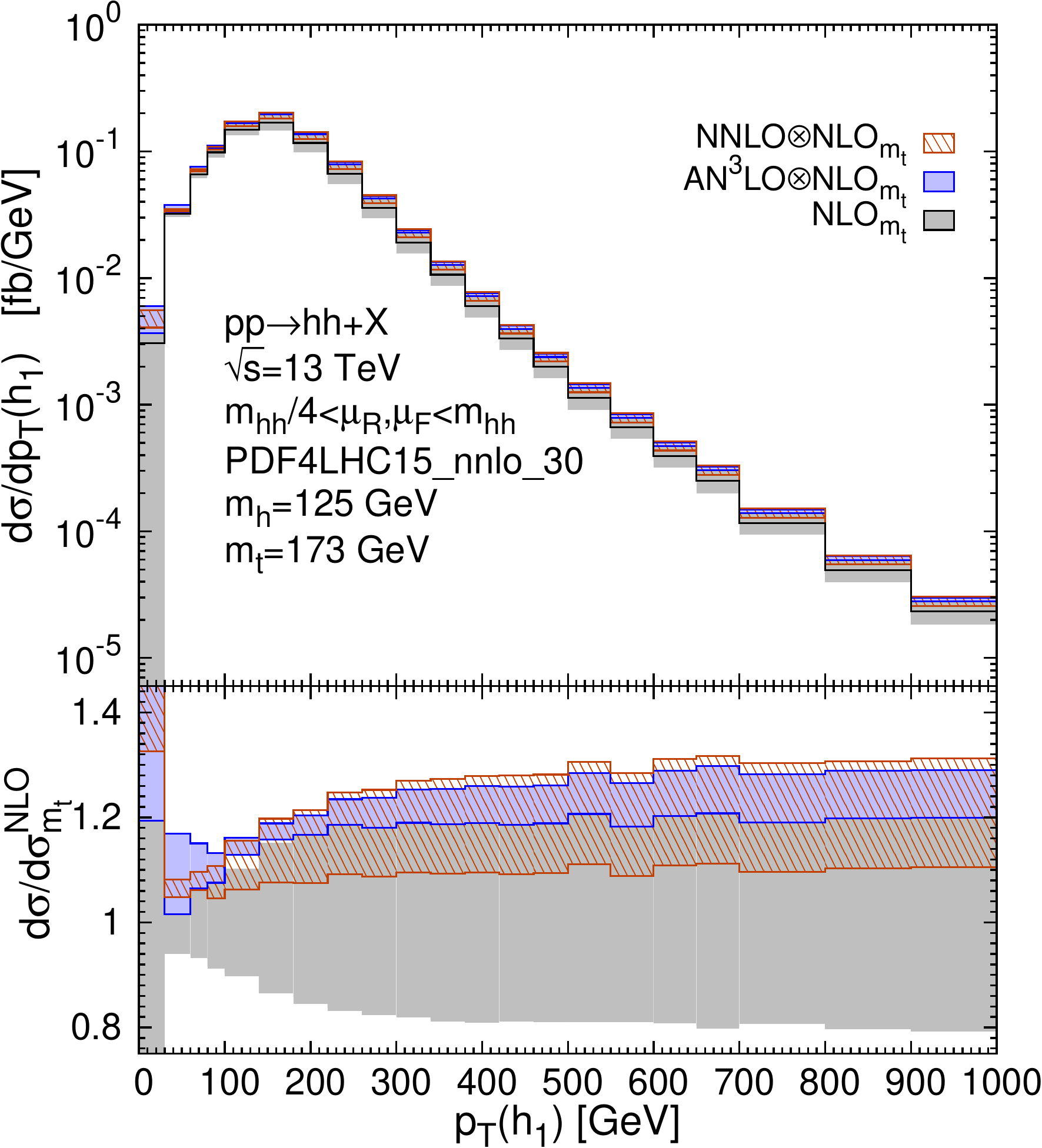}
    \includegraphics[scale=.38,draft=false,trim = 0mm 0mm 0mm 0mm,clip]{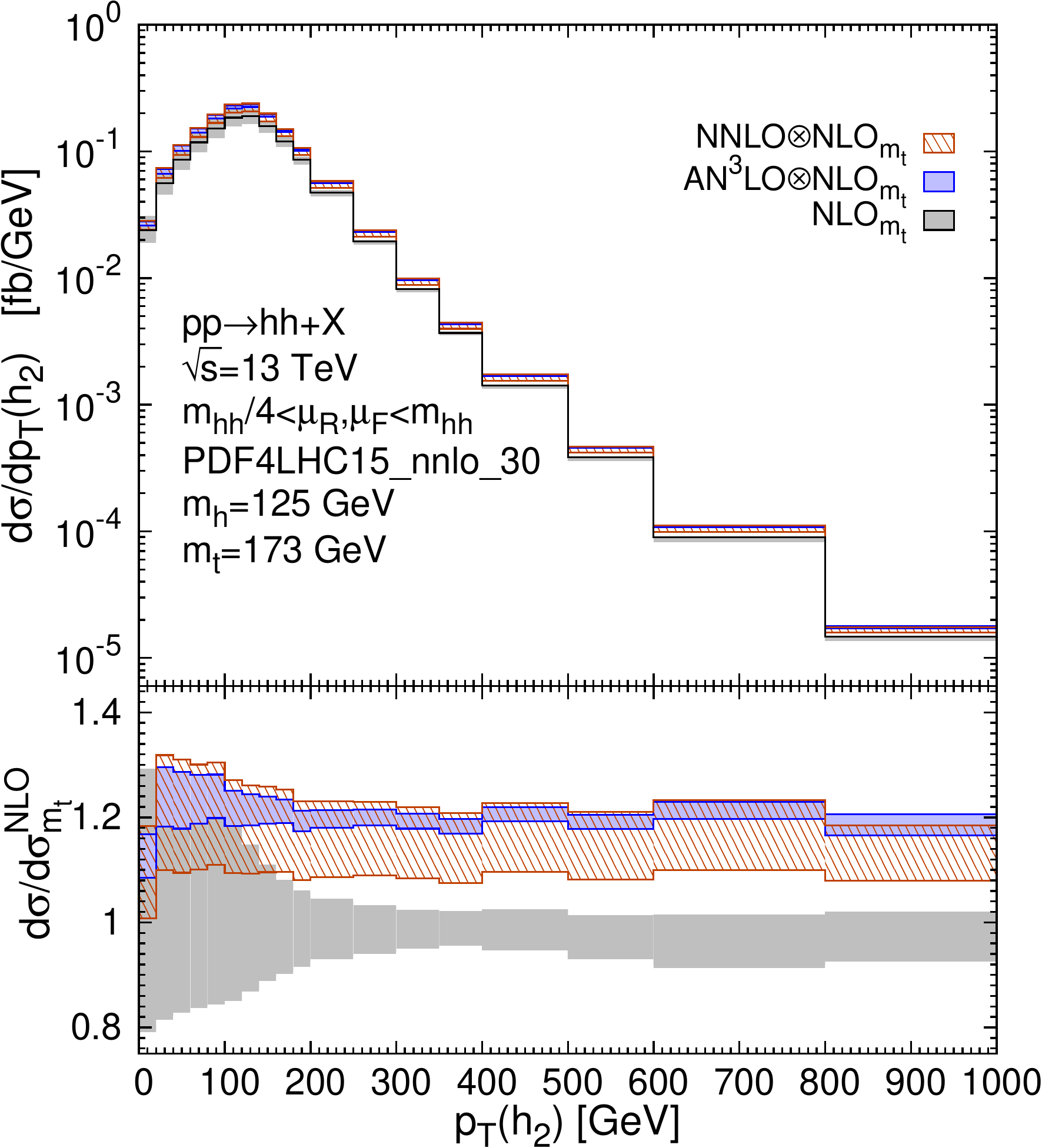}\\
    \includegraphics[scale=.38,draft=false,trim = 0mm 0mm 0mm 0mm,clip]{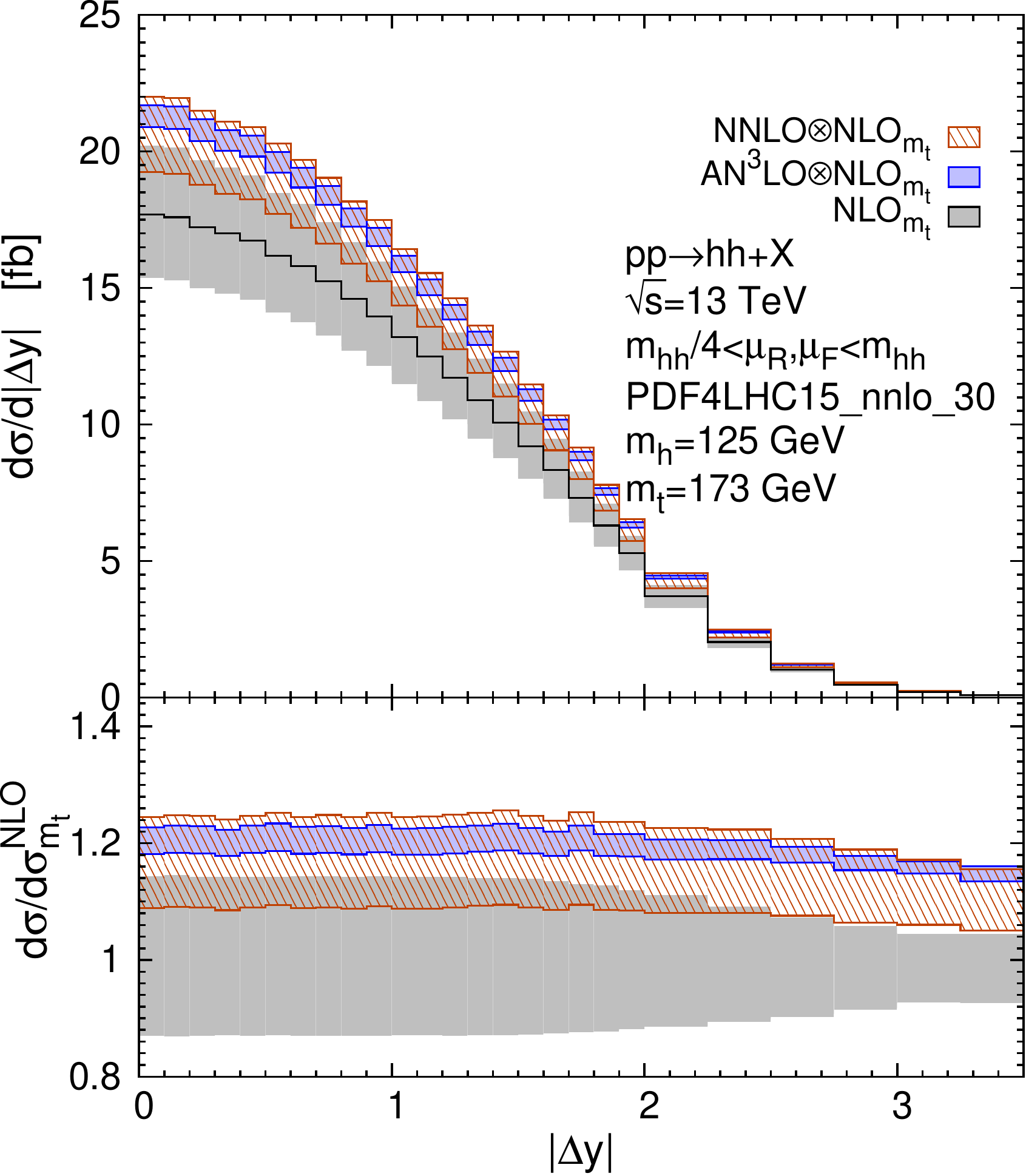}
    \includegraphics[scale=.38,draft=false,trim = 0mm 0mm 0mm 0mm,clip]{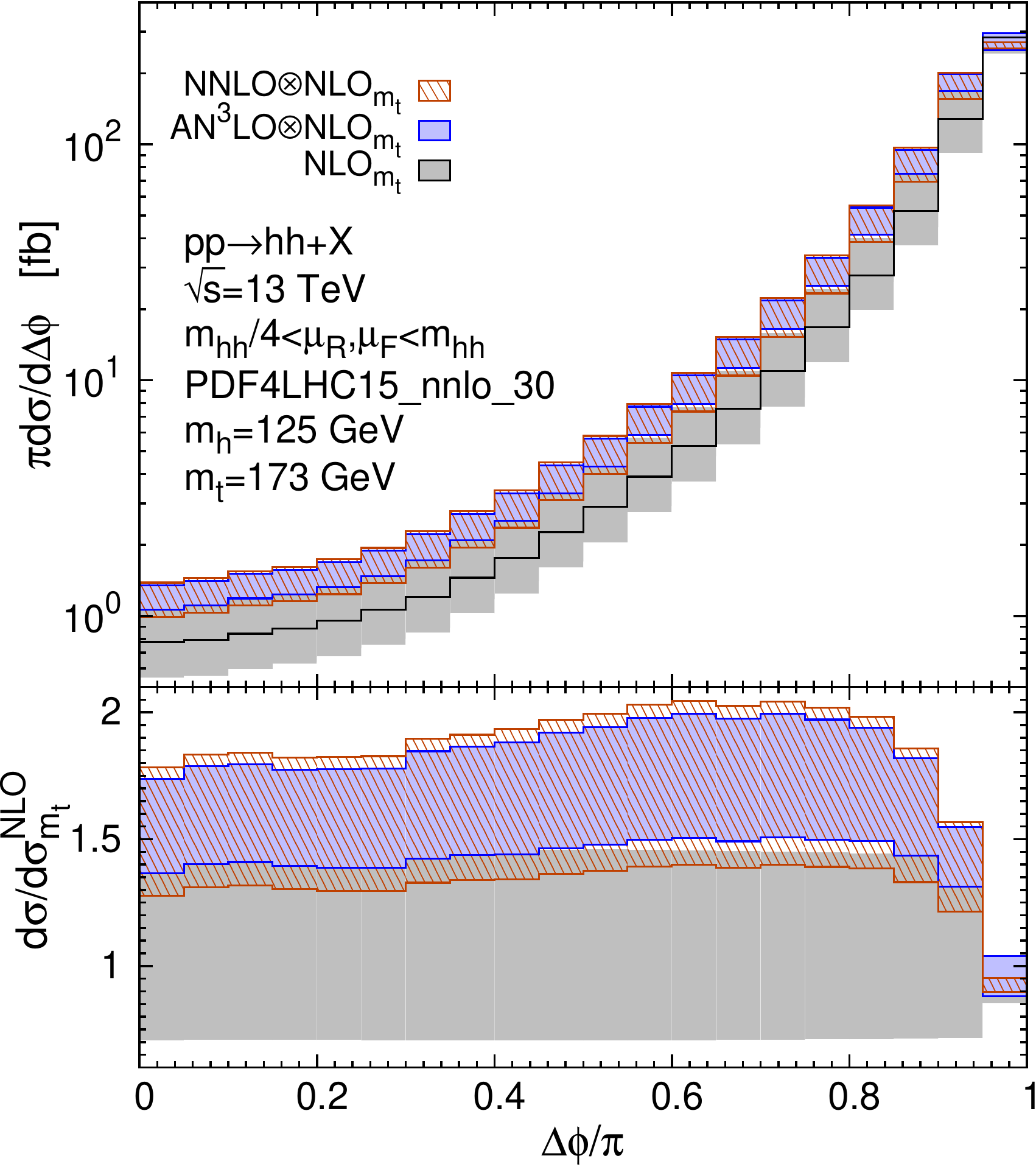}\\
    \vspace{0cm}
    \caption{Same as in figure~\ref{fig:othersMTLHC14} but at $\sqrt{s}=13$ TeV.}
    \label{fig:othersMTLHC13}
\end{figure}

\begin{figure}[h]
    \centering
    \includegraphics[scale=.38,draft=false,trim = 0mm 0mm 0mm 0mm,clip]{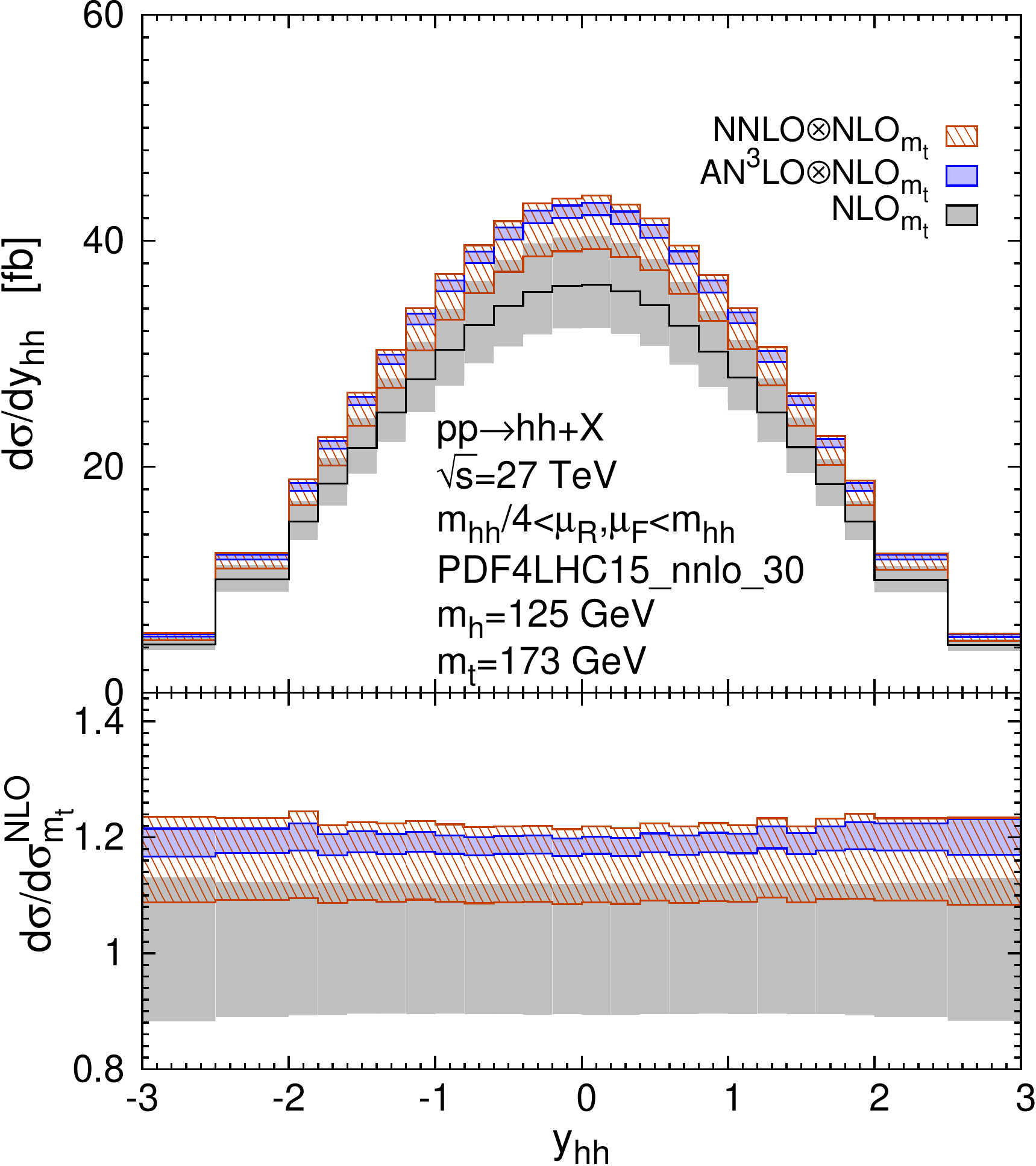}
    \includegraphics[scale=.38,draft=false,trim = 0mm 0mm 0mm 0mm,clip]{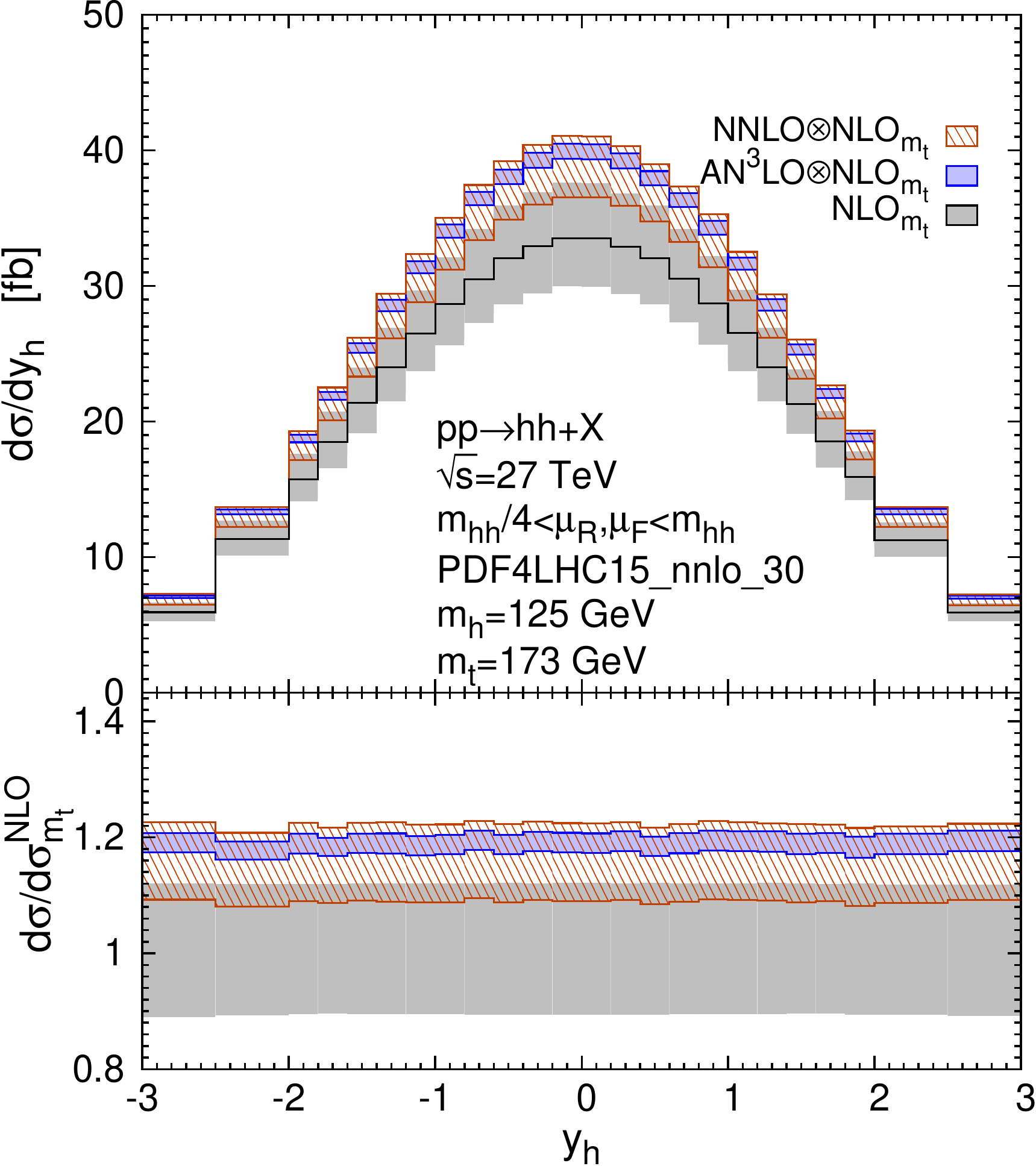}\\
    \includegraphics[scale=.38,draft=false,trim = 0mm 0mm 0mm 0mm,clip]{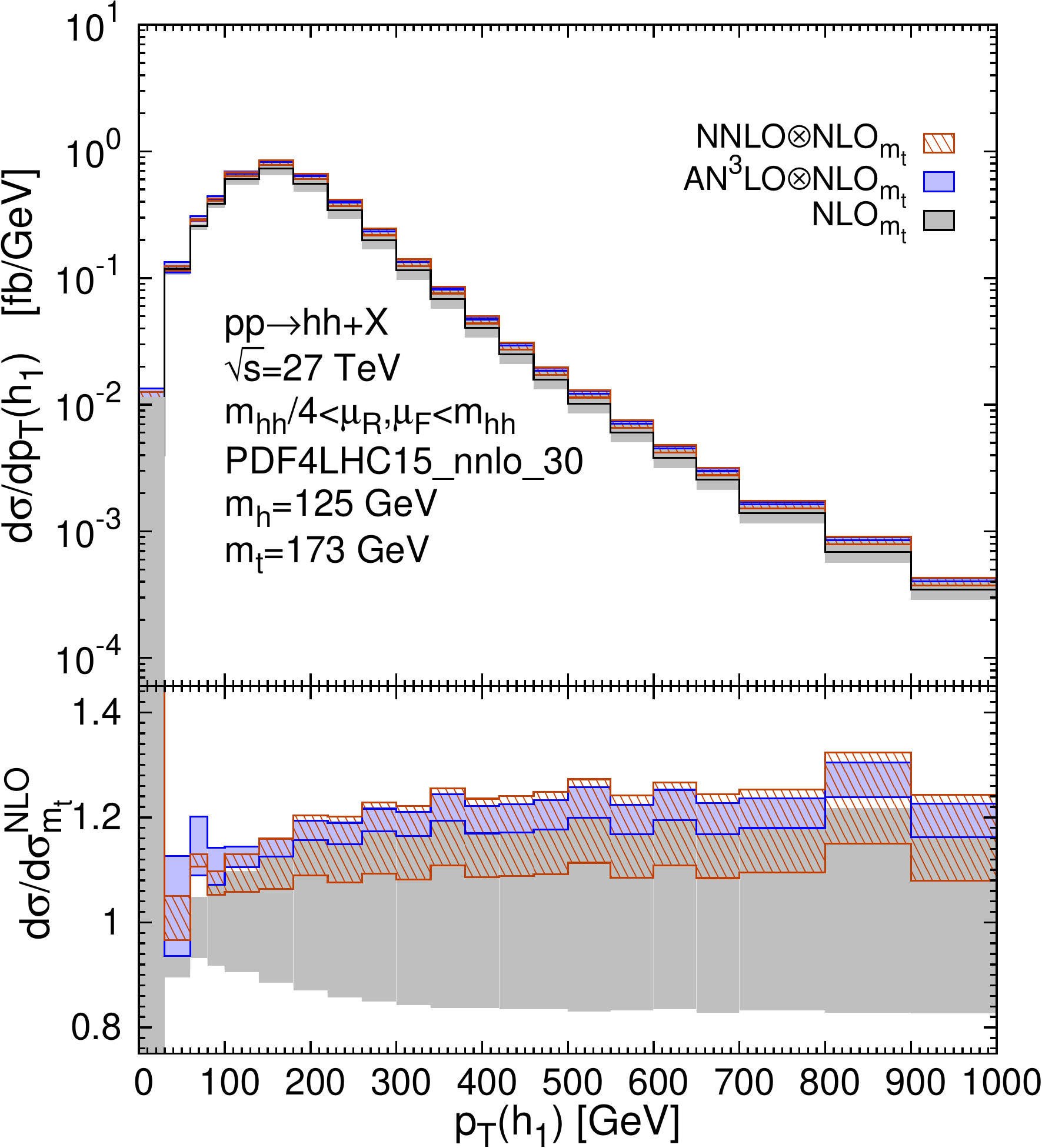}
    \includegraphics[scale=.38,draft=false,trim = 0mm 0mm 0mm 0mm,clip]{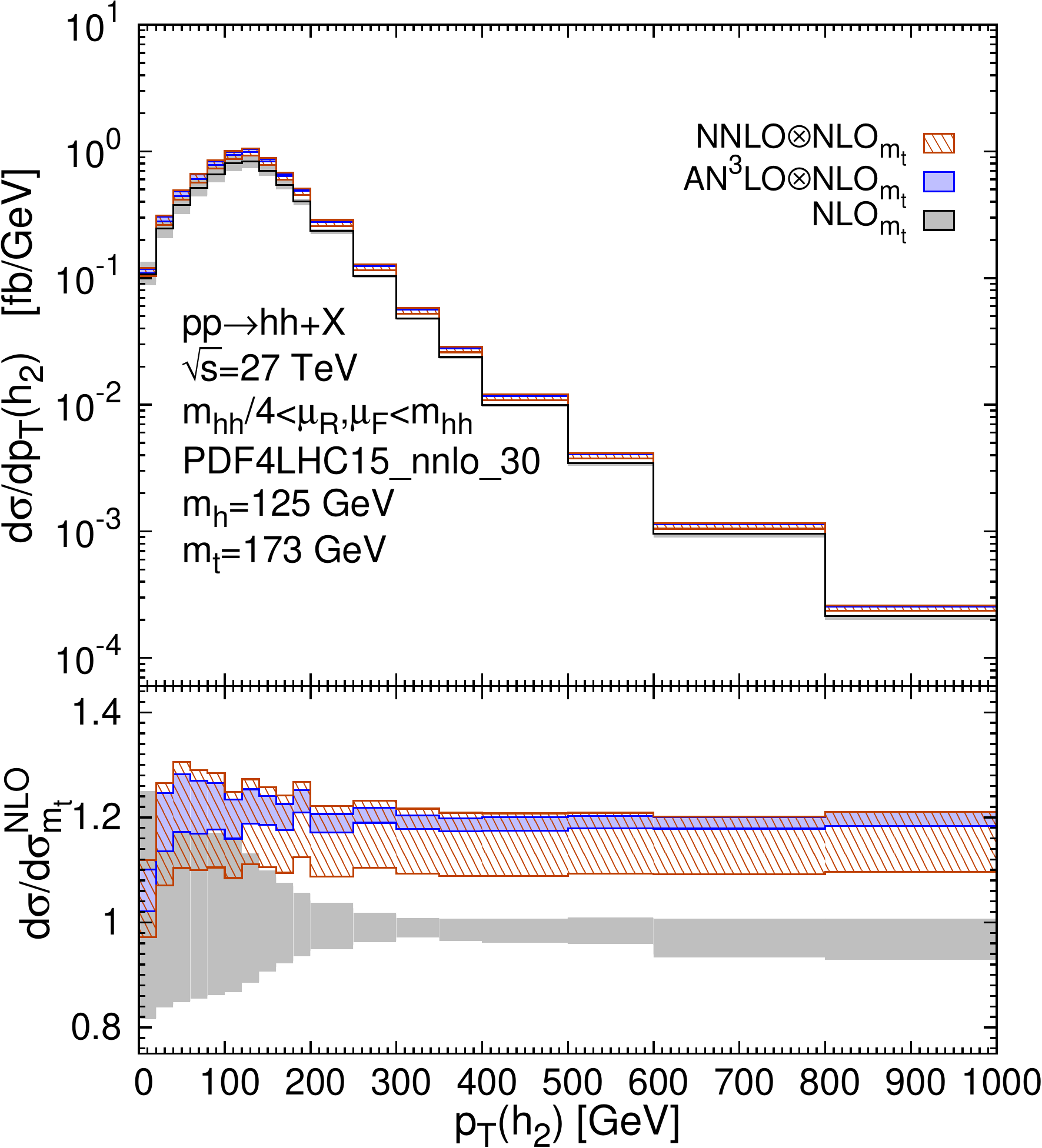}\\
    \includegraphics[scale=.38,draft=false,trim = 0mm 0mm 0mm 0mm,clip]{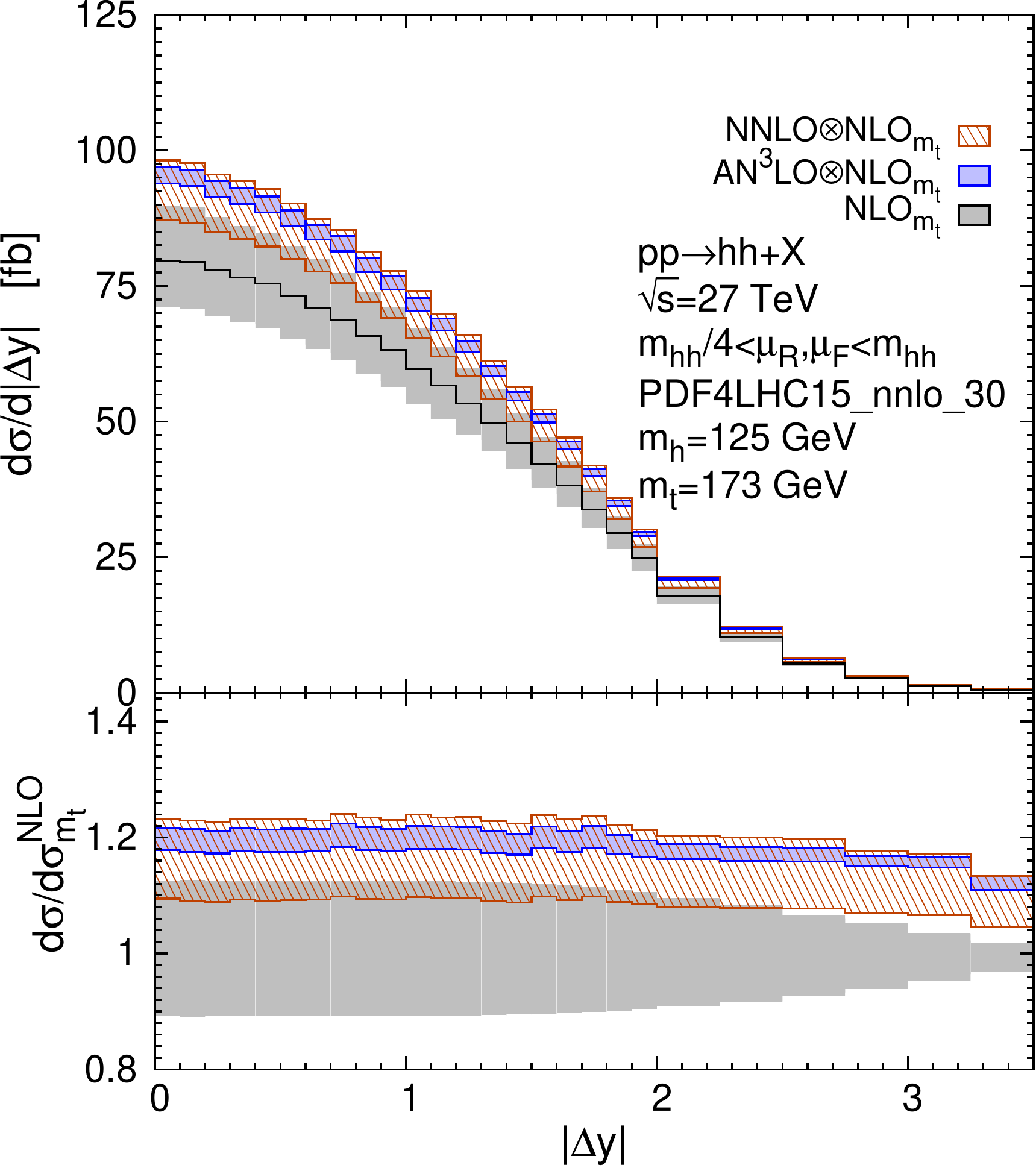}
    \includegraphics[scale=.38,draft=false,trim = 0mm 0mm 0mm 0mm,clip]{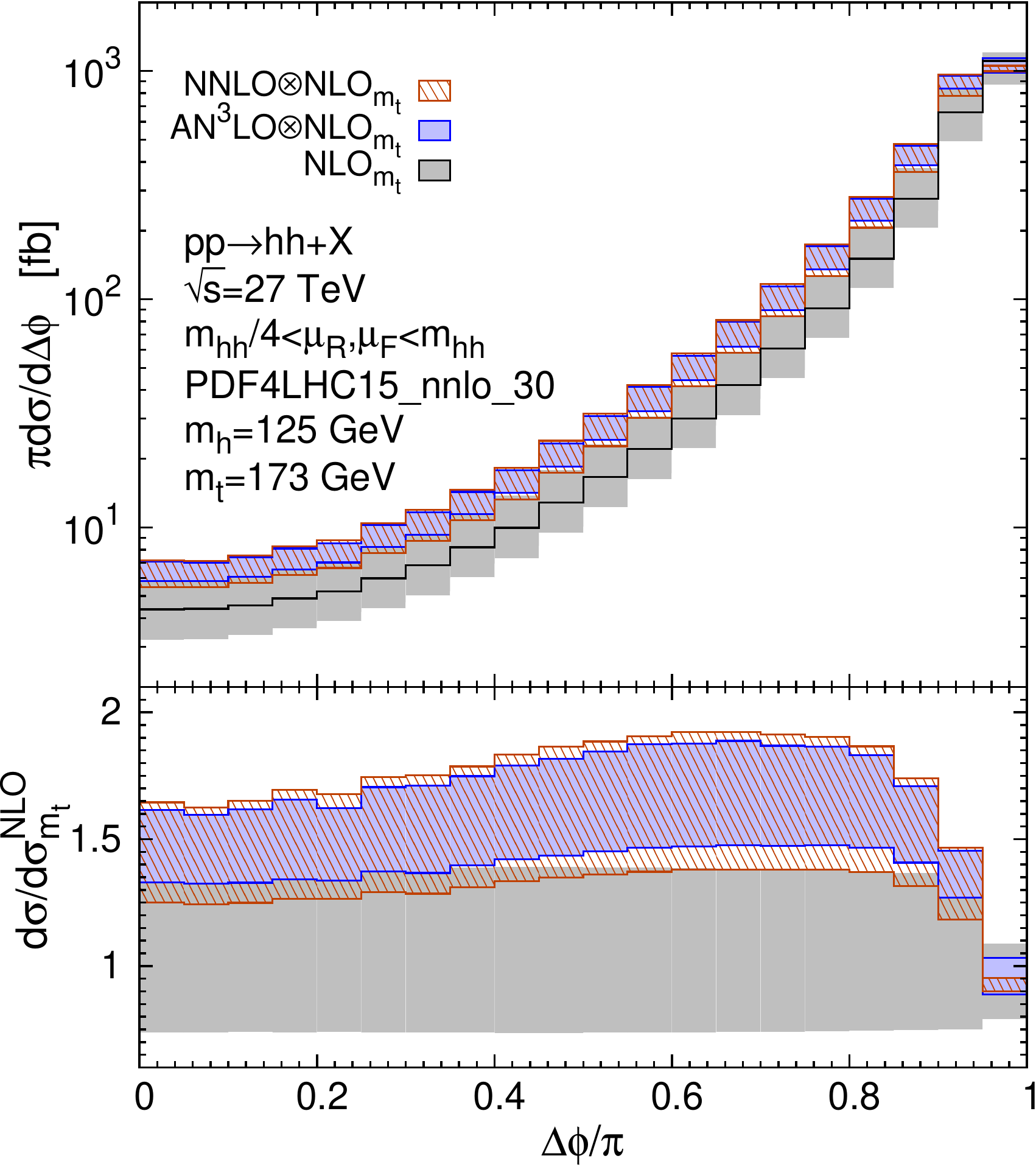}\\
    \vspace{0cm}
    \caption{Same as in figure~\ref{fig:othersMTLHC14} but at $\sqrt{s}=27$ TeV.}
    \label{fig:othersMTLHC27}
\end{figure}

\begin{figure}[h]
    \centering
    \includegraphics[scale=.38,draft=false,trim = 0mm 0mm 0mm 0mm,clip]{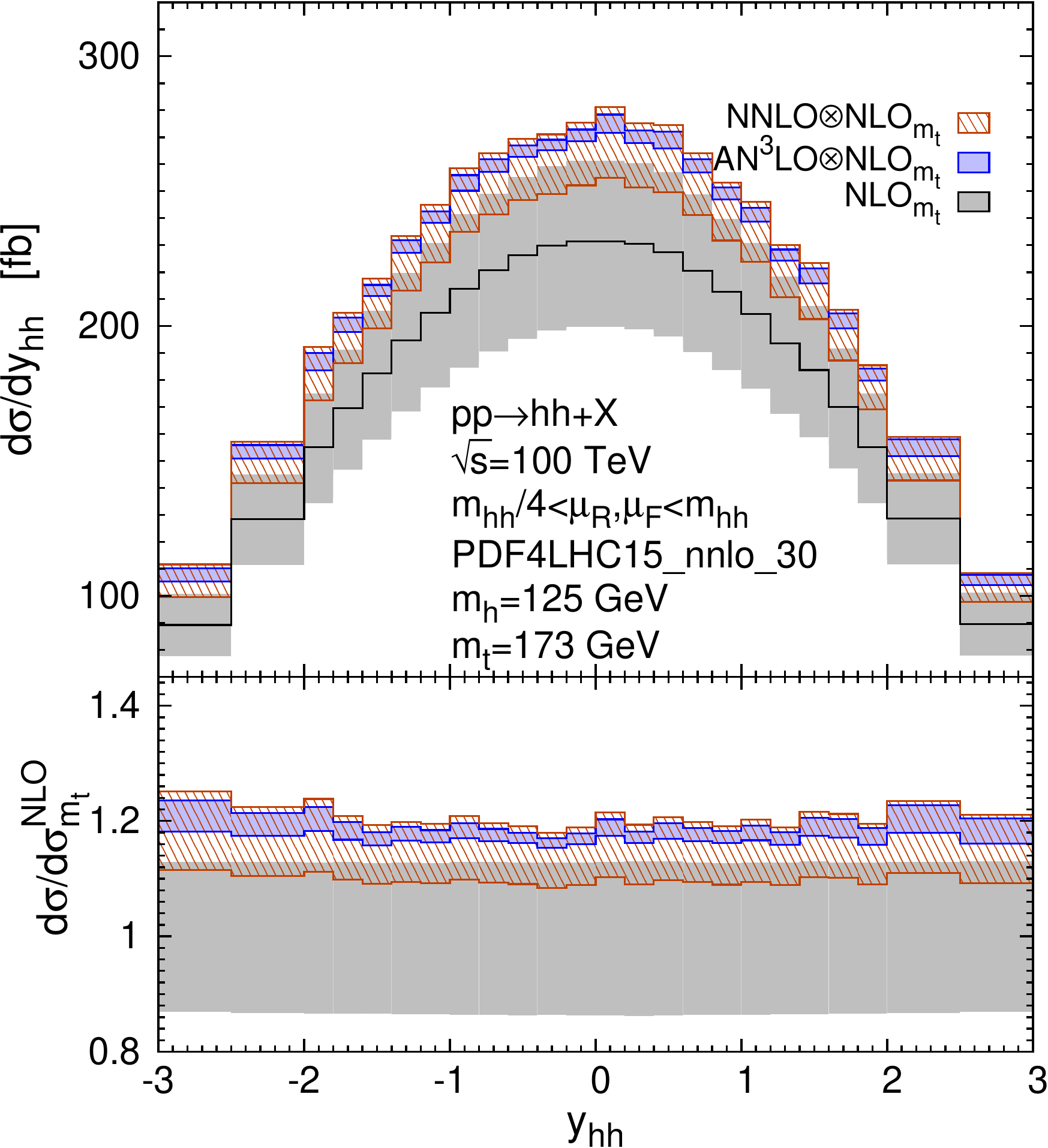}
    \includegraphics[scale=.38,draft=false,trim = 0mm 0mm 0mm 0mm,clip]{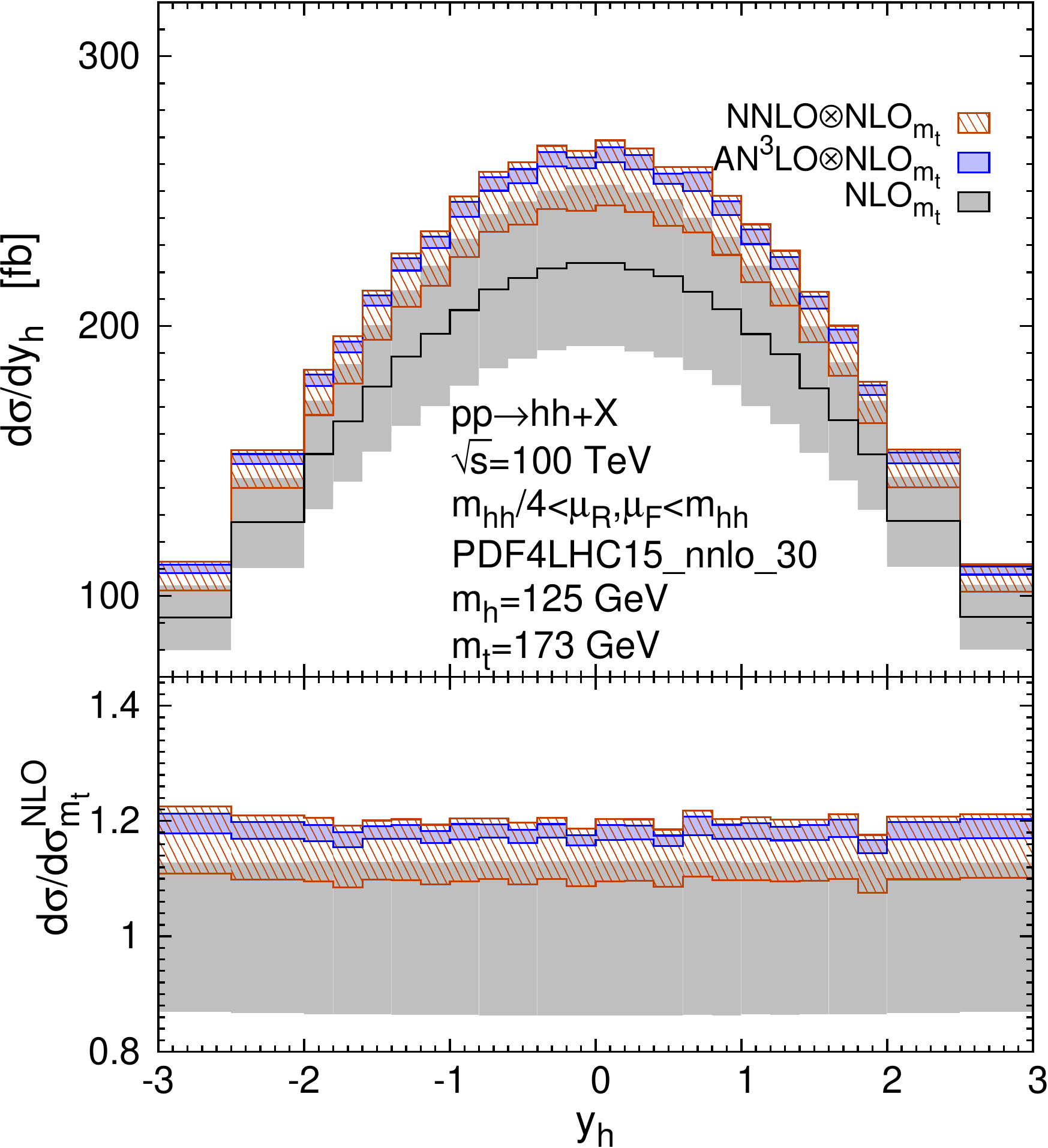}\\
    \includegraphics[scale=.38,draft=false,trim = 0mm 0mm 0mm 0mm,clip]{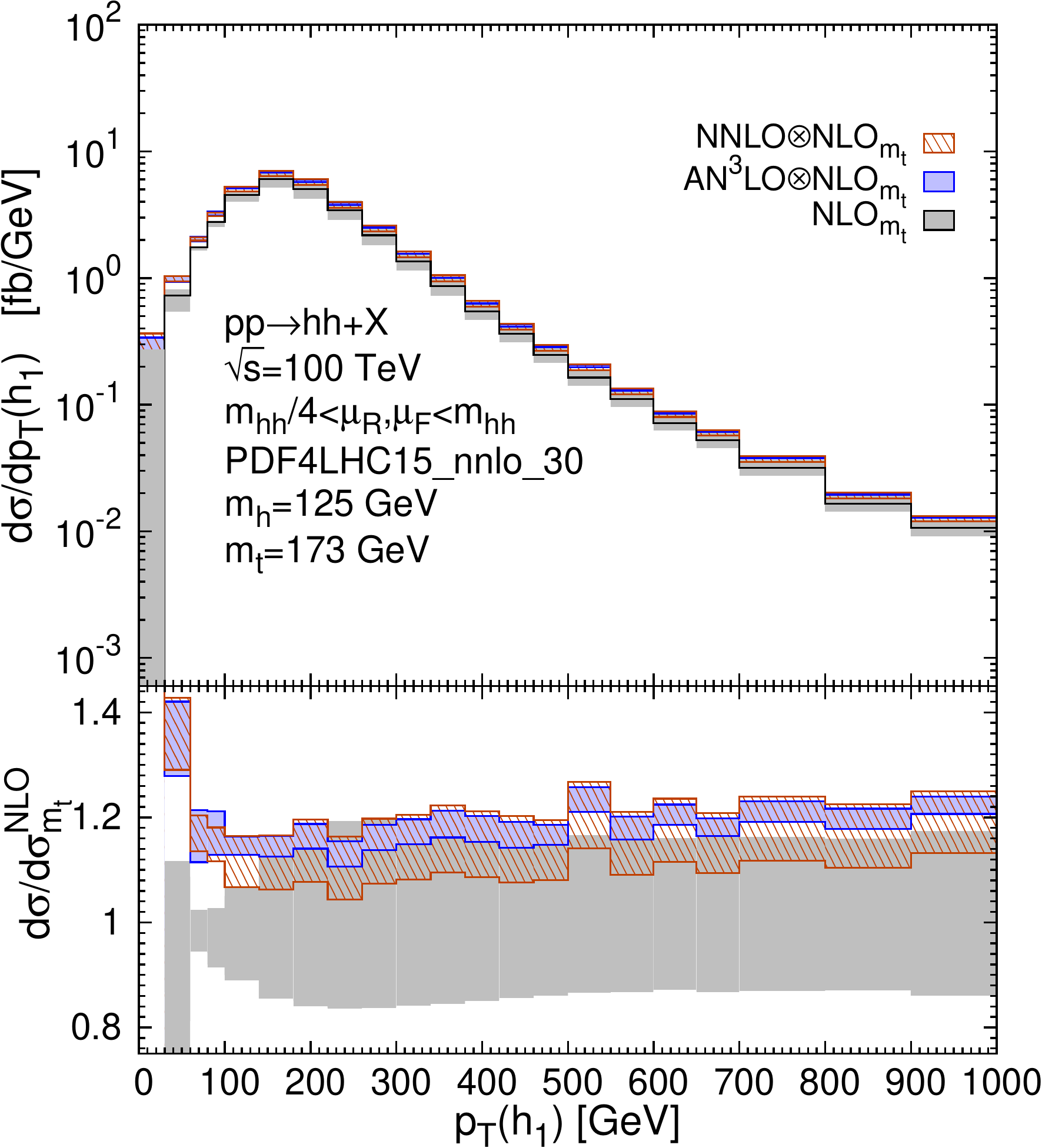}
    \includegraphics[scale=.38,draft=false,trim = 0mm 0mm 0mm 0mm,clip]{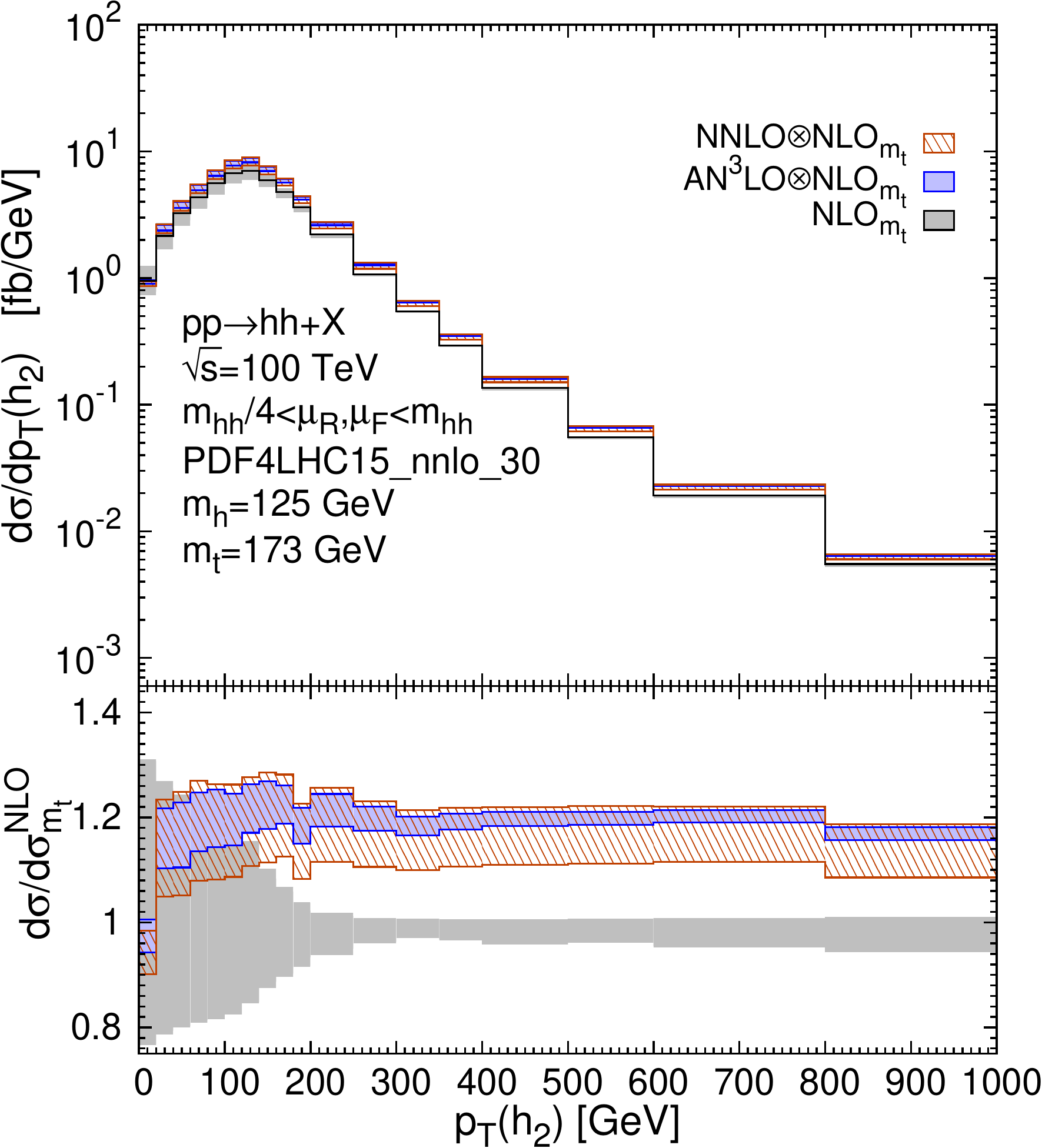}\\
    \includegraphics[scale=.38,draft=false,trim = 0mm 0mm 0mm 0mm,clip]{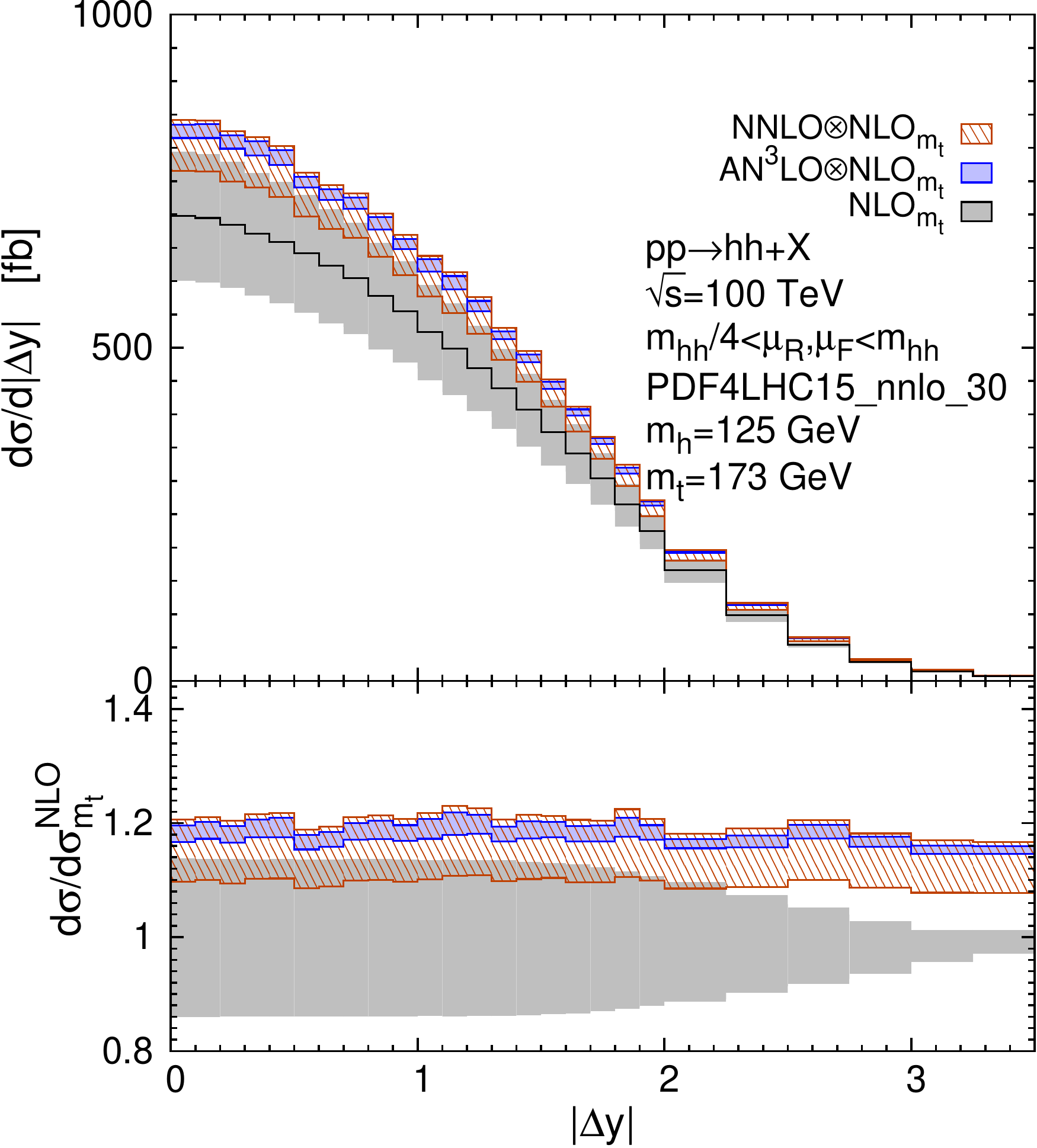}
    \includegraphics[scale=.38,draft=false,trim = 0mm 0mm 0mm 0mm,clip]{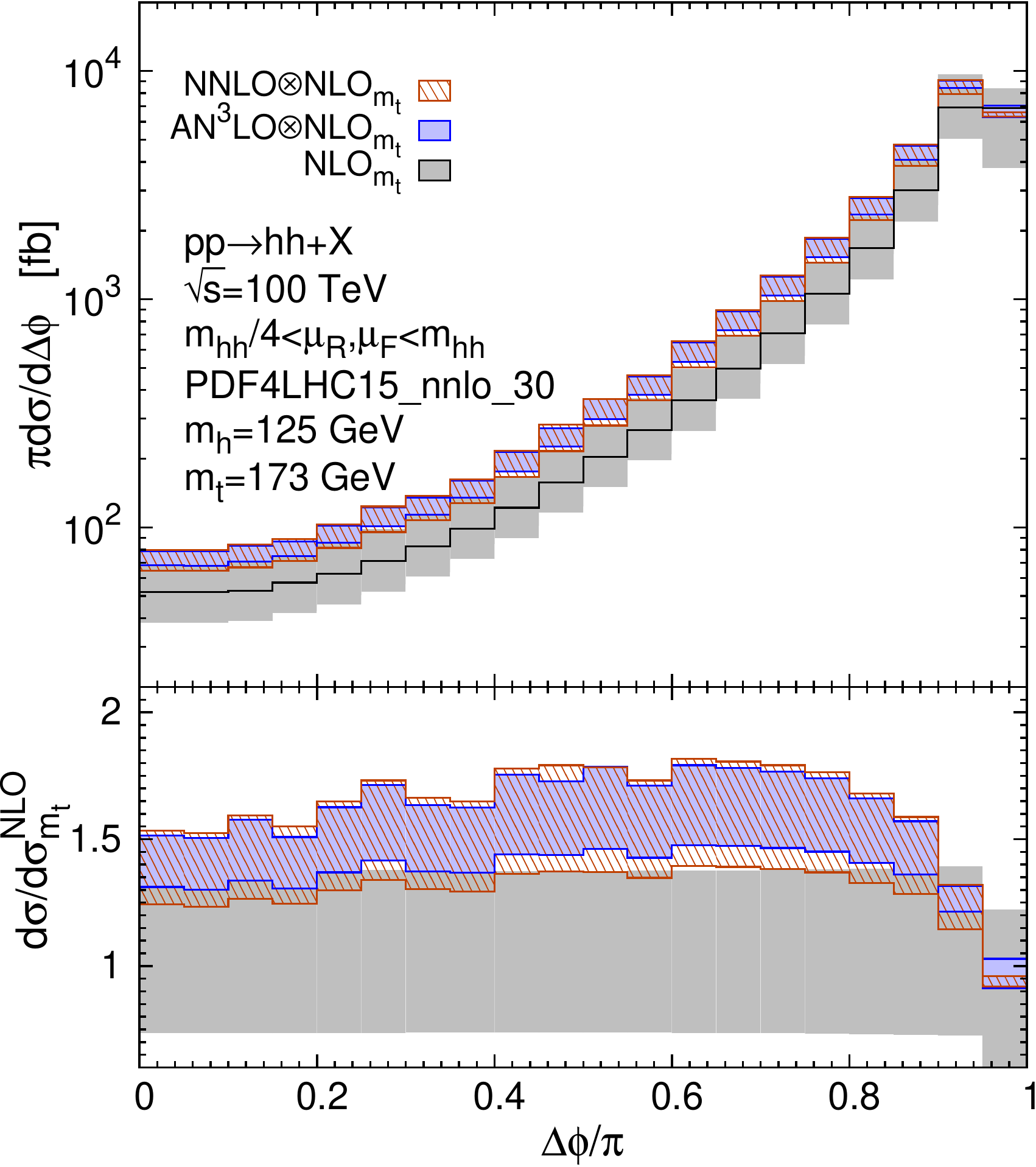}\\
    \vspace{0cm}
    \caption{Same as in figure~\ref{fig:othersMTLHC14} but at $\sqrt{s}=100$ TeV.}
    \label{fig:othersMTLHC100}
\end{figure}

We collect additional plots of 6 differential distributions from LO to AN$^3$LO in the infinite top-quark mass limit and with top-quark mass effects in this appendix. The distributions without finite $m_t$ corrections at $\sqrt{s}=13$ TeV (27 TeV, 100 TeV) can be found in figure~\ref{fig:othersLHC13} (figure~\ref{fig:othersLHC27}, figure~\ref{fig:othersLHC100}), while those with the NNLO$\otimes$ NLO$_{m_t}$ and AN$^3$LO$\otimes$NLO$_{m_t}$ approximations for finite $m_t$ corrections can be found in figures~\ref{fig:othersMTLHC13}, \ref{fig:othersMTLHC27}, \ref{fig:othersMTLHC100}.

\bibliographystyle{JHEP}
\bibliography{bib}





\end{document}